\documentclass[a4paper,11pt]{article}

\usepackage{jheppub}
\usepackage[T1]{fontenc}
\usepackage[utf8]{inputenc}

\usepackage{mathrsfs}                   
\usepackage[dvipsnames]{xcolor}         
\usepackage{booktabs}                   
\usepackage{multirow}                   
\usepackage{tikz}                       
\usepackage[compat=1.1.0]{tikz-feynman} 
\usepackage{lipsum}                     
\usepackage{rotating}                   
\usepackage{slashed}                    
\usepackage{mathtools}
\usepackage{simplewick}
\usepackage{simpler-wick}
\usepackage{adjustbox}
\usepackage[section]{placeins}
\usepackage{fontawesome}
\usepackage{multirow}

\usepackage[alsoload=hep,detect-all]{siunitx}
\newcommand{\mynum}[1]{
  \num[
  scientific-notation = true,
  round-mode = figures,
  round-precision = 1,
  exponent-product = \cdot
  ]{#1}
}

\usepackage{array}
\usepackage{makecell}

\usepackage{pifont}

\newcommand{\xmark}{\text{\ding{55}}}

\usepackage{caption}
\usepackage{subcaption}
\usepackage{longtable}

\usetikzlibrary{decorations.markings}
\tikzset{W->-/.style={decoration={
  markings,
  mark=at position 0.5*\pgfdecoratedpathlength+2pt with
  {\draw[-latex] (-2pt,0pt) -- (1pt,0pt);}},postaction={decorate}},
  W-<-/.style={decoration={
  markings,
  mark=at position 0.5*\pgfdecoratedpathlength with
  {\draw[latex-] (-2pt,0pt) -- (1pt,0pt);}},postaction={decorate}}
  }
\pgfkeys{
  /simplerwick/.cd,
  arrows/.store in=\LstWickArrows,
  arrows/.initial={-,-,-,-,-,-,-,-,-}, 
  arrows={-,-,-,-,-,-,-,-,-},
  positions/.store in=\LstWickPositions,
  positions={+1,+1,+1,+1,+1,+1,+1,+1,+1},
  positions/.initial={+1,+1,+1,+1,+1,+1,+1,+1,+1},
}

\makeatletter
\newcounter{Wick@up}
\newcounter{Wick@down}
\def\swick@end#1#2{
  \swick@setfalse@#1
  \tikzexternaldisable
  \begin{tikzpicture}[remember picture, baseline=(swick-close#1.base)]
    \node[use as bounding box, inner sep=0pt, outer sep=0pt] (swick-close#1) {$\displaystyle #2$};
  \end{tikzpicture}
  \tikz[remember picture, overlay]
{
\xdef\myW@style{\empty}
\foreach \W@X[count=\W@C] in \LstWickArrows
{\ifnum\W@C=#1
\xdef\myW@style{\W@X}
\fi}
\ifx\myW@style\empty
\PackageWarning{simpler-wick}{%
The list arrows has not enough entries!%
}{}
\xdef\myW@style{-}
\fi
\xdef\myW@pos{-77}
\foreach \W@X[count=\W@C] in \LstWickPositions
{\ifnum\W@C=#1
\xdef\myW@pos{\W@X}
\fi}
\ifnum\myW@pos=-77
\PackageWarning{simpler-wick}{%
The list positions has not enough entries!%
}{}
\xdef\myW@pos{+1}
\fi
\ifnum\myW@pos=-1
    \draw[\myW@style] ($(swick-open#1.south) + (0, -3pt)$)
          -- ($(swick-open#1.base) + (0, -\swick@offset) + \theWick@down*(0, -\swick@sep)$)
          -- ($(swick-close#1.base) + (0, -\swick@offset) + \theWick@down*(0, -\swick@sep)$)
          -- ($(swick-close#1.south) + (0, -3pt)$);
\stepcounter{Wick@down}
\else
\stepcounter{Wick@up}
    \draw[\myW@style] ($(swick-open#1.north) + (0, 3pt)$)
          -- ($(swick-open#1.base) + (0, \swick@offset) + \theWick@up*(0, \swick@sep)$)
          -- ($(swick-close#1.base) + (0, \swick@offset) + \theWick@up*(0, \swick@sep)$)
          -- ($(swick-close#1.north) + (0, 3pt)$);
\fi}
  \tikzexternalenable}
\def\wick@[#1]#2{\setcounter{Wick@up}{0}
\setcounter{Wick@down}{-1}
  \ifmmode
    \begingroup
    \pgfkeys{
        simplerwick,
        #1}
    \swick@cond@reset
    \swick@count=0
    \def\swick@max{0}
    \def\c{\swick@smart}
    #2
    \dimen0=\swick@sep
    \multiply\dimen0 by \swick@max
    \advance\dimen0 by \swick@offset
    \vbox to \dimen0{}
    \swick@cond@any{
      \PackageWarning{simpler-wick}{%
        I have reached the end of \protect\wick\space with some unclosed
        contractions%
      }{}
    }{}
    \endgroup
  \else
    \PackageWarning{simpler-wick}{%
      \protect\wich\space has been called outside a math environment, this will
      be ignore%
    }
  \fi
}
\makeatother

\usepackage{array}
\newcolumntype{B}{>{\centering\arraybackslash}m{6cm}}
\newcolumntype{M}{>{\centering\arraybackslash}m{3cm}}
\newcolumntype{S}{>{\centering\arraybackslash}m{1.5cm}}

\title{Exploding operators for Majorana neutrino masses and beyond}

\author{John Gargalionis}
\author{and Raymond R. Volkas}
\affiliation{ARC Centre of Excellence for Particle Physics at the
  Terascale,\\School of Physics, The University of Melbourne, Victoria 3010, Australia}

\emailAdd{garj@student.unimelb.edu.au}
\emailAdd{raymondv@unimelb.edu.au}

\abstract{Building UV completions of lepton-number-violating effective operators
  has proved to be a useful way of studying and classifying models of Majorana
  neutrino mass. In this paper we describe and implement an algorithm that
  systematises this model-building procedure. We use the algorithm to generate
  computational representations of all of the tree-level completions of the
  operators up to and including mass-dimension 11. Almost all of these
  correspond to models of radiative neutrino mass. Our work includes operators
  involving derivatives, updated estimates for the bounds on the new-physics
  scale associated with each operator, an analysis of various features of the
  models, and a look at some examples. We find that a number of operators do not
  admit any completions not also generating lower-dimensional operators or
  larger contributions to the neutrino mass, ruling them out as playing a
  dominant role in the neutrino-mass generation. Additionally, we show that
  there are at most five models containing three or fewer exotic multiplets that
  predict new physics that must lie below 100 TeV. Accompanying this work we
  also make available a searchable database containing all of our results and
  the code used to find the completions. We emphasise that our methods extend
  beyond the study of neutrino-mass models, and may be useful for generating
  completions of high-dimensional operators in other effective field theories.{
    \begin{center}
      Example code: \href{https://github.com/johngarg/neutrinomass}{\large\faGithub}
    \end{center}
  }
}

\begin{document}
\maketitle
\flushbottom

\section{Introduction}
\label{sec:intro}

Laboratory experiments to date have firmly established the predictive power of
the Standard Model (SM). Mass generation for the weak gauge bosons and charged
fermions is by now a familiar narrative, and the only clear terrestrial
measurements pointing to physics beyond the SM are those of neutrino flavour
transformations. Even here, observations are consistent with the prevailing
orthodoxy: the neutrino flavour eigenstates are unitary superpositions of
mass eigenstates and the probability of detecting a neutrino of a given flavour
oscillates with distance. On the origin of these neutrino oscillations --- and
the non-zero neutrino masses they imply --- the SM has nothing to say.
Oscillation experiments have shown that the mixing in the lepton sector is of a
different structure and extent to that seen in quarks; and measurements from
cosmology, neutrinoless double-beta decay and tritium beta decay strongly
constrain the absolute scale of the neutrino masses. These facts lend themselves
to the possibility that an alternate mass-generating mechanism is operating for
the uncharged leptons.

A characteristic feature of the neutrinos is that they are the only fermions in
the SM that could acquire a Majorana mass, as long as lepton number isn't
endowed with any special significance. Many models pursue this line of
reasoning, with the neutrinos acquiring a Majorana mass through the
lepton-number-violating (LNV) interactions of heavy exotica. The most famous
examples are the three canonical seesaw models~\cite{MINKOWSKI1977421,
  Yanagida:1979as, GellMann:1980vs, PhysRevLett.44.912, Glashow:1979nm,
  Magg:1980ut, PhysRevD.22.2227, LAZARIDES1981287, Wetterich:1981bx,
  PhysRevD.23.165, Foot:1988aq} that generate the dimension-five Weinberg
operator $(L^iL^j)H^kH^l \epsilon_{ik}\epsilon_{jl}$ at tree-level upon
integrating out the heavy fields. Additionally, the historically important
Zee~\cite{Zee:1980ai} and Zee--Babu~\cite{Zee:1985id, Babu:1988ki} models have
come to be archetypal radiative scenarios in which interactions violating
lepton-number by two units ($\Delta L = 2$) generate a Majorana mass for the
neutrinos at loop level. Such models are economic, since they do not require the
imposition of \textit{ad hoc} symmetries, and in many cases make a connection to
other unsolved problems of the SM such as the nature of dark matter or the
matter--antimatter asymmetry of the Universe. They are also elegant, since the
smallness of the neutrino masses emerges as a natural consequence, rather than
through the imposed requirement of exceedingly small coupling constants. For
recent reviews of radiative models see Refs.~\cite{Boucenna:2014zba,
  Cai:2017jrq}.

Although the seesaw models are attractive solutions to the neutrino mass
problem, they are difficult to test experimentally. The region of their
parameter space in which the seesaw field's couplings to the SM are very small
can be probed at colliders~\cite{Cai:2017mow}, although for $\mathcal{O}(1)$
couplings the seesaw scale is predicted to be $\sim 10^{12}~\text{TeV}$.
Radiative models are easier to probe experimentally since the additional loop
suppression and products of couplings bring down the allowed scale of the new
physics, in some cases to LHC-accessible energy ranges~\cite{deGouvea:2007qla}.
The two-loop Zee--Babu model, for example, is non-trivially constrained by
same-sign dilepton searches performed by ATLAS~\cite{ATLAS:2012hi,
  ATLAS:2014kca, Aaboud:2017qph} and CMS~\cite{Chatrchyan:2012ya, CMS:2016cpz,
  CMS:2017pet}, but it is only one of a very large number of radiative models,
none of which are \textit{a priori} more likely to be true than any other. In
the context of such a large theory-space, it is useful to have an organising
principle to aid in the study and classification of these models, and beginning
with $\Delta L = 2$ effective operators has been shown to be an effective
strategy.

One approach to this model taxonomy involves studying loop-level completions of
the Weinberg operator, and its dimension-($5+2n$) generalisations
$\mathcal{O}_1^{\prime \cdots \prime} = (LL)HH(H^\dagger H)^n$. Here, models can
be systematically written down by studying the various topologies able to be
accommodated by the operator with increasing number of loops. This is done in
such a way that models implying lower-order contributions to the neutrino mass
can be discarded~\cite{Farzan:2012ev}. Such an approach has been applied to the
Weinberg operator up to three loops~\cite{Bonnet:2012kz, Sierra:2014rxa,
  Cepedello:2018rfh} and to its dimension-seven generalisation at one
loop~\cite{Cepedello:2017eqf}. An alternative and complementary method begins by
considering all of the gauge-invariant $\Delta L = 2$ operators in the SM
effective field theory (SMEFT), first listed in this context by Babu and Leung
(BL)~\cite{Babu:2001ex} and extended by de Gouv\^{e}a and Jenkins
(dGJ)~\cite{deGouvea:2007qla}. Supposing that the tree-level coefficient of one
of these is non-zero at the high scale, neutrino masses will be generated from
loop graphs contributing to the mixing of this operator and the Weinberg-like
operators $\mathcal{O}_1^{\prime \cdots \prime}$. The process of expanding the
operator into a series of UV-complete, renormalisable models that generate the
parent operator at tree-level is called \emph{opening up} or \emph{exploding}
the operator. The remaining external fields must be looped-off, with additional
loops of SM gauge bosons or Higgs fields added as necessary in order to obtain a
neutrino self-energy diagram. A model-building formula along these lines has
been formulated in Ref.~\cite{PhysRevD.87.073007}, and it has been used to write
down all of the minimal, tree-level UV-completions of $\Delta L = 2$ operators
at dimension seven~\cite{Cai:2014kra} corresponding to tree-level and radiative
neutrino-mass models. The tree-level completions of the Weinberg-like operators
have been written down up to dimension eleven~\cite{Cai:2014kra, Bonnet:2009ej,
  Anamiati:2018cuq}.

Our analysis continues in the tradition Refs.~\cite{Babu:2001ex,
  deGouvea:2007qla, PhysRevD.87.073007, Cai:2014kra}, but where appropriate we
make a connection to the results from loop-level matching~\cite{Bonnet:2012kz,
  Sierra:2014rxa, Cepedello:2017eqf, Cepedello:2018rfh} for completeness. We
consider that there is complementary insight to be gained from thorough and
complete analyses involving both approaches. Building models from tree-level
completions of the $\Delta L = 2$ operators allows for a direct connection to be
made between the neutrino-mass mechanism and other lepton-number-violating
phenomena. The models derived in this way are also minimal in the sense that
they involve the fewest number of exotic fields required to furnish a given
loop-level topology, since the neutrino self-energy graphs always involve some
SM fields. This has a number of important implications. First, the neutrino
masses depend on SM parameters, and their rough scale can therefore be readily
estimated from the effective operator alone. Second, neutrino-mass mechanisms
containing SM gauge bosons are included automatically, and these constitute a
large fraction of the models. Finally, it also means that our approach never
produces models that contain loops of only exotic fields, although these can be
added easily (see, for example, section IV.C of Ref.~\cite{PhysRevD.87.073007}).
The appeal of these models notwithstanding, a benefit of giving up heavy loops
is that the transformation properties of the beyond-the-standard-model particle
content of each model are now uniquely determined, and therefore the total
number of minimal models is finite. Minimal exotic particle content, in the
aforementioned sense, is an attractive feature of this approach. Indeed, there
are many examples of operators whose insertion and closure lead to neutrino
masses at dimension nine and higher, but for which the number of exotic degrees
of freedom introduced are not more than those of a garden-variety model
generating the Weinberg operator at the low scale. The consideration of such
equally simple models in the loop-level matching paradigm would require a
detailed analysis of the dimension-seven and dimension-nine analogues of the
Weinberg operator\footnote{One can always generate the dimension-five Weinberg
  operator from its analogues at dimensions seven, nine and eleven with
  additional Higgs loops, but these models usually contain more than three
  loops.} up to a large number of loops.

An economic classification scheme, separate from an EFT framework, was presented
in Ref.~\cite{Klein:2019iws} based on the number of exotic degrees of freedom by
which the SM is extended. There, the method is applied to the case of radiative
models with two exotics\footnote{Including models with one scalar and one Dirac
  fermion.}, and has also been used to study minimal neutrino-mass models
compatible with $\mathrm{SU}(5)$ unification~\cite{Klein:2019jgb}.

Here, we sharpen the model building prescription developed in
Ref.~\cite{PhysRevD.87.073007} and extend it to the case of operators involving
field-strength tensors and derivatives. This procedure is automated and applied
to all $\Delta L = 2$ operators in the SM effective field theory up to dimension
eleven. We classify the neutrino-mass topologies, completions and their exotic
fields. We also make available a database containing our main results and
example code used to generate the operators along with their completions and
Lagrangians~\cite{neutrinomass2020}. We emphasise that the usefulness of these
methods and tools extends beyond the study of neutrino mass and
lepton-number-violating phenomena. To illustrate this point we reproduce some
recent results of work listing completions of SMEFT
operators~\cite{deBlas:2017xtg}.

The remainder of the paper is structured as follows. Section~\ref{sec:notation}
sets out our mathematical conventions and notation.
Section~\ref{sec:treelevelmatching} contains a review of tree-level matching and
a description of the methods we use to find the tree-level completions of the
operators. Neutrino mass model building is described in
section~\ref{sec:modelbuilding}, while section~\ref{sec:models} presents a
preliminary analysis of the models along with some examples.

\section{Conventions}
\label{sec:conventions}

In this section we establish the conventions we employ throughout the rest of
the paper: the nomenclature of fields and indices, our operational semantics and
the classification of the lepton-number-violating operators on which our
analysis is based. We highlight that this classification differs mildly from
that found in earlier work, since our list includes additional structures as
well as operators containing derivatives. We find the operators containing
field-strength tensors to be uninteresting from the perspective of model
building --- a point justified in detail in
Sec.~\ref{sec:exploding-derivative-operators} --- and choose not to include them
in our classification in this section.

\subsection{Mathematical notation}
\label{sec:notation}

Throughout the paper we choose to label representations by their dimension,
which we typeset in bold. Multiplets are labelled by their transformation
properties under the Lorentz group and the SM gauge group
$\mathrm{SU}(3)_{c} \otimes \mathrm{SU}(2)_{L} \otimes \mathrm{U}(1)_{Y}$, and
we often refer to them simply as fields. All spinors are treated as
two-component objects transforming as either $(\mathbf{2}, \mathbf{1})$
(left-handed) or $(\mathbf{1}, \mathbf{2})$ (right-handed) under the Lorentz
group, written as $\mathrm{SU}(2)_{+} \otimes \mathrm{SU}(2)_{-}$. The
left-handed spinors carry undotted spinor indices
$\alpha, \beta, \ldots \in \{1, 2\}$, while the right-handed spinors carry
dotted indices $\dot{\alpha}, \dot{\beta}, \ldots \in \{\dot{1}, \dot{2}\}$.
Wherever possible we attempt to conform to the conventions of
Ref.~\cite{Dreiner:2008tw} when working with spinor fields (see appendix G for
the correspondence to four-component notation and appendix J for SM-fermion
nomenclature). For objects carrying a single spacetime index $V_\mu$ we define
\begin{equation}
  V_{\alpha \dot{\beta}} = \sigma^\mu_{\alpha \dot{\beta}} V_\mu \quad \text{
    and
  } \quad \bar{V}_{\dot{\alpha}\beta } = \bar{\sigma}^\mu_{\dot{\alpha}\beta} V_{\mu}.
\end{equation}
Note that in this notation
\begin{equation}
  \Box = \partial_{\mu} \partial^{\mu} = \tfrac{1}{2}\text{Tr}[\partial \bar{\partial}] = \tfrac{1}{2}\text{Tr}[\bar{\partial} \partial],
\end{equation}
and we will sometimes just use $\Box$ to represent the contraction of two
covariant derivatives $D_{\mu}D^{\mu}$ where this is clear from context. For
field-strength tensors, generically $X_{\mu\nu}$, we work with the irreducible
representations (irreps) $X_{\alpha \beta}$ and
$\bar{X}_{\dot{\alpha} \dot{\beta}}$, where
\begin{equation}
  X_{\{\alpha \beta\}} = 2i [\sigma^{\mu \nu}]^{~\gamma}_\alpha \epsilon_{\gamma \beta} X_{\mu \nu} \quad \text{ and } \quad
  \bar{X}_{\{\dot{\alpha} \dot{\beta}\}} = 2i [\bar{\sigma}^{\mu \nu}]^{\dot{\gamma}}_{~\dot{\beta}} \epsilon_{\dot{\alpha} \dot{\gamma}} X_{\mu \nu},
\end{equation}
or the alternate forms with one raised and one lowered index.

Indices for $\mathrm{SU}(2)_{L}$ (isospin) are taken from the middle of the
Latin alphabet. These are kept lowercase for the fundamental representation for
which $i, j, k, \ldots \in \{1, 2\}$ and the indices of the adjoint are
capitalised $I, J, K, \ldots \in \{1, 2, 3\}$. Colour indices are taken from the
beginning of the Latin alphabet and the same distinction between lowercase and
uppercase letters is made. For both $\mathrm{SU}(2)$ and $\mathrm{SU}(3)$, a
distinction between raised and lowered indices is maintained such that, for
example, $(\psi^i)^\dagger = (\psi^\dagger)_i$ for an isodoublet field $\psi$.
However, we often specialise to the case of only raised, symmetrised indices for
$\mathrm{SU}(2)$, and use a tilde to denote a conjugate field whose
$\mathrm{SU}(2)_{L}$ indices have been raised:
\begin{equation}
  \label{eq:su2l-conj}
  \tilde{\psi}^{i} \equiv  \epsilon^{i j}\psi^{\dagger}_{j}.
\end{equation}
We adopt this notation from the usual definition of $\tilde{H}$, and note that
throughout the paper we freely interchange between $\tilde{\psi}^{i}$ and
$\psi^{\dagger}_{i}$. For the sake of tidiness, we sometimes use parentheses
$(\cdots)$ to indicate the contraction of suppressed indices. Curly braces are
reserved to indicate symmetrised indices $\{\cdots\}$ and square brackets
enclose antisymmetrised indices $[\cdots]$, but this notation is avoided when
the permutation symmetry between indices is clear. We use $\tau^I$ and
$\lambda^A$ for the Pauli and Gell-Mann matrices, and normalise the non-abelian
vector potentials of the SM such that
\begin{equation}
  (W_{\alpha \dot{\beta}})^i_{\ j} = \frac{1}{2} (\tau^I)^i_{\ j} W^I_{\alpha
    \dot{\beta}} \quad \text{ and } \quad (G_{\alpha \dot{\beta}})^a_{\ b} =
  \frac{1}{2} (\lambda^A)^a_{\ b} G^A_{\alpha \dot{\beta}}.
\end{equation}
Flavour (or family) indices of the SM fermions are represented by the lowercase
Latin letters $\{r, s, t, u, v, w\}$.

For the non-gauge degrees of freedom in the SM we capitalise isospin doublets
($Q$, $L$, $H$), while the left-handed isosinglets are written in lowercase with
a bar featuring as a part of the name of the field ($\bar{u}$, $\bar{d}$,
$\bar{e}$). The representations and hypercharges for the SM field content are
summarised in Table~\ref{tbl:sm}. Our definition of the SM gauge-covariant
derivative is exemplified by
\begin{equation}
  \label{eq:covdi}
  \bar{D}_{\dot{\alpha}\beta} Q^{\beta a i}_r = \left[ \delta^a_b \delta^i_j (\bar{\partial}_{\dot{\alpha}\beta} + i g_1 Y_Q \bar{B}_{\dot{\alpha} \beta}) + i g_2 \delta^a_b (\bar{W}_{\dot{\alpha} \beta})^i_{\ j} + i g_3 \delta^i_j (\bar{G}_{\dot{\alpha}\beta})^a_{\ b} \right] Q_r^{\beta b j} \ .
\end{equation}
Note that the derivative implicitly carries $\mathrm{SU}(2)_{L}$ and
$\mathrm{SU}(3)_{c}$ indices [explicit on the right-hand side of
Eq.~\eqref{eq:covdi}] which are suppressed on the left-hand side to reduce
clutter. Where appropriate we show these indices explicitly.

\begin{table}[t]
  \centering
  \begin{tabular}{ccc}
    \toprule
    Field                        & $\mathrm{SU}(3)_{c} \otimes \mathrm{SU}(2)_{L} \otimes \mathrm{U}(1)_{Y}$ & $\mathrm{SU}(2)_{+} \otimes \mathrm{SU}(2)_{-}$ \\
    \midrule
    $Q^{\alpha a i}$             & $(\mathbf{3}, \mathbf{2}, \tfrac{1}{6})$                                  & $(\mathbf{2}, \mathbf{1})$                      \\
    $L^{\alpha i}$               & $(\mathbf{1}, \mathbf{2}, -\tfrac{1}{2})$                                 & $(\mathbf{2}, \mathbf{1})$                      \\
    $\bar{u}^{\alpha}_a$                  & $(\bar{\mathbf{3}}, \mathbf{1}, -\tfrac{2}{3})$                           & $(\mathbf{2}, \mathbf{1})$                      \\
    $\bar{d}^{\alpha}_a$                  & $(\bar{\mathbf{3}}, \mathbf{1}, \tfrac{1}{3})$                            & $(\mathbf{2}, \mathbf{1})$                      \\
    $\bar{e}^{\alpha}$                    & $(\mathbf{1}, \mathbf{1}, 1)$                                             & $(\mathbf{2}, \mathbf{1})$                      \\
    $(G_{\alpha \beta})^a_{\ b}$ & $(\mathbf{8}, \mathbf{1}, 0)$                                             & $(\mathbf{3}, \mathbf{1})$                      \\
    $(W_{\alpha \beta})^i_{\ j}$ & $(\mathbf{1}, \mathbf{3}, 0)$                                             & $(\mathbf{3}, \mathbf{1})$                      \\
    $B_{\alpha \beta}$           & $(\mathbf{1}, \mathbf{1}, 0)$                                             & $(\mathbf{3}, \mathbf{1})$                      \\
    $H^{i}$                      & $(\mathbf{1}, \mathbf{2}, \tfrac{1}{2})$                                  & $(\mathbf{1}, \mathbf{1})$                      \\
    \bottomrule
  \end{tabular}
  \caption{The SM fields and their transformation properties under the SM gauge
    group $G_{\text{SM}}$ and the Lorentz group. The final unbolded number in
    the 3-tuples of the $G_{\text{SM}}$ column represents the $\mathrm{U}(1)_Y$
    charge of the field, normalised such that $Q = I_{3} + Y$. For the fermions
    a generational index has been suppressed.}
  \label{tbl:sm}
\end{table}

We represent the SM quantum numbers of fields as a 3-tuple
$(\mathbf{C}, \mathbf{I}, Y)_{L}$, with $\mathbf{C}$ and $\mathbf{I}$ the
dimension of the colour and isospin representations, $Y$ the hypercharge of the
field, and $L$ an (often omitted) label of the Lorentz representation: $S$
(scalar), $F$ (fermion) or $V$ (vector), although sometimes we use the irrep,
\textit{e.g.} $(\mathbf{2}, \mathbf{1})$. We normalise the hypercharge such that
$Q = I_{3} + Y$. Finally, for exotic fields that contribute to dimension-six
operators at tree-level, we try and adopt names consistent with Tables 1 and 2
of Ref.~\cite{deBlas:2017xtg}, which we reproduce here in
Table~\ref{tab:field-labels}.

\begin{table}[t]
  \begin{center}
    {\small
      \begin{tabular}{lcccccccc}
        \toprule
        Name &
        ${\cal S}$ &
        ${\cal S}_1$ &
        ${\cal S}_2$ &
        $\varphi$ &
        $\Xi$ &
        $\Xi_1$ &
        $\Theta_1$ &
        $\Theta_3$ \\
        Irrep &
        $(\mathbf{1},\mathbf{1},0)$ &
        $(\mathbf{1},\mathbf{1},1)$ &
        $(\mathbf{1},\mathbf{1},2)$ &
        $(\mathbf{1},\mathbf{2},{\tfrac 12})$ &
        $(\mathbf{1},\mathbf{3},0)$ &
        $(\mathbf{1},\mathbf{3},1)$ &
        $(\mathbf{1},\mathbf{4},{\tfrac 12})$ &
        $(\mathbf{1},\mathbf{4},{\tfrac 32})$ \\[1.3mm]
        \midrule
        Name &
        ${\omega}_{1}$ &
        ${\omega}_{2}$ &
        ${\omega}_{4}$ &
        $\Pi_1$ &
        $\Pi_7$ &
        $\zeta$ &
        & \\
        Irrep &
        $(\mathbf{\bar{3}},\mathbf{1},{\tfrac 13})$ &
        $(\mathbf{3},\mathbf{1},{\tfrac 23})$ &
        $(\mathbf{\bar{3}},\mathbf{1},{\tfrac 43})$ &
        $(\mathbf{3},\mathbf{2},{\tfrac 16})$ &
        $(\mathbf{3},\mathbf{2},{\tfrac 76})$ &
        $(\mathbf{\bar{3}},\mathbf{3},{\tfrac 13})$ \\[1.3mm]
        \midrule
        Name &
        $\Omega_{1}$ &
        $\Omega_{2}$ &
        $\Omega_{4}$ &
        $\Upsilon$ &
        $\Phi$ &
        &
        & \\
        Irrep &
        $(\mathbf{6},\mathbf{1},{\tfrac 13})$ &
        $(\mathbf{\bar{6}},\mathbf{1},{\tfrac 23})$ &
        $(\mathbf{6},\mathbf{1},{\tfrac 43})$ &
        $(\mathbf{6},\mathbf{3},{\tfrac 13})$ &
        $(\mathbf{8},\mathbf{2},{\tfrac 12})$ \\[1.3mm]
        \bottomrule
        \toprule
        Name &
        $N$ & $E$ & $\Delta_1$ & $\Delta_3$ & $\Sigma$ & $\Sigma_1$ & \\
        Irrep &
        $(\mathbf{1}, \mathbf{1},0)$ &
        $(\mathbf{1}, \mathbf{1},{1})$ &
        $(\mathbf{1}, \mathbf{2},{\frac{1}{2}})$ &
        $(\mathbf{1}, \mathbf{2},{\frac{3}{2}})$ &
        $(\mathbf{1}, \mathbf{3},0)$ &
        $(\mathbf{1}, \mathbf{3},{1})$ & \\[1.3mm]
        \midrule
        Name &
        $U$ & $D$ & $Q_1$ & $Q_5$ & $Q_7$ & $T_1$ & $T_2$ \\
        Irrep &
        $(\mathbf{3}, \mathbf{1},{\frac{2}{3}}),$ &
        $(\mathbf{\bar{3}}, \mathbf{1},{\frac{1}{3}})$ &
        $(\mathbf{3}, \mathbf{2},{\frac{1}{6}})$ &
        $(\mathbf{3}, \mathbf{2},{-\frac{5}{6}})$ &
        $(\mathbf{3}, \mathbf{2},{\frac{7}{6}})$ &
        $(\mathbf{\bar{3}}, \mathbf{3},{\frac{1}{3}})$ &
        $(\mathbf{3}, \mathbf{3},{\frac{2}{3}})$ \\ [1.3mm]
        \bottomrule
      \end{tabular}
    }
    \end{center}
    \caption{The table shows the exotic scalars (top) and vectorlike or Majorana
      fermions (bottom) contributing to the dimension-six SMEFT at
      tree-level~\cite{deBlas:2017xtg}. We sometimes use the label of a field as
      presented in the table to represent its conjugate, although we always
      define the transformation properties each time a field is mentioned to
      avoid confusion. For the leptoquarks (second row), we add a prime to the
      field name presented here if the baryon-number assignment is such that
      only the diquark couplings are allowed.}
    \label{tab:field-labels}
  \end{table}

\subsection{On operators and tree-level completions}
\label{sec:operatorsandcompletions}

Below we discuss our use of the terms operator and completion. We establish
naming conventions of types of operators that we use throughout the paper, and
illustrate the sense in which we talk about models as completions of operators
with the use of a simple example from the dimension-six SMEFT.

The term operator is used in the literature to loosely denote one of
three\footnote{These correspond to \textit{operators}, \textit{terms} and
  (roughly) \textit{types of operators} in the convention of
  Ref.~\cite{Fonseca:2019yya}.} things:
\begin{enumerate}
  \item A gauge- and Lorentz-invariant product of fields of specified flavour
  and their derivatives. Understood in this sense, the Weinberg `operator'
  $\mathcal{O}_{1}^{\{rs\}} = (L_{r}^{i}L_{s}^{j})H^{k}H^{l}\epsilon_{ik}\epsilon_{jl}$
  is really $n_{f}(n_{f}+1)/2$ complex operators for $n_{f}$ SM-fermion
  generations.
  \item A gauge- and Lorentz-invariant product of fields of unspecified flavour
  and their derivatives. According to this definition,
  $\mathcal{O}_{1}^{\{rs\}}$ is counted as a single operator.
  \item A collection of fields and their derivatives whose product contains a
  Lorentz- and gauge-singlet part. In this sense, the string of fields $LLHH$
  could be called an operator. In this category we also include operators of
  an intermediate type for which some gauge or Lorentz structure is specified
  but the rest is implied. For example, a term like\footnote{Although the
    colour structure is unique here, this is not true of the Lorentz structure.}
  $\mathcal{O}_{3a} = L^{i}L^{j}Q^{k}\bar{d}H^{l} \epsilon_{ij}\epsilon_{kl}$,
  for which colour and Lorentz structure are implicit.
\end{enumerate}

The catalogues of $\Delta L = 2$ operators are lists of operators of type 3 in
the above sense, since they are only distinguished on the basis of field content
and $\mathrm{SU}(2)_{L}$ structure. Thus, the operators $\mathcal{O}_{3a}$ and
$\mathcal{O}_{3b} = L^{i}L^{j}Q^{k}\bar{d}H^{l} \epsilon_{ik}\epsilon_{jl}$, for
example, are understood to stand in for a large family of operators of types 1
and 2. In this case these differ in Lorentz structure (since the colour
contraction is unique), and almost all of them are linearly dependent. They are
related to each other by Fierz and $\mathrm{SU}(2)$-Schouten identities, and can
in general be related to other dimension-seven operators such as
$(\bar{d} L) (L D \bar{u}^{\dagger})$ and $(LL)H\Box H$ through field
redefinitions involving the classical equations of motion (EOM) of SM-fermion
and Higgs fields. (Operators related by these kinds of field redefinitions lead
to identical $S$-matrix elements~\cite{Arzt:1993gz}.) The total number of
independent operators of type 1 can be found using Hilbert-series
techniques~\cite{Lehman:2015via, Henning:2015daa, Lehman:2015coa,
  Henning:2015alf, Henning:2017fpj}, which give $2n_{f}^{4}$ independent
operators with field content $L^{2} Q \bar{d} H$ with the methods of
Ref.~\cite{Henning:2015alf}. These can be arranged into two terms with the
Lorentz structure of the operators chosen such that the flavour indices don't
have any permutation symmetries~\cite{Lehman:2014jma}:
\begin{subequations}
  \label{eq:o3a3b-complete}
  \begin{align}
    \underset{rstu}{\mathcal{O}_{3a}^{(LQ)(Ld)}} &= (L^{i}_{r}Q^{k}_{t})(L^{j}_{s}\bar{d}_{u})H^{l} \epsilon_{ij} \epsilon_{kl},\\
    \underset{rstu}{\mathcal{O}_{3b}^{(LQ)(Ld)}} &= (L^{i}_{r}Q^{k}_{t})(L^{j}_{s}\bar{d}_{u})H^{l} \epsilon_{ik} \epsilon_{jl}.
  \end{align}
\end{subequations}
From the perspective of $\Delta L = 2$ phenomenology, the $\mathrm{SU}(2)_L$
structure of the operators is most important. This can be seen in the following
way: given a non-zero value for the coefficient of such an operator, the
$\mathrm{SU}(2)_{L}$ structure is sufficient to tell at how many loops the
neutrino self-energy or neutrinoless-double-beta-decay diagrams will arise, and
what they will look like. Considering the example of operators
$\mathcal{O}_{3a}$ and $\mathcal{O}_{3b}$ introduced above, it is clear that no
component of $\mathcal{O}_{3a}$ contains two neutrino fields. Therefore, the
Weinberg operator will be generated by one-loop graphs involving $W$ bosons,
which are additionally suppressed by powers of the weak coupling $g$. This
coupling and loop suppression leads to inferred values of the new-physics scale
characterising the operators $\mathcal{O}_{3a}$ and $\mathcal{O}_{3b}$ that
differ by three orders of magnitude. On the other hand, predictions for the
neutrino-mass scale from operators with different Lorentz structures differ only
by $\mathcal{O}(1)$ factors~\cite{deGouvea:2007qla}.

Thus, our main goal is to find particle content in the UV that generates
particular $\mathrm{SU}(2)_L$ structures of $\Delta L = 2$ operators at the low
scale through tree graphs. In this way, we organise the catalogue of radiative
neutrino-mass models by the number of loops in the neutrino self-energy diagram,
or equivalently, by the implied scale of the new physics. In this sense,
exploding the operator $\mathcal{O}_{3a}$, for instance, means finding the
combinations of heavy field content that generate an operator of type 2 with
$\mathrm{SU}(2)_{L}$ structure $3a$. This generated operator will not in general
be $\mathcal{O}_{3a}^{(LQ)(Ld)}$ of Eq.~\eqref{eq:o3a3b-complete}, but will be
expressible as a linear combination of $\mathcal{O}_{3a}^{(LQ)(Ld)}$ and
$\mathcal{O}_{3b}^{(LQ)(Ld)}$, or any other chosen spanning set of operators.

This last point highlights the importance of the operator basis in talking about
the completions of operators. A completion of an operator $\mathcal{O}$ is a
model generating a non-zero value for the operator coefficient $C_{\mathcal{O}}$
at the high scale. Even a change of basis that leaves $\mathcal{O}$ unchanged
will in general change $C_{\mathcal{O}}$, so one cannot talk about the
completions of $\mathcal{O}$ \textit{in vacuo}, apart from the other operators
which together constitute the EFT. Restricting to the case of tree-level
matching, after eliminating the heavy fields through their EOM, a UV model will
generate some structure organically, which we call the \textit{organic}
operator, and this must then be matched onto the operator basis to extract
coefficients. Our goal here is not to perform this matching onto a complete set
of operators. Instead, we work with an implicitly overcomplete set of operators
and define a convention that allows us to speak unambiguously about the UV
models that might give rise to an operator in the set.

The existing catalogues of $\Delta L = 2$ operators enumerate operators of type
3 with definite $\mathrm{SU}(2)_{L}$-structure. The different isospin
contractions are constructed by contracting indices in all possible ways with
the invariant $\epsilon$ tensor. Operators with symmetric combinations of
indices [which come about from non-trivial exotic irreps of
$\mathrm{SU}(2)_{L}$] generate organic operators in general expressible as many
linear combinations of different operators in the spanning set. One such
combination is sufficient for our purposes, and we choose the one implied by the
convention that non-trivial irreps never give rise to fields contracted with an
$\epsilon$ symbol. We now illustrate this with an example from the dimension-six
SMEFT below.

An overcomplete spanning set of two-Higgs--two-derivative operators is
\begin{subequations}
  \begin{align}
    \mathcal{O}_{H^{2}D^{2}}^{(1)} &= \tilde{H}^{i}\tilde{H}^{j} \Box H^{k} H^{l} \epsilon_{ik} \epsilon_{jl}, \\
    \mathcal{O}_{H^{2}D^{2}}^{(2)} &= \tilde{H}^{i}H^{j} \Box \tilde{H}^{k} H^{l} \epsilon_{ij} \epsilon_{kl}, \\
    \mathcal{O}_{H^{2}D^{2}}^{(3)} &= \tilde{H}^{i}H^{j} \Box \tilde{H}^{k} H^{l} \epsilon_{ik} \epsilon_{jl}, \\
    \mathcal{O}_{H^{2}D^{2}}^{(4)} &= \tilde{H}^{i}H^{j} \Box \tilde{H}^{k} H^{l} \epsilon_{il} \epsilon_{jk}.
  \end{align}
\end{subequations}
The renormalisable UV models of interest are a scalar $\mathrm{SU}(2)_{L}$
triplet with unit hypercharge
$\Xi_{1} \sim (\mathbf{1}, \mathbf{3}, 1)_{S}$, as well as a triplet and a
singlet with vanishing hypercharge:
$\Xi\sim (\mathbf{1}, \mathbf{3}, 0)_{S}$ and
$\mathcal{S} \sim (\mathbf{1}, \mathbf{1}, 0)_{S}$. We envisage integrating
these out from an interaction Lagrangian like
\begin{equation}
  -\mathscr{L} \supset \tilde{H}^{i} H^{j} (x \mathcal{S} \epsilon_{ij} + y \Xi^{\{kl\}}\epsilon_{ik}\epsilon_{jl}) + (z H^{i}H^{j} \tilde{\Xi}_{1}^{\{kl\}} \epsilon_{ik} \epsilon_{jl} + \text{h.c.}),
\end{equation}
with couplings $x, y, z \in \mathbb{C}$. They will generate organic operators
that can be written as linear combinations of the operators listed above
\begin{subequations}
  \label{eq:organic-ops}
  \begin{align}
    \mathcal{S}: &\quad \frac{x^{2}}{M_{\mathcal{S}}^{2}} \mathcal{O}_{H^{2}D^{2}}^{(2)}, \\
    \Xi: &\quad \frac{y^{2}}{M_{\Xi}^{2}} \left[ \mathcal{O}_{H^{2}D^{2}}^{(3)} + \mathcal{O}_{H^{2}D^{2}}^{(4)} \right], \label{eq:organic-ops-2} \\
    \Xi_{1}: &\quad \frac{|z|^{2}}{M_{\Xi_{1}}^{2}} \mathcal{O}_{H^{2}D^{2}}^{(1)},
  \end{align}
\end{subequations}
up to $\mathcal{O}(1)$ factors. Of course, these can then be matched onto a
genuine basis of operators like
\begin{subequations}\label{eq:2h2d-basis}
  \begin{align}
    \mathcal{O}_{\phi \Box} = \mathcal{O}_{H^{2}D^{2}}^{(2)} &= \tilde{H}^{i}H^{j} \Box \tilde{H}^{k}H^{l} \epsilon_{ij} \epsilon_{kl}, \\
    \mathcal{O}_{\phi D} \overset{\text{IBP}}{\sim} \mathcal{O}_{H^{2}D^{2}}^{(3)} &= \tilde{H}^{i}H^{j} \Box \tilde{H}^{k}H^{l} \epsilon_{ik} \epsilon_{jl},
  \end{align}
\end{subequations}
but this is unnecessary for our purposes. (Note here that IBP stands for
integration by parts.) The construction of the organic operator is in general
not unique, since we work with an overcomplete set of operators. Here, for
example,
$\mathcal{O}_{H^{2}D^{2}}^{(3)} + \mathcal{O}_{H^{2}D^{2}}^{(4)} = 2\mathcal{O}_{H^{2}D^{2}}^{(3)} - \mathcal{O}_{H^{2}D^{2}}^{(2)}$,
indicating clearly the redundancy of one of the operators. The convention that
non-trivial representations never give rise to fields contracted with an
$\epsilon$ symbol implies $\mathcal{O}^{(2)}_{H^{2}D^{2}}$ should not be chosen
to feature in Eq.~\eqref{eq:organic-ops-2}. Thus, we call $\Xi$ a completion of
operators $\mathcal{O}_{H^{2}D^{2}}^{(3)}$ and $\mathcal{O}_{H^{2}D^{2}}^{(4)}$,
even though the operator it generates can also be expressed as a linear
combination of $\mathcal{O}^{(2)}_{H^{2}D^{2}}$ and
$\mathcal{O}^{(3)}_{H^{2}D^{2}}$. This convention allows us to talk
unambiguously about completions of the $\Delta L = 2$ operators in a way that
makes their implications for neutrino mass most clear, while avoiding
constructing a complete basis all the way up to dimension eleven.

We remark that this discussion can be extended to operators of type 3 with
explicit $\mathrm{SU}(3)_{c}$-structure with minor modifications. Here,
irreducible representations are furnished by traceless tensors with raised and
lowered symmetrised indices, which can be written as sums of operators in which
contractions between raised and lowered indices are written with the $\delta$
symbol. The tracelessness condition can be enforced by additionally allowing
contractions with the three-index $\epsilon$ symbol, and choosing that
non-trivial representations never give rise to fields contracted with a
$\delta$, \textit{i.e.} always choosing
$[\lambda^{A}]_{c}^{a} [\lambda^{A}]_{d}^{b} = \tfrac{4}{3}\delta^{a}_{d}\delta^{b}_{c} - \tfrac{2}{3}\epsilon_{cde}\epsilon^{abe}$
over
$[\lambda^{A}]_{c}^{a} [\lambda^{A}]_{d}^{b} = 2\delta^{a}_{d}\delta^{b}_{c} - \tfrac{2}{3}\delta^{a}_{c}\delta^{b}_{d}$.
Explicit examples involving non-trivial colour contractions are presented in
Sec.~\ref{sec:exploding-operators} and in the publicly available notebook we
introduce in Sec.~\ref{sec:algorithm}, which contains complete matching
calculations for some of the dimension-six operators in the SMEFT.

\subsection{Operator taxonomy}
\label{sec:operatortaxonomy}

The list of gauge-invariant, $\Delta L = 2$ operators first provided by BL runs
from $\mathcal{O}_1$ to $\mathcal{O}_{60}$~\cite{Babu:2001ex}. Each numbered
operator is distinguished on the basis of field content, although each in
general corresponds to a family of operators differing in $\mathrm{SU}(2)_{L}$-,
Lorentz-, and flavour-structure. The operators are constructed from SM fermion
fields and Higgs fields only and no internal global symmetries are imposed on
the operators aside from baryon number. To violate lepton number by two units,
each operator must contain at least one $\Delta L = 2$ fermion bilinear: one of
$\{LL, L\bar{e}^\dagger, \bar{e}^\dagger \bar{e}^\dagger\}$. The operators enter
the list at odd mass dimension~\cite{Kobach:2016ami} and only up to dimension
eleven, since it was thought that higher dimensional operators generally imply
neutrinos insufficiently heavy to meet the atmospheric lower bound. (It seems
that a truly exhaustive treatment requires operators of higher
mass-dimension~\cite{Gargalionis:2019drk}, and this is discussed in detail in
Sec.~\ref{sec:operator-closures}.) An additional 15 operators (acknowledged by
BL, but left implicit) of mass dimension nine and eleven were added to the list
by dGJ, increasing the total number to 75. These are constructed as products of
lower-dimensional operators with the dimension-four Yukawa operators of the SM.
Thus, they have the same field content as other operators in the list but carry
different numerical labels. Latin subscripts were introduced by the same authors
to distinguish different $\mathrm{SU}(2)_L$ contractions. The number of type-3
operators counted in this way is 129. Inclusion of the \textit{all-singlets}
operator
$\bar{e}^\dagger\bar{e}^\dagger \bar{u}^\dagger \bar{u}^\dagger \bar{d} \bar{d}$,
whose tree-level completions were recently written down~\cite{deGouvea:2019xzm},
brings the tally to 130. Even in the extended dGJ scheme, product operators of
the form $\mathcal{O} \cdot H^\dagger_i H^i$ are left implicit.

Here we work with a modified classification scheme which differs mildly from
those used in the previous analyses. We list all operators explicitly, including
product operators built from lower-dimensional ones and SM Yukawas or
$H^\dagger H$, and enforce that operators with the same field content carry the
same numerical labels. We adopt the convention of labelling
$\mathrm{SU}(2)_L$-structures with an additional Latin subscript\footnote{We
  note that this introduces a notational ambiguity with colour indices, the
  resolution of which must be based on context.}. We have a greater number of
such structures for each numbered operator than the other catalogues because we
include product-type operators and new structures which may have been missed
previously. We attempt to
ensure that these new operators have labels that do
not break compatibility with these and other previous works using lepton-number
violating operators. A small exception is the case where only one structure is
listed by BL and dGJ. In such situations this corresponds to operator $a$ in our
classification.

We find some new non-product operators not appearing in previous classifications
even implicitly. These include new $\mathrm{SU}(2)_L$-structures but also new
numbered operators. Dimension-eleven product-type operators built from a
lower-dimensional operator and factors of $H^\dagger H$ that are not given
numerical labels in the previous catalogues are given primed labels here, a
common convention in the literature. In cases where a number of such operators
carry the same field content, we prefer to use a new numerical label. For
example, operators $\mathcal{O}_{5a}^\prime = \mathcal{O}_{5a} (H^\dagger H)$
and $\mathcal{O}_{3a}^{\prime \prime} = \mathcal{O}_{3a}(H^\dagger H)^2$ have
the same field content. They appear in our list as different
$\mathrm{SU}(2)_L$-structures of the new numbered operator $\mathcal{O}_{80}$.

This means that the 75 numbered type-3 operator classes presented by dGJ now
correspond to 82 classes and additional $\mathrm{SU}(2)_{L}$-structures
$\{a, b, c, \ldots\}$. We present our list of $\Delta L = 2$ operators
containing SM fermion and Higgs fields in Table~\ref{tab:long}. Product
operators as presented in our tables must be read with care. This is just a
convenient shorthand to represent the field-content of an operator and
illustrate that isospin indices are internally contracted. For example, by
writing
$\mathcal{O}_{5b} = \mathcal{O}_{1} Q^{i} \bar{d} \tilde{H}^{j} \epsilon_{ij}$,
we do not mean to suggest that Lorentz indices must be contracted internally to
$\mathcal{O}_{1}$ and the down-type Yukawa. We discuss the additional
information presented in Table~\ref{tab:long} as it is introduced throughout the
paper.

The table also includes a list of $\Delta L = 2$ operators involving derivatives
up to dimension nine. The pertinent operators at dimension seven were mentioned
in Ref.~\cite{Babu:2001ex} and listed in the context of a complete basis of
operators for the dimension-seven SMEFT in Ref.~\cite{Lehman:2014jma}. The
operators of higher dimension were excluded from the earlier catalogues of
$\Delta L = 2$ operators on the basis that they may be less important for
neutrino-mass model building, although they have appeared
recently~\cite{Li:2020xlh}. We find that opening up these operators does yield
novel neutrino-mass models, although this is not clear at dimension seven. The
derivative operators are also interesting from a broader phenomenological
perspective, for example in the study of lepton-number-violating hadron decays,
see \textit{e.g.} Ref.~\cite{Cata:2019wbu}. The procedure we use for identifying
these operators draws from the earlier $\Delta L = 2$ catalogues, Hilbert series
techniques~\cite{Lehman:2015via, Henning:2015daa, Lehman:2015coa,
  Henning:2015alf, Henning:2017fpj} as well as more recent automated
approaches~\cite{Gripaios:2018zrz, Criado:2019ugp, Fonseca:2011sy,
  Fonseca:2017lem, Fonseca:2019yya, Banerjee:2020bym}.

Although operators related by field redefinitions through the classical EOM lead
to identical $S$-matrix elements, we do not account for these redundancies in
our catalogue of operators containing derivatives. This is done for two reasons:
(1) we are ultimately interested in comparing Green's functions in the effective
theory to those in various compatible UV theories; and (2) we are only
interested in tree-level completions of effective operators, and EOM
redundancies may relate operators generated from tree graphs to those generated
by loops~\cite{Arzt:1994gp, Einhorn:2013kja}. Redundancies arising from
integration by parts (IBP) are also not accounted for, and it should be
understood that derivatives act on the operators listed in Table~\ref{tab:long}
in all possible ways. In our listing, we prefer to act them in whichever way
maximises the number of non-vanishing $\mathrm{SU}(2)_{L}$ structures, so that
they can all be labelled. Often this means that derivatives will be carried by
Higgs fields.

\section{Tree-level matching forwards and backwards}
\label{sec:treelevelmatching}

In this section we outline the procedure we use for opening up operators of the
sort introduced in Sec.~\ref{sec:operatorsandcompletions} and
Sec.~\ref{sec:operatortaxonomy} for the purpose of exploratory model building.
It also includes prefatory comments on tree-level matching for scalars and
fermions, and a discussion of the tree-level completions of operators containing
derivatives and field-strength tensors. We highlight that the results of this
section are not specific to $\Delta L = 2$ physics, and the model-building
prescription can be applied (high-dimensional) operators in other EFTs. To
illustrate the point, we apply the methods to an EFT unrelated to neutrino
masses: the SMEFT at dimension-six.

The model-building framework introduced and used in Ref.~\cite{Angel:2012ug}
assumes that the new heavy fields introduced in the UV completions are only
scalars, vector-like Dirac fermions or Majorana fermions. This particle content
ensures the models are genuinely UV complete in the sense that their predictions
can be extrapolated to arbitrarily high energies. Chiral fermions will in
general introduce gauge anomalies, and the generation of their masses may
introduce unnecessary complications. This treatment of exotic fermion fields is
also used in Ref.~\cite{deBlas:2017xtg}, where a tree-level dictionary of the
dimension-six SMEFT is written down. Exotic Proca fields will still need to be
interpreted in the context of some larger UV framework (\textit{e.g.} an
extended gauge group), and so these are not introduced in our approach. Thus for
the remainder of the paper we limit the discussion of building UV-complete
models to those containing only scalars and non-chiral fermions.

\subsection{Effective Lagrangians and tree-level completions}
\label{sec:efflag}

Suppose one has a theory with light particle states described by fields
$\pi_{i}$ and heavy states described by $\Pi_{i}$ with a Lagrangian of the form
\begin{equation}
  \begin{aligned}
  \mathscr{L}_{\text{UV}}[\pi, \Pi] &= \mathscr{L}_{\text{kin}}[\pi, \Pi] + \mathscr{L}_{\text{int}}[\pi, \Pi], \text{ with } \\
  \mathscr{L}_{\text{int}}[\pi, \Pi] &= \mathscr{L}^{l}[\pi] + \mathscr{L}^{h}[\Pi] +  \mathscr{L}^{lh}[\pi, \Pi].
  \end{aligned}
\end{equation}
Below the threshold for $\Pi_{i}$ production, an effective description of the
theory can be used that involves interactions only between the light fields.
This effective theory is described by a Lagrangian
$\mathscr{L}_{\text{eff}}[\pi]$ involving interactions between the
$\pi_{i}$ that correspond to diagrams in the full theory containing only heavy
internal propagators and light external states. At the classical level,
$\mathscr{L}_{\text{eff}}$ can be written down by integrating out the
$\Pi_{i}$. Perturbatively this corresponds to expanding the heavy propagators
$\Delta$ in powers of momenta on the heavy mass scale\footnote{We note that some
  UV scenarios may have more than one characteristic scale. In this case
  $\Lambda$ can be understood as an effective scale which may not necessarily
  correspond to the mass of a specific particle.} $\Lambda$, such
that
\begin{equation}
  \label{eq:prop}
  \Delta = \begin{dcases}
    -\frac{1}{\Lambda^{2}}\left(1 + \frac{p^2}{\Lambda^{2}} + \cdots \right) & \text{for
    } \includegraphics[width=0.13\linewidth]{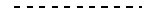} \\
    -\frac{\delta_{\alpha}^{\ \beta}}{\Lambda} \left( 1 + \frac{p^2}{\Lambda^{2}} + \cdots \right) & \text{for
    } \includegraphics[width=0.13\linewidth]{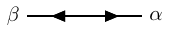} \\
    -\frac{i p \cdot \bar{\sigma}^{\dot{\alpha}\beta}}{\Lambda^{2}} \left( 1 + \frac{p^2}{\Lambda^{2}} + \cdots \right) & \text{for
    } \includegraphics[width=0.13\linewidth]{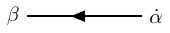}
  \end{dcases}
  \ .
\end{equation}
In this notation, the arrow-preserving propagator corresponds to the part of the
regular four-component fermion propagator proportional to momentum, while the
arrow-violating one is the part proportional to the fermion mass. Expressions
for the fermion propagators with reversed arrows follow from
$\bar{\sigma}^{\mu} \to \sigma^{\mu}$ and interchanging dotted and undotted
indices (see Ref.~\cite{Dreiner:2008tw} Sec.~4.2 for the Lorentz structure).

Equivalently, the integration can be performed using the classical EOM of the
$\Pi_{i}$. For some heavy field $\Pi$, the linearised solution to its classical
EOM can be used to remove it from the Lagrangian completely. This procedure is
mildly different for scalars and fermions, and we briefly outline these
separately below. In both cases, we begin with a Lagrangian
$\mathscr{L}_{\text{UV}}$ for which we imagine that kinetic and mass mixing
terms between heavy and light fields have been removed.

There are tree-level contributions to $\mathscr{L}^{\text{eff}}$ as long as
there are interaction terms linear in $\Pi$. For scalar $\Pi$, the UV Lagrangian
contains the terms
\begin{equation}
  \label{eq:scalar-lag}
 \mathscr{L}_{\text{UV}}[\Pi, \pi] \supset \Pi^{\dagger} (- D^{2} - m_{\Pi}^{2}) \Pi + \left(\Pi \frac{\partial \mathscr{L}^{lh}}{\partial \Pi} + \text{h.c.} \right) \ ,
\end{equation}
where $\partial \mathscr{L}^{lh} / \partial \Pi$ is a function only of
light fields, and we are neglecting interactions of the form
$\Pi^{\dagger} \Pi f(\pi)$ for the sake of conciseness. The EOM are
\begin{equation}
  \label{eq:scalar-eom}
 (- D^{2} - m_{\Pi}^{2}) \Pi = - \frac{\partial \mathscr{L}^{lh}}{\partial \Pi^{\dagger}} + \mathcal{O}(\Pi^{2}) \ ,
\end{equation}
which can be solved for $\Pi^{\text{cl}}$, the classical field configuration, by
inverting the differential operator on the LHS of Eq.~\eqref{eq:scalar-eom} and
expanding in $D^{2} / m_{\Pi}^{2}$:
\begin{equation}
  \label{eq:scalar-repl}
  \Pi^{\text{cl}} = - \frac{1}{m_{\Pi}^{2}} \left( 1 - \frac{D^{2}}{m_{\Pi}^{2}}  + \cdots \right) \frac{\partial \mathscr{L}^{lh}}{\partial \Pi^{\dagger}} \ .
\end{equation}
This solution can be substituted back into Eq.~\eqref{eq:scalar-lag} to give
interactions between light fields in the tree-level effective Lagrangian:
\begin{equation}
  \label{eq:classical-efflag-scalar}
  \mathscr{L}_{\text{eff}}[\pi] \supset - \frac{\partial \mathscr{L}^{lh}}{\partial \Pi} \frac{1}{m_{\Pi}^{2}} \left( 1 - \frac{D^{2}}{m_{\Pi}^{2}} + \cdots \right) \frac{\partial \mathscr{L}^{lh}}{\partial \Pi^{\dagger}} \ .
\end{equation}
Many concrete examples of this procedure can be found in the literature, see
\textit{e.g.} Ref.~\cite{Henning:2014wua}. The expansion in $D^{2}/m_{\Pi}^{2}$
corresponds to the expansion in $p^{2} / \Lambda^{2}$ in the first case of
Eq.~\eqref{eq:prop}, showing the expansion of the scalar propagator.

Next we sketch out the procedure for a Dirac fermion
$\Pi + \bar{\Pi}^{\dagger}$, where $\Pi$ and $\bar{\Pi}$ are separate
two-component spin-$\tfrac{1}{2}$ fields transforming oppositely under
$G_{\text{SM}}$. In this case, the UV theory is described by a Lagrangian like
\begin{equation}
  \label{eq:fermion-lag}
  \mathscr{L}_{\text{UV}}[\Pi, \pi] \supset i \Pi^{\dagger} \bar{\sigma}^{\mu} D_{\mu} \Pi + i \bar{\Pi}^{\dagger} \bar{\sigma}^{\mu} D_{\mu} \bar{\Pi} + \left( \Pi \frac{\partial \mathscr{L}^{lh}}{\partial \Pi} + \bar{\Pi} \frac{\partial \mathscr{L}^{lh}}{\partial \bar{\Pi}} - m_{\Pi} \bar{\Pi} \Pi  + \text{h.c.} \right)
\end{equation}
Varying the action with respect to the heavy fields gives two coupled EOM:
\begin{align}
 i \bar{\sigma}^{\mu} D_{\mu} \Pi - m \bar{\Pi}^{\dagger} + \frac{\partial \mathscr{L}^{lh}}{\partial \Pi^{\dagger}} &= 0 \ , \label{eq:pi-fermion-eom} \\
 i \bar{\sigma}^{\mu} D_{\mu} \bar{\Pi} - m \Pi^{\dagger} + \frac{\partial \mathscr{L}^{lh}}{\partial \bar{\Pi}^{\dagger}} &= 0 \ . \label{eq:pibar-fermion-eom}
\end{align}
Substituting Eq.~\eqref{eq:pi-fermion-eom} into Eq.~\eqref{eq:pibar-fermion-eom}
gives a second-order partial differential equation in $\Pi$, analogous to
Eq.~\eqref{eq:scalar-eom}. Inverting the differential operator in a similar way
gives
\begin{equation}
  \label{eq:fermion-repl}
  \Pi^{\text{cl}}_{\beta} = \frac{1}{m_{\Pi}^{2}} \left( \epsilon_{\alpha \beta} + \frac{ \tfrac{1}{2} X_{\alpha \beta} - D^{2} \epsilon_{\alpha \beta}}{m_{\Pi}^{2}} + \cdots \right) \left( i D^{\alpha \dot{\alpha}} \frac{\partial \mathscr{L}^{lh}}{\partial \Pi^{\dagger}_{\dot{\beta}}} \epsilon_{\dot{\alpha} \dot{\beta}} + m_{\Pi} \frac{\partial \mathscr{L}^{lh}}{\partial \bar{\Pi}_{\alpha}} \right) \ ,
\end{equation}
where the field-strength tensor comes about from a structure like
\begin{align}
  [\sigma^{\mu} \bar{\sigma}^{\nu}]_{\alpha}^{\ \beta} D_{\mu} D_{\nu} &= \eta^{\mu\nu} D_{\mu} D_{\nu} \delta_{\alpha}^{\ \beta} - 2i [\sigma^{\mu\nu}]_{\alpha}^{\ \beta} D_{\mu} D_{\nu} \\
  &= D^{2} \delta_{\alpha}^{\ \beta} - \tfrac{1}{2} X_{\alpha}^{\ \beta} \ .
\end{align}
Here, and later in this section, the replacement $\bar{\Pi} \to \Pi$ should be
understood for Majorana $\Pi$. Each contribution corresponds to a particular
kind of propagator in the perturbative picture. The first term in the last
parenthesis of Eq.~\eqref{eq:fermion-repl} results from the fermion propagator
proportional to momentum: the arrow-preserving fermion propagator shown as the
last case of Eq.~\eqref{eq:prop}. The second term in the same parentheses stems
from the fermion propagator proportional to the mass, corresponding to the
arrow-violating propagator shown in the middle case of Eq.~\eqref{eq:prop}.
Replacing $\Pi$ in Eq.~\eqref{eq:fermion-lag} gives the tree-level effective
Lagrangian with the heavy fermion integrated out:
\begin{equation}
  \begin{aligned}
  \label{eq:classical-efflag-fermion}
  \mathscr{L}_{\text{eff}}[\pi] &\supset \frac{\partial \mathscr{L}^{lh}}{\partial \Pi_{\beta}} \frac{1}{m_{\Pi}^{2}} \left(\epsilon_{\alpha \beta} + \frac{ \tfrac{1}{2} X_{\alpha \beta} - D^{2} \epsilon_{\alpha \beta}}{m_{\Pi}^{2}} +  \cdots \right) i D^{\alpha \dot{\alpha}} \frac{\partial \mathscr{L}^{lh}}{\partial \Pi^{\dagger}_{\dot{\beta}}} \epsilon_{\dot{\alpha} \dot{\beta}}\\
  &\quad + \frac{\partial \mathscr{L}^{lh}}{\partial \Pi_{\beta}} \frac{1}{m_{\Pi}^{2}} \left(\epsilon_{\alpha \beta} + \frac{ \tfrac{1}{2} X_{\alpha \beta} - D^{2} \epsilon_{\alpha \beta}}{m_{\Pi}^{2}} + \cdots \right) \frac{\partial \mathscr{L}^{lh}}{\partial \bar{\Pi}_{\alpha}} \ .
  \end{aligned}
\end{equation}

As shown in Eqs.~\eqref{eq:classical-efflag-scalar} and
\eqref{eq:classical-efflag-fermion}, expanding in powers of derivatives on heavy
masses leads to a tower of local operators of increasing mass dimension $d_{i}$
organised as a power series in the inverse heavy scale:
\begin{equation}
  \mathscr{L}_{\text{eff}}[\pi] = \mathscr{L}^{l}[\pi] + \sum_{i} \frac{C_{i}}{\Lambda^{d_{i}-4}} \mathcal{O}_{i}[\pi].
\end{equation}
The $C_{i}$ are dimensionless coefficients which are in general calculable if
one knows the high-energy theory. We are interested in the case where the UV
theory is unknown. Here, the EFT is a useful way to encapsulate the effects of
the entire class of possible UV theories in a model-agnostic way. We advocate
that it is also a practical model-building tool, since the operators provide
information about the types of UV models from which the EFT may arise. Subject
to a number of assumptions, the possible UV models implied by an effective
operator can be enumerated by building all possible tree graphs with an
external-leg structure reflecting that of the operator. The quantum numbers of
the heavy propagators can then be read off by imposing Lorentz- and
gauge-invariance at every vertex, starting with vertices with two or three (for
scalars) external edges. This is equivalent to exploring all of the possible
ways the light fields may have been grouped into terms in
$\mathscr{L}[\pi, \Pi]$ and distributed in the products of
Eqs.~\eqref{eq:classical-efflag-scalar} and \eqref{eq:classical-efflag-fermion}.
In the following we develop this picture into a precise algorithm.

\subsubsection{Exploding operators}
\label{sec:exploding-operators}

\begin{figure}[t]
  \centering
  \includegraphics[width=0.3\linewidth]{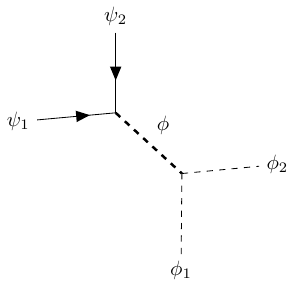}
  \includegraphics[width=0.3\linewidth]{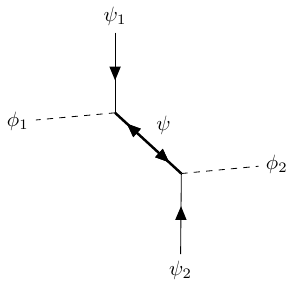}\\
  \includegraphics[width=0.3\linewidth]{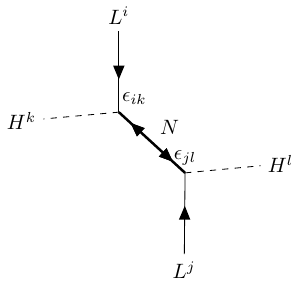}
  \includegraphics[width=0.3\linewidth]{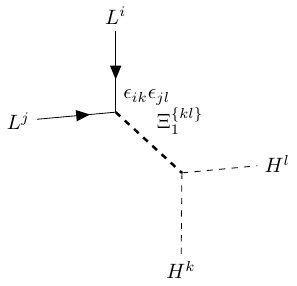}
  \includegraphics[width=0.3\linewidth]{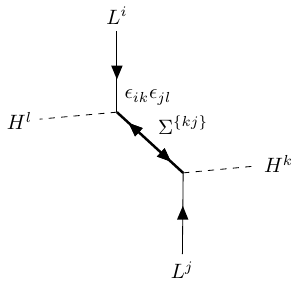}
  \caption{\textit{Above:} Scalar-only and fermion-only topologies which
    complete dimension-five two-scalar--two-fermion operators, like the Weinberg
    operator $\mathcal{O}_{1}$. \textit{Below:} The three minimal tree-level
    completions of $\mathcal{O}_{1}$, each corresponding to a different
    permutation of the fields on the external lines of the topologies. These are
    traditionally called (read from left to right) the type-I, type-II and
    type-III seesaw models. The $\mathrm{SU}(2)_{L}$ indices are included
    explicitly to distinguish type-I and type-III, while making a more clear
    connection to Eq.~\eqref{eq:seesaw-contractions}. The exotic propagators are
    shown in bold.}
  \label{fig:seesaw-figs}
\end{figure}

As an introductory example we use the Weinberg operator
$\mathcal{O}_{1} = (L^{i}L^{j})H^{k}H^{l} \epsilon_{ik} \epsilon_{jl}$, whose
minimal tree-level completions are the canonical seesaw models:
$N \sim (\mathbf{1}, \mathbf{1}, 0)_{(\mathbf{2}, \mathbf{1})}$,
$\Xi_{1} \sim (\mathbf{1}, \mathbf{3}, 1)_{S}$ and
$\Sigma \sim (\mathbf{1}, \mathbf{3}, 0)_{(\mathbf{2}, \mathbf{1})}$. These can
be derived by considering the allowed ways of decorating the two tree-level
two-scalar--two-fermion topologies with the field content of the operator. These
topologies are shown in Fig.~\ref{fig:seesaw-figs} along with the possible ways
of furnishing the topologies into Feynman diagrams, each corresponding to a
seesaw model. As discussed above, this is equivalent to grouping fields together
as they may have arisen in the partial derivatives of
Eqs.~\eqref{eq:classical-efflag-scalar} and \eqref{eq:classical-efflag-fermion}.
For the Weinberg operator, these groupings are:
\begin{subequations}
  \label{eq:seesaw-contractions}
  \begin{align}
    &\wick{\c L^{i} L^{j} \c H^{k} H^{l}} \epsilon_{ik} \epsilon_{jl} \quad \Rightarrow \quad \frac{\partial \mathscr{L}^{lh}}{\partial N_{\alpha}} \supseteq x_{r} L_{r}^{\alpha i} H^{k} \epsilon_{ik} \sim N, \\
    \label{eq:typeii}
    &\wick{\c L^{i} \c L^{j} H^{k} H^{l}} \epsilon_{ik} \epsilon_{jl} \quad \Rightarrow \quad \frac{\partial \mathscr{L}^{lh}}{\partial \Xi_{1 \alpha}^{kl}} \supseteq [y_{rs} (L_{r}^{\{i}L_{s}^{j\}}) + \kappa \tilde{H}^{i} \tilde{H}^{j}] \epsilon_{ik} \epsilon_{jl} \sim \Xi_{1}^{\dagger},\\
    \label{eq:typeiii}
    &\wick{\c L^{i} L^{j} H^{k} \c H^{l}} \epsilon_{ik} \epsilon_{jl} \quad \Rightarrow \quad \frac{\partial \mathscr{L}^{lh}}{\partial \Sigma_{\alpha}^{kj}} \supseteq z_{r} L_{r}^{\alpha \{i} H^{l\}} \epsilon_{ik} \epsilon_{jl} \sim \Sigma,
  \end{align}
\end{subequations}
where we use $\sim$ to mean `transforms as' under
$\mathrm{SU}(2)_{+} \otimes \mathrm{SU}(2)_{-} \otimes G_{\text{SM}}$. Each
pattern of contractions corresponds to a topology, with each individual grouping
of the fields corresponding to a vertex, or equivalently, a term in the
$\Delta L = 2$ UV Lagrangian. The explicit form of these terms can be written
down by keeping track of the isospin indices as in
Eq.~\eqref{eq:seesaw-contractions}, and expanding implicit index structures in
all possible ways (\textit{i.e.} decomposing products of fields into irreducible
representations), consistent with our model building assumptions. (In our case
this means keeping only scalar and fermion Lorentz irreps.) In
Eq.~\eqref{eq:typeiii}, the indices $i,l$ are symmetrised since this is the only
way the component $L^{i}H^{l}$ (with $i,l$ not antisymmetric under exchange) can
appear in the Yukawa interaction $L \Sigma H$. Note that we adopt the convention
that the conjugate exotic field couples to the contracted fields in the
operator. This means that $\Xi_{1}^{\dagger}$ transforms like $L^{\{i} L^{j\}}$,
as implied in Eq.~\eqref{eq:typeii}, but the renormalisable term in the UV
theory which corresponds to the vertex is $L\Xi_{1}L$. For Majorana fermions
there is only one state which can couple in both cases, while for a Dirac
fermion $\psi + \bar{\psi}^{\dagger}$ we arbitrarily choose $\bar{\psi}$ to
couple to the contracted fields.

This process of grouping fields into renormalisable interaction terms can be
conveniently expressed with the following replacement rules:
\begin{equation}
  \label{eq:non-deriv-rules}
  \wick{\c \psi^{\alpha}_{1} \c \psi_{2 \alpha}} \to \Phi^{\dagger}, \quad \wick{\c \phi_{1} \c \phi_{2}} \to \Phi^{\dagger}, \quad \wick[positions={-1, 1}, offset=1em]{\c2 \phi_{1} \c1 {\c2 \phi_{2}} \c1 \phi_{3}} \to \Phi^{\dagger} , \quad \wick{\c \phi \c \psi^{\alpha}} \to \bar{\Psi}^{\alpha}, \quad \wick{\c \psi_{1}^{\alpha} \c \psi_{2}^{\dagger \dot{\alpha}}} \to \xmark,
\end{equation}
with free raised or lowered gauge-indices (suppressed above) of the same type
always symmetrised on the right-hand side. We are using $\Phi$ and $\bar{\Psi}$
to represent a heavy scalar and fermion; while the lowercase $\phi_{i}$
and $\psi_{i}$ represent scalar and fermion fields that may be light or heavy.
Note that $\bar{\Psi} = \Psi$ for a Majorana fermion. The mark $\xmark$ signals
that the completion should be discarded, in this case because it represents a
model involving a heavy vector field. The repeated application of these rules
allows us to build explicit computational representations of the $\Delta L = 2$
Lagrangian and diagram topology for a completion.

\begin{figure}[t]
  \centering
  \subcaptionbox{\label{fig:6f-scalar-only}}{
    \includegraphics[width=0.35\linewidth]{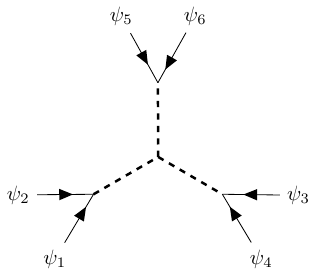}
  }
  \subcaptionbox{\label{fig:6f-scalar-plus-fermion}}{
    \includegraphics[width=0.4\linewidth]{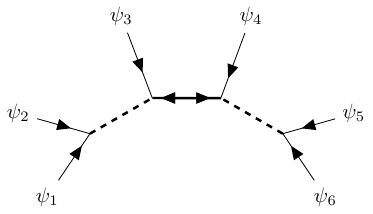}
  }
  \caption{The two tree-level topologies relevant to six-fermion operators. For
    some operators, some fermion arrows may be reversed. The exotic propagators
    are shown in bold. (a) The scalar-only topology. (b) The scalar-plus-fermion
    or central-fermion topology.}
  \label{fig:6f-topologies}
\end{figure}

We move on with a more involved example that also involves colour structure: a
completion of
$\mathcal{O}_{12} = LLQ^{\dagger}Q^{\dagger}\bar{u}^{\dagger}\bar{u}^{\dagger}$.
According to Table~\ref{tab:long} there are two $\mathrm{SU}(2)$
structures. Both of these structures need to be opened up to enumerate all of
the completions, and models will in general generate sums of these with a
specific Lorentz structure, as per the discussion in
Sec.~\ref{sec:operatorsandcompletions}. We choose to look at
\begin{equation}
  \underset{rstuvw}{\mathcal{O}_{12a}} = \underset{r}{L}^{i} \underset{s}{L}^{j} \underset{t}{\tilde{Q}}^{k} \underset{u}{\tilde{Q}}^{l} \underset{v}{\bar{u}}^{\dagger} \underset{w}{\bar{u}}^{\dagger} \epsilon_{ik}\epsilon_{jl}
\end{equation}
and begin with some preliminary comments. There are only two topologies that
accommodate tree-level completions for six-fermion operators. A scalar-only
topology (shown in Fig.~\ref{fig:6f-scalar-only}), where pairs of fermions are
contracted into scalars which meet at a trilinear vertex, and a
scalar-plus-fermion topology (shown in Fig.~\ref{fig:6f-scalar-plus-fermion}) in
which two exotic scalars come about by fermion contractions and each meets
another SM fermion. Since we are not interested in introducing exotic vector
fields, contractions between fermions must come about by grouping only fields
with dotted or undotted indices, \textit{i.e.} from
$(\mathbf{2}, \mathbf{1}) \otimes (\mathbf{2}, \mathbf{1})$ or
$(\mathbf{1}, \mathbf{2}) \otimes (\mathbf{1}, \mathbf{2})$ contracted into a
$\mathrm{SU}(2)_{\pm}$-scalar representation with an epsilon tensor. These
contractions fix the Lorentz-structure of the generated type-2 operator. For
$\mathcal{O}_{12a}$ it is clear that all scalar-only completions will contain
the triplet scalar $\Xi_{1}$, since the two $L$ fields in the operator are the
only fermions carrying undotted indices, making the contraction
\begin{equation}
  \overset{\tiny \hspace{-8em} \Xi^{\dagger}_{1} \sim (\mathbf{1}, \mathbf{3}, -1)}{\wick{\c L^{i} \c L^{j} \tilde{Q}^{k} \tilde{Q}^{l} \bar{u}^{\dagger} \bar{u}^{\dagger} \epsilon_{ik}\epsilon_{jl}}}
\end{equation}
unique. For the quark fields there are a number of choices to be made. First,
the choice of grouping. There are only two choices for how to group the quark
fields: as $(\tilde{Q}\tilde{Q})(\bar{u}^{\dagger} \bar{u}^{\dagger})$ or
$(\tilde{Q}\bar{u}^{\dagger})^{2}$. The second choice is of the colour
representations. These can be explored recursively, or all invariants can be
constructed and each opened up separately, following the conventions of
Sec.~\ref{sec:operatorsandcompletions}. We opt for the latter case, and
enumerate the colour contractions
\begin{subequations}
  \begin{align}
    \mathcal{O}_{12a\epsilon} &= L^{i} L^{j} \tilde{Q}^{k}_{a} \tilde{Q}^{l}_{b} \bar{u}^{\dagger c} \bar{u}^{\dagger d} \epsilon_{ik}\epsilon_{jl} \epsilon^{abe}\epsilon_{cde}, \\
    \mathcal{O}_{12a\delta} &= L^{i} L^{j} \tilde{Q}^{k}_{a} \tilde{Q}^{l}_{b} \bar{u}^{\dagger c} \bar{u}^{\dagger d} \epsilon_{ik}\epsilon_{jl} \delta_{c}^{a} \delta_{d}^{b}.
  \end{align}
\end{subequations}
The colour sextet combinations
$Q^{\dagger}_{\{a} Q^{\dagger}_{b\}} \bar{u}^{\dagger a} \bar{u}^{\dagger b}$
come about as a sum of flavour permutations of the left-handed quark doublets in
$\mathcal{O}_{12a\delta}$, and the octet combinations
$(Q^{\dagger} \lambda^{A} \bar{u}^{\dagger})^{2}$ as a linear combination of
$\mathcal{O}_{12a\delta}$ and $\mathcal{O}_{12a\epsilon}$. Thus, we understand
contractions like
$Q^{\dagger}_{a} Q^{\dagger}_{b} \delta^{a}_{c} \delta^{b}_{d}$ as coming about
from colour-sextet scalars, and
$Q^{\dagger}_{a} \bar{u}^{\dagger b} \delta^{a}_{c} \delta^{d}_{b}$ or
$Q^{\dagger}_{a} \bar{u}^{\dagger b} \epsilon_{bce}\epsilon^{ade}$ as coming
about from octets.

\begin{figure}
  \centering
  \includegraphics[width=0.81\linewidth]{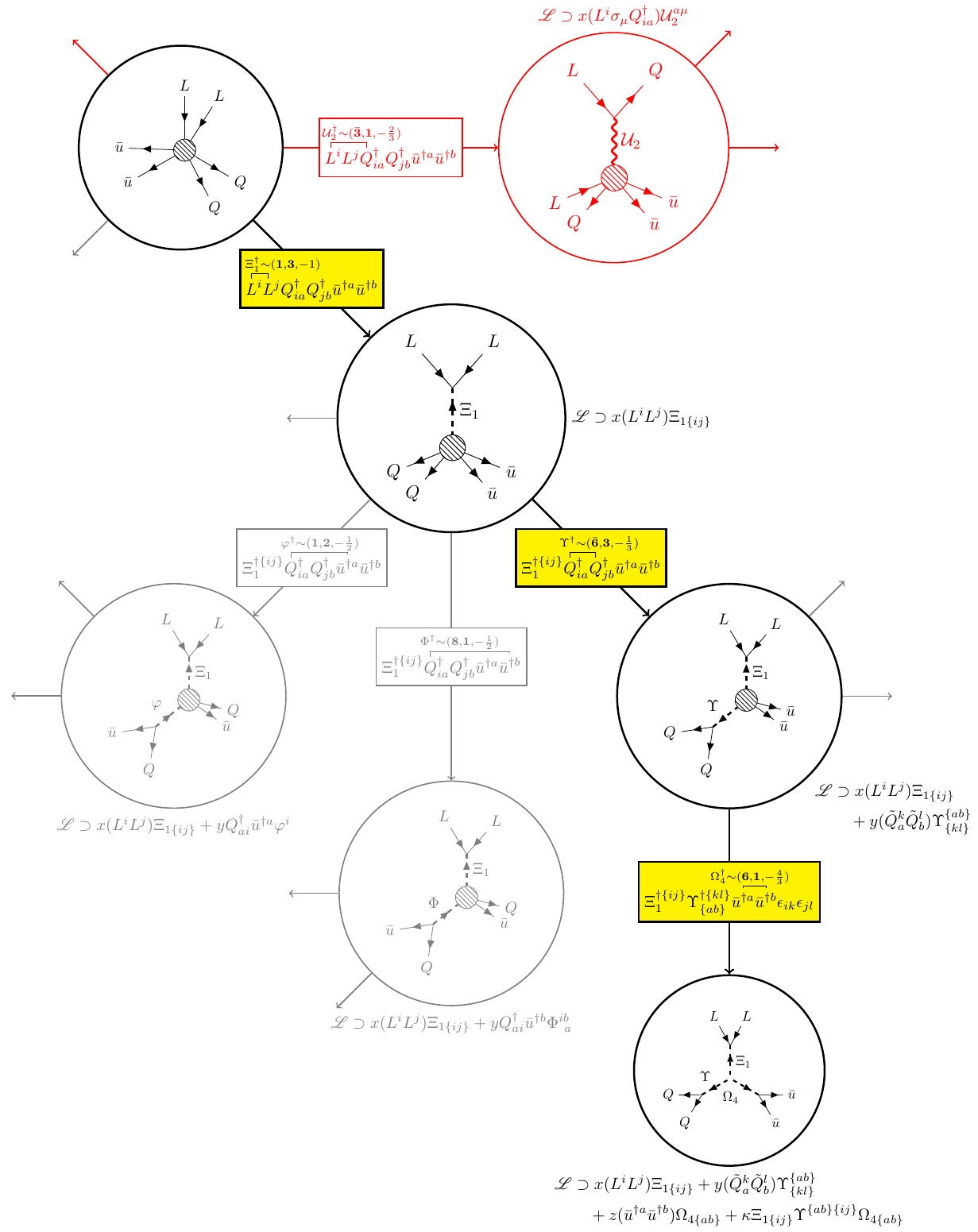}
  \caption{The graph visualises our completion procedure by showing some of the
    possible ways to explode the operator
    $\mathcal{O}_{12a\delta} = L^{i} L^{j} Q^{\dagger}_{ia} Q^{\dagger}_{jb} \bar{u}^{\dagger a} \bar{u}^{\dagger b}$.
    The options are only followed to a fully UV-complete model on one branch,
    shown in bold with yellow edge labels. In each step groups of fields are
    contracted at a vertex, fixing the properties of the exotic field as well as
    the structure of the term describing the interaction, shown alongside each
    diagram. The effective operator is gradually opened up until each vertex in
    the diagram corresponds to a term with mass-dimension $d \leq 4$. Opening up
    the operator fully requires repeating this procedure for all possible
    contractions. In this case this includes other scalar-only completions and
    scalar-plus-fermion models. We also show steps that we choose to forbid in
    our approach in red, like the vector contraction giving rise to the vector
    leptoquark $\mathcal{U}_{2}$ in the figure. Flavour indices have been
    suppressed.}
  \label{fig:example-completion-graph}
\end{figure}

Finding all of the completions of $\mathcal{O}_{12a}$ involves contracting all
fields in all possible ways for each colour contraction. We work through the
example of a particular scalar-only completion of $\mathcal{O}_{12a\delta}$ in
Fig.~\ref{fig:example-completion-graph}. Each step follows the grouping of
fields into a vertex, the Lagrangian term this grouping corresponds to, and the
evolving topology of the completion under the replacement rules of
Eq.~\eqref{eq:non-deriv-rules}. At intermediate stages in the explosion of the
operator, the theory described is still effective because some vertices still
correspond to irrelevant operators\footnote{We note that one can make a
  connection here to the framework of Ref.~\cite{Herrero-Garcia:2019czj}, where
  neutrino-mass models are classified and studied in the context of single-field
  extensions of the SM, corresponding to the first intermediate step in our
  completions procedure. Similar approaches to SMEFT extensions have also been
  considered elsewhere in the literature,
  \textit{e.g.}~\cite{Banerjee:2020jun}.}. The procedure stops once all vertices
have mass-dimension $d \leq 4$. We replace the contracted fields in the operator
with the irreducible representation that, following the restrictions described
in Sec.~\ref{sec:operatorsandcompletions}, could give rise to the contraction.
This will in general require the addition of other structures\footnote{The
  organic operator of the model can be written as a linear combination of these
  other operators and the operator being opened up, and all of these share the
  model as a completion in our sense.}, although this is not the case here. The
operator generated by the model highlighted in
Fig.~\ref{fig:example-completion-graph} is
\begin{equation}
  (\underset{r}{L}^{\{i}\underset{s}{L}^{j\}})(\underset{t}{Q^{\dagger}_{i a}} \underset{u}{Q^{\dagger}_{j b}})(\underset{v}{\bar{u}}^{\dagger \{a} \underset{w}{\bar{u}}^{\dagger b\}}) = \underset{rstuvw}{\mathcal{O}_{12a\delta}} + \underset{srtuvw}{\mathcal{O}_{12a\delta}} + \underset{rsutvw}{\mathcal{O}_{12a\delta}} + \underset{srutvw}{\mathcal{O}_{12a\delta}},
\end{equation}
with the same Lorentz structure carried through $\mathcal{O}_{12a\delta}$. The
relevant part of the $\Delta L = 2$ Lagrangian of the model can be read directly off each contraction
\begin{equation}
  \begin{split}
    - \mathscr{L}_{\Delta L = 2} &\supseteq x_{\{rs\}} (\underset{r}{L^{i}} \underset{s}{L^{j}}) \Xi_{1 \{ij\}} + y_{\{tu\}}(\underset{t}{\tilde{Q}_{a}^{k}} \underset{u}{\tilde{Q}_{b}^{l}}) \Upsilon_{\{kl\}}^{\{ab\}} + z_{[vw]}(\underset{v}{\bar{u}}^{\dagger a} \underset{w}{\bar{u}}^{\dagger b}) \Omega_{4 \{ab\}}\\
    &\quad + \kappa \Xi_{1 \{ij\}} \Upsilon^{\{ab\}}_{\{kl\}} \Omega_{4 \{a b\}} \epsilon_{ik} \epsilon_{jl} + \text{h.c.},
  \end{split}
\end{equation}
although the generation of the entire Lagrangian implied by the field content
requires a program implementing group-theory methods, spin-statistics and tensor
algebra (see Sec.~\ref{sec:algorithm}). This particular model inherits the high
level of symmetry in the effective operator. This introduces symmetries in the
Yukawa couplings of the model, reducing the total number of free parameters.

Given an effective operator, we have established a simple rule for reducing it
to a renormalisable interaction through a processes of contracting fields into
each other, corresponding diagrammatically to pairing the fields off into Yukawa
or scalar interaction vertices according to a system of rewrite rules. Applying
these groupings in all possible ways and following quantum numbers through index
structure allows one to efficiently write down not only the particle content
generating the operator at tree-level, but also the pertinent interaction terms
in the Lagrangian. In the next section, we discuss how to expand this rule to
reducing operators containing derivatives.

\subsection{Tree-level completions of derivative operators}
\label{sec:derivatives}

In the following we broaden the discussion to exploratory model building through
effective operators containing (covariant) derivatives and field-strength
tensors. We begin by summarising the main results of this section. We argue that
(if only scalars and fermions are introduced) a large class of such operators do
not contribute new completions to the pool of models. That is, models derived
from these operators could be found by opening up operators without derivatives
and field strengths. With notable exceptions, it is usually sufficient to study
only single-derivative operators. Some of the derivative operators also admit
fermion-only completions, which are otherwise only found for the Weinberg-like
operators~\cite{Anamiati:2018cuq}. The completion of operators containing
derivatives has been studied before in the context of $\Delta L = 2$
physics~\cite{delAguila:2011gr, delAguila:2012nu, Herrero-Garcia:2016uab}, and
our work expands on this.

\subsubsection{Exploding derivative operators}
\label{sec:exploding-derivative-operators}

In our setup, derivatives in effective operators arise at tree-level by the
expansions given in Eqs.~\eqref{eq:classical-efflag-fermion} and
\eqref{eq:classical-efflag-scalar}. It is clear that derivatives occur in one of
two ways: (1) in pairs as $D^{2}$ or $X$ from next-to-leading order terms in the
EFT expansion, or (2) as single derivatives contracted with fermions
($\slashed{D}$ in traditional notation) coming about from arrow-preserving
fermion propagators. The job of finding the completions of operators containing
derivatives is therefore equivalent to enumerating all possible tree graphs with
the appropriate external-leg structure including arrow-preserving propagators
proportional to momentum for heavy fermion fields and taking powers of momentum
from past the leading order in the expansion of all propagators. As in the
non-derivative case, the quantum numbers of the heavy fields can then be deduced
by imposing Lorentz and gauge invariance at each vertex.

It is not always guaranteed that a tree-level topology with internal fermion and
scalar lines exists for an effective operator containing derivatives. This is in
contrast to the non-derivative case, where this is guaranteed for all operators
of mass dimension larger than four. For example, at dimension seven there are
$\Delta L = 2$ effective operators like
$\bar{d}_{\alpha} \bar{u}^{\dagger}_{\dot{\alpha}} L^{i\beta} D^{\alpha \dot{\alpha}} L_{\beta}^{i} \epsilon_{ij}$
containing four fermions: three with undotted indices and one with a dotted
index. In this case there is no tree-level topology that allows a
arrow-preserving fermion propagator to give rise to the derivative, and so the
operator can only be generated with loops. We call such operators
\textit{non-explosive}. This distinction between tree and loop operators has
been discussed in the literature in the context of the dimension-six operators
of the SMEFT, see \textit{e.g.}~\cite{Arzt:1994gp, Einhorn:2013kja,
  deBlas:2017xtg}, and more recently for the dimension-eight
operators~\cite{Craig:2019wmo}.

\begin{figure}[t]
  \centering
  \subcaptionbox{\label{fig:example-deriv-1}}{
    \includegraphics[width=0.4\linewidth]{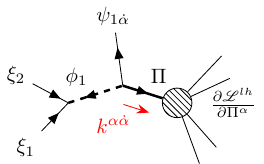}
  }
  \subcaptionbox{\label{fig:example-deriv-2}}{
    \includegraphics[width=0.4\linewidth]{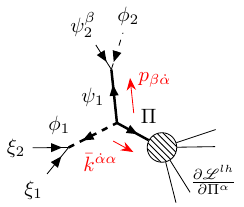}
  }
  \caption{(a) The diagram shows an example opening of an operator containing at
    least one derivative. The derivative can be understood as arising from the
    leading-order term in the expansion of the arrow-preserving fermion
    propagator, emphasised in the diagram. As shown, the fields $\xi_{i}$ and
    $\psi_{1}$ are external and therefore light, but in general they could
    themselves be heavy propagators. (b) The case where the fermion $\psi_{1}$
    is heavy, coupling to the light fields $\psi_{2}$ and $\phi_{2}$. The
    $\sigma$-matrix structure of the propagators is in accordance with the
    conventions of Ref.~\cite{Dreiner:2008tw}. Here, the Lorentz structure is
    such that the momenta are contracted, which arises from contractions of
    derivatives which share one contracted index.}
  \label{fig:example-deriv}
\end{figure}

The derivatives originating from arrow-preserving fermion propagators in the UV
theory enter the effective Lagrangian through the first term in
Eq.~\eqref{eq:classical-efflag-fermion}. Here, the derivative acts on an object
with which it shares a contracted index, \textit{i.e.} it is contracted as
$(\mathbf{2}, \mathbf{2}) \otimes (\mathbf{1}, \mathbf{2}) = (\mathbf{2}, \mathbf{1})$
with the object carrying the index $\dot{\beta}$. This object must be a
$(\mathbf{1}, \mathbf{2})$-fermion if it comes from a renormalisable
interaction, which in our case is uniquely a Yukawa interaction. Thus,
\begin{equation}
  \label{eq:deriv-indices}
  \frac{\partial \mathscr{L}^{lh}}{\partial \Pi_{\dot{\beta}}^{\dagger}} = \sum_{i} \psi^{\dagger \dot{\beta}}_{i} \phi_{i} ,
\end{equation}
with $\psi_{i}$ and $\phi_{i}$ defined as in Eq.~\eqref{eq:non-deriv-rules}. For
example, a structure like
$D^{\alpha \dot{\alpha}} \psi_{1}^{\dagger \dot{\beta}} \phi_{1} \epsilon_{\dot{\alpha} \dot{\beta}}$
could enter an effective operator by integrating out a heavy fermion $\Pi$ that
couples through $\mathscr{L} \supset \Pi^{\dagger} \psi^{\dagger}_{1} \phi_{1}$.
For clarity, the effective Lagrangian looks like
\begin{align}
  \mathscr{L}_{\text{eff}}[\pi] &\supset \frac{\partial \mathscr{L}^{lh}}{\partial \Pi_{\beta}} \frac{1}{m_{\Pi}^{2}} D^{\alpha \dot{\alpha}} \frac{\partial \mathscr{L}^{lh}}{\partial \Pi^{\dagger}_{\dot{\beta}}}   \epsilon_{\alpha \beta} \epsilon_{\dot{\alpha} \dot{\beta}} + \cdots \\
  &= \frac{\partial \mathscr{L}^{lh}}{\partial \Pi_{\beta}} \frac{1}{m_{\Pi}^{2}} D^{\alpha \dot{\alpha}} \left( \psi_{1}^{\dagger \dot{\beta}} \phi_{1} + \cdots \right) \epsilon_{\alpha \beta} \epsilon_{\dot{\alpha} \dot{\beta}} \label{eq:deriv-efflag-2}
\end{align}
in this case. The fields $\phi_{i}$ and $\psi_{i}$ need not be light, and could
have arisen from the contraction of fields in a complicated way. For example,
$\phi_{1}$ may have come from the contraction of two light fermions
$\phi_{1} \sim \xi_{1} \xi_{2}$. This situation is visualised diagrammatically
in Fig.~\ref{fig:example-deriv-1}. The figure shows the $\xi_{i}$ fermions
coupling to the heavy $\phi$ propagator, which in turn couples to
$\psi_{1}^{\dagger}$ leading to the arrow-preserving fermion propagator for the
heavy $\Pi$ carrying momentum $k^{\alpha \dot{\alpha}}$. It is clear from
Eq.~\eqref{eq:deriv-efflag-2} that the derivative acts on both the fermion and
the scalar, reflecting the fact that in the diagram $k$ is the sum of the $\psi$
and $\phi$ momenta. So, derivatives acting on fermions or scalars can be grouped
off into a Yukawa interaction in this way, leaving a arrow-preserving fermion
propagator in their wake. This corresponds to the replacement rules
\begin{equation}
  \label{eq:deriv-rules}
  \wick{D^{\alpha \dot{\alpha}} (\c \psi^{\dagger}_{\dot{\alpha}}) \c \phi} \to \Pi^{\alpha}, \qquad \wick{D^{\alpha \dot{\alpha}} (\c \psi^{\dagger}_{\dot{\alpha}} \c \phi)} \to \Pi^{\alpha}, \qquad \wick{D^{\alpha \dot{\alpha}} (\c \phi) \c \psi^{\dagger}_{\dot{\alpha}}} \to \Pi^{\alpha} .
\end{equation}
We highlight that the arrow-preserving propagator implies that only one
chirality of the Dirac fermion $\Pi$ is necessary for LNV in these models.
However, we still only work with vector-like fermions in our completions to
guarantee anomaly cancellation and straightforwardly give them large masses.

In an effective operator the derivative may act on a fermion with which it does
not share a contracted index. For example, in the model shown in
Fig.~\ref{fig:example-deriv-1}, the effective operator at the low scale looks
something like
\begin{equation}
  \label{eq:distribute-deriv}
D^{\beta \dot{\beta}}(\xi^{\alpha}_{1} \xi_{2 \alpha} \psi^{\dagger}_{1 \dot{\beta}}) \frac{\partial \mathscr{L}^{lh}}{\partial \Pi^{\beta}} = D^{\beta \dot{\beta}}(\xi^{\alpha}_{1}) \xi_{2 \alpha} \psi^{\dagger}_{1 \dot{\beta}} \frac{\partial \mathscr{L}^{lh}}{\partial \Pi^{\beta}} + \cdots
\end{equation}
although as long as the operator is generated at tree-level, the term with the
derivative acting on $\psi^{\dagger}$ will always also be present as long as it
is not removed by a field redefinition involving its classical EOM. Our approach
is the following: act the derivative in all possible ways on the fields
constituting the effective operator and discard the topologies in which a
contraction like $\wick{(D^{\alpha \dot{\alpha}} \c \psi_{1}^{\beta}) \c \phi}$
is made. After a UV-complete model is derived, the operator it implies will
still have the form of the one on the left-hand side of
Eq.~\eqref{eq:distribute-deriv}, so no information is lost. This implies the
rules
\begin{equation}
  \label{eq:bad-deriv-repl-rules}
  \wick{(D^{\alpha \dot{\alpha}} \c \psi^{\beta}) \c \phi} \to \xmark, \qquad \wick{(D^{\alpha \dot{\alpha}} \c \psi^{\beta} \c \phi)} \to \xmark, \qquad \wick{(D^{\alpha \dot{\alpha}} \c \phi) \c \psi^{\beta}} \to \xmark, \qquad \wick{(D^{\alpha \dot{\alpha}} \c \phi_{1}) \c \phi_{2}} \to \xmark \ .
\end{equation}

The first parentheses of Eqs.~\eqref{eq:classical-efflag-fermion} and
\eqref{eq:classical-efflag-scalar} contribute powers of $D^{2}$ or $X$ to
operators in the effective Lagrangian. They contribute the rules
\begin{equation}
  \begin{aligned}
    \label{eq:box-rules}
    \wick{(D^{\alpha \dot{\alpha}} \c \psi_{1}^{\beta}) (D_{\alpha \dot{\alpha}} \c \psi_{2 \beta})} \to \Phi^{\dagger}, \quad \wick{(D^{\alpha \dot{\alpha}} \c \phi_{1}) (D_{\alpha \dot{\alpha}} \c \phi_{2})} \to \Phi^{\dagger} \\
    \wick{(D^{\alpha \dot{\alpha}} \c \psi^{\beta}) (D_{\alpha \dot{\alpha}} \c \phi)} \to \bar{\Psi}^{\beta},\quad \wick[positions={-1, 1}, offset=1em]{(D^{\alpha \dot{\alpha}} \c2 \phi_{1}) (D_{\alpha \dot{\alpha}} \c1 {\c2 \phi_{2}}) \c1 \phi_{3}} \to \Phi^{\dagger}
  \end{aligned}
\end{equation}
to those discussed previously. We intend these to stand in for similar rules
like \textit{e.g.} $\wick{\c \phi_{1} \Box \c \phi_{2}}$ as well. For the
field-strength contractions, there is the additional requirement that one or
both of the fields in the contraction be charged under the corresponding gauge
interaction, but these cannot be contracted into a gauge singlet, since the
field-strength tensor comes about from the anticommutator of the covariant
derivatives acting on the exotic fermion. These rules are
\begin{equation}
  \contraction[1ex]{}{\psi}{{}^{\beta}}{X}
  \bcontraction[1.3ex]{\psi^{\beta}}{X}{{}_{\alpha \mathsf{j}}^{\mathsf{i} \beta}}{\phi}
  \psi^{\alpha}_{\mathsf{i}} {X_{\alpha \mathsf{j}}^{\mathsf{i} \beta}} \phi \to \bar{\Psi}^{\beta}_{\mathsf{j}} , \quad
  \contraction[1ex]{}{\phi}{}{X}
  \bcontraction[1.3ex]{\phi}{X}{{}_{\alpha}^{\beta}}{\phi}
  \phi {X_{\alpha}^{\beta}} \phi \to \xmark , \quad
  \contraction[1ex]{}{\psi}{{}^{\beta}}{X}
  \bcontraction[1.3ex]{\psi^{\beta}}{X}{{}_{\alpha \mathsf{j}}^{\mathsf{i} \beta}}{\phi}
  \psi^{\alpha}_{\mathsf{i}} {X_{\alpha \mathsf{j}}^{\mathsf{i} \beta}} \phi^{\mathsf{j}} \to \xmark , \quad
\end{equation}
where $\mathsf{i}$ and $\mathsf{j}$ stand in for fundamental indices of
$\mathrm{SU}(2)_{L}$, $\mathrm{SU}(3)_{c}$, or no indices at all for the
field-strength tensor of $\mathrm{U}(1)_{Y}$.

Operators with derivatives coming about as this way, \textit{i.e.} as $D^{2}$ or
$X$, are often redundant from the perspective of model discovery, since they
imply the existence of the leading-order operator in which these derivatives do
not appear. Thus, the tree-level completions of these operators can be found by
studying the lower-dimensional operators without those derivatives or
field-strength tensors. It may however be the case that the leading-order
operator is absent, in which case these operators may be important. For the
$n_{f} = 3$ SMEFT with one Higgs doublet, we conjecture this can only come about
from operators with a structure like
\begin{equation}
  \label{eq:hdh}
  \mathcal{O}^{\mu} H^{i} \partial_{\mu} H^{j} \epsilon_{ij},
\end{equation}
which vanishes when the derivative is removed. (Similar structures like
$L^{i}_{r} L^{j}_{s} \epsilon_{ij}$ are non-vanishing since there is an
additional space of flavour indices to carry the antisymmetry.) This exception
does not apply to the case of field-strength tensors, since
$[X^{\mu\nu}, H] = 0$ for all field strengths $X$. This is the justification for
our earlier comments that operators containing field-strength tensors are not
interesting from the perspective of model discovery.

The replacement rules given in Eq.~\eqref{eq:box-rules} do not exhaust the
possible Lorentz-structures for two derivatives, scalars and fermions. The
additional structures involve single indices contracted between the derivatives,
and others contracted into fermions. Diagrammatically, we find that these
combinations come about from fermion lines containing two arrow-preserving
propagators, each contributing a factor of momentum. This would be the case, for
example, if $\psi$ in Fig.~\ref{fig:example-deriv} were a heavy arrow-preserving
propagator, as shown in Fig.~\ref{fig:example-deriv-2}. Here the rules are
\begin{equation}
  \label{eq:two-deriv-other-contractions}
  \wick{(D^{\alpha \dot{\alpha}} \c \psi_{1 \alpha}) (D^{\beta \dot{\beta}} \c \psi_{2 \beta}) \epsilon_{\dot{\alpha} \dot{\beta}}} \to \Phi^{\dagger}, \quad  \wick{(D^{\alpha \dot{\alpha}} \c \psi) (D^{\beta \dot{\beta}} \c \phi) \epsilon_{\dot{\alpha} \dot{\beta}}} \to \bar{\Psi}^{\beta},\quad \text{other combinations} \to \xmark .
\end{equation}

In summary, exploding derivative operators can lead to novel models that would
not be found by exploding non-derivative operators. We have already seen that
this happens when a structure such as Eq.~\eqref{eq:hdh} is present in the
operator. It can also happen when the presence of an odd number of derivatives
allows new topologies with novel chirality structures. The presence of an even
number of derivatives implies either that the derivatives arose as $D^{2}$ or
$X$, which usually do not contribute new models, or else from the contractions
of structures like those in Eq.~\eqref{eq:two-deriv-other-contractions}. It is
clear from Fig.~\ref{fig:example-deriv-2} that in such cases, the two
arrow-preserving fermion propagators can be replaced with arrow-violating
propagators, and indeed these will generically be present since we work with
vector-like fermions. So, with the exception of operators with structures like
Eq.~\eqref{eq:hdh}, studying single derivative operators is sufficient for model
discovery.

\subsubsection{Derivative operator examples}
\label{sec:deriv-op-examples}

\begin{figure}[t]
  \centering
  \subcaptionbox{\label{fig:od1-completion-1}}{
    \includegraphics[width=0.4\linewidth]{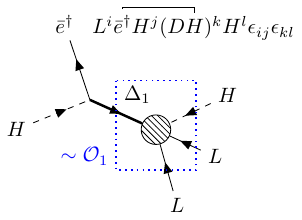}
  }
  \subcaptionbox{\label{fig:od1-completion-2}}{
    \includegraphics[width=0.35\linewidth]{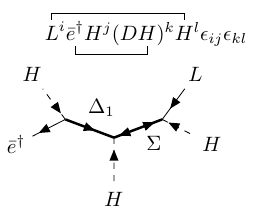}
    \qquad
    \includegraphics[width=0.35\linewidth]{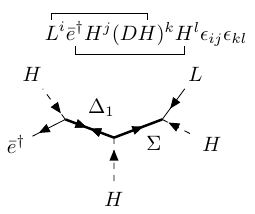}
  }
  \caption{(a) An intermediate topology representing the operator
    $\mathcal{O}_{D3}$ with all heavy fields except $\Delta_{1}$ integrated out.
    The contraction $\bar{e}^{\dagger} H \sim \Delta^{\dagger}_{1}$ gives rise
    to an effective operator similar to the Weinberg operator $\mathcal{O}_{1}$,
    shown in blue. This branch of the completion tree therefore involves models
    featuring $\Delta_{1}$ along with one of the seesaw fields. (b) The model
    with field content $\{\Sigma, \Delta_{1} + \bar{\Delta}^{\dagger}\}$ arises
    from two similar diagrams, shown here. These correspond to the two ways the
    arrow-preserving and arrow-violating fermion propagators can be placed in
    the graphs for this furnishing of the topology.}
  \label{fig:od1-completion}
\end{figure}

Among the simplest derivative operators in the $\Delta L = 2$ SMEFT is the
dimension seven operator
\begin{equation}
 \mathcal{O}_{D 3} = L^{i}_{\alpha} \bar{e}^{\dagger}_{\dot{\beta}} H^{j} (DH)^{\alpha \dot{\beta} k} H^{l} \epsilon_{ij} \epsilon_{kl}
\end{equation}
which we use as a paradigm for showing how single-derivative operators can be
opened up. We note that the operator's tree-level completions have also been
discussed in Ref.~\cite{delAguila:2012nu}. The placement of the derivative on
the Higgs field is enforced by the unique $\mathrm{SU}(2)_{L}$ contraction. This
is not generally true, and the derivative should be acted in all possible ways
if it can be. The contraction of $(DH)$ into another Higgs is forbidden by
Eq.~\eqref{eq:bad-deriv-repl-rules}. Thus, the $(DH)$ must be contracted into a
fermion. The options are
\begin{equation}
  \label{eq:contraction-choices}
  \wick{ \c L^{i}_{\alpha} (D {\c H})^{\alpha \dot{\beta} k} } \epsilon_{ij} \epsilon_{kl} \sim \Sigma^{\dagger \dot{\beta}}_{jl} \quad \text{ and } \quad \wick{ \c {\bar{e}}_{\dot{\beta}}^{\dagger} {(D {\c H})^{\alpha \dot{\beta} k}} } \sim \bar{\Delta}_{1}^{\alpha k}
\end{equation}
with the Dirac fermion
$\Delta_{1} + \bar{\Delta}_{1}^{\dagger} \sim (\mathbf{1}, \mathbf{2}, -\tfrac{1}{2})$
transforming like $L$ under $G_{\text{SM}}$. The field $\Sigma$ is the
protagonist in the type-III seesaw model, and further contractions on the
resulting operator $\Sigma^{\dagger}_{jl}\bar{e}^{\dagger}H^{j}H^{l}$ lead to
the models $\{\Sigma, \Delta_{1} + \bar{\Delta}^{\dagger}_{1}\}$ (from
$\wick{\c {\bar{e}^{\dagger}} \c H}$) and $\{\Sigma, \Xi_{1}\}$ (from
$\wick{\c H \c H}$) in that case. The second option in
Eq.~\eqref{eq:contraction-choices} leads to the operator
$L^{i} \Delta_{1}^{k} H^{j} H^{l} \epsilon_{ij} \epsilon_{jl}$, which is the
Weinberg operator with the second $L$ replaced with the exotic vector-like
lepton. This contraction is illustrated diagrammatically in
Fig.~\ref{fig:od1-completion-1}. It thus implies the same
completions\footnote{This phenomenon is discussed in more detail in
  Sec.~\ref{sec:filtering}.} as $\mathcal{O}_{1}$, each along with
$\Delta_{1} + \bar{\Delta}^{\dagger}_{1}$. This is expected since
$\bar{e}^{\dagger} (D H)$ transforms like $L$. There are then a total of five
completions, but four models, since two have the same particle content:
$\{\Sigma, \Delta_{1} + \bar{\Delta}^{\dagger}_{1}\}$. In
Fig.~\ref{fig:od1-completion-2} we show how this can be seen as coming about
from the fact that the chirality structure of the diagram allows two positions
for the arrow-preserving fermion propagator. Note that this is not the case for
the completion with the singlet fermion $N$. Interestingly, there are two
fermion-only models found: $\{N, \Delta_{1} + \bar{\Delta}^{\dagger}_{1}\}$ and
$\{\Sigma, \Delta_{1} + \bar{\Delta}^{\dagger}_{1}\}$. Both of them contain
seesaw fields, which is consistent with the proof of Ref.~\cite{Klein:2019iws}
that models containing two exotic fermion fields must contain one of $N$ or
$\Sigma$ if they violate lepton-number by two units. Since the structure of the
operator $\mathcal{O}_{D3}$ is unique, there is no work to be done in writing
down the organic operator generated by these models at the low scale.

We move on to a two-derivative operator example by studying a completion of
\begin{equation}
\mathcal{O}_{18 d} = L^{i} L^{j} H^{k} H^{l} (D^{\mu} H)^{m} (D_{\mu}\tilde{H})^{n} \epsilon_{ij} \epsilon_{km} \epsilon_{ln} ,
\end{equation}
which has the property that it vanishes when the derivatives are removed. (Note
that, comparing to the operator in Table~\ref{tab:long}, the first derivative
has been moved onto a Higgs field.) Applying the only allowed replacement rule
on the derivatives first implies the presence of the real triplet
scalar\footnote{We remind the reader that this is not the seesaw field present
  in the type-II scenario, which has unit hypercharge.}
$\Xi \sim (\mathbf{1}, \mathbf{3}, 0)_{S}$ in the theory
\begin{equation}
  \wick{L^{i} L^{j} H^{k} H^{l} (D^{\mu} \c H )^{m} (D_{\mu} \c {\tilde{H}} )^{n}} \epsilon_{ij} \epsilon_{km} \epsilon_{ln} \to   L^{i} L^{j} H^{k} H^{l} \Xi^{mn} \epsilon_{ij} \epsilon_{km} \epsilon_{ln} .
\end{equation}
From here there are a number of choices. We choose to look at a particular
scalar-only completion involving the unit-hypercharge isosinglet scalar present
in the Zee model $\mathcal{S}_{1} \sim (\mathbf{1}, \mathbf{1}, 1)$:
\begin{equation}
  \wick{\c L^{i} \c L^{j} H^{k} H^{l} \Xi^{mn}} \epsilon_{ij} \epsilon_{km} \epsilon_{ln} \to  \mathcal{S}^{\dagger}_{1} H^{k} H^{l} \Xi^{\dagger}_{kl},
\end{equation}
implying the interaction Lagrangian
\begin{equation}
  \label{eq:18d-comp-lag}
  \mathscr{L}_{\text{int}} \supset x_{[rs]} L_{r} L_{s} \mathcal{S}_{1} + \kappa H^{i} \tilde{H}^{j} \Xi_{\{ij\}} + \lambda \mathcal{S}_{1} \Xi^{ij} H^{\dagger}_{i} H^{\dagger}_{j} + \text{h.c.}
\end{equation}
This model was studied in Ref.~\cite{Law:2013dya} and identified as the simplest
neutrino mass model according to their assumptions. It has remarkably few free
parameters since the scalar $\Xi$ does not have Yukawa couplings to SM fermions,
and the couplings of $\mathcal{S}_{1}$ to leptons are antisymmetric in flavour.
As in the minimal Zee--Wolfenstein scenario~\cite{Wolfenstein:1980sy}, this
model implies a neutrino-mass matrix with zeros down the diagonal and is
therefore incompatible with neutrino oscillation data~\cite{He:2003ih}. It is,
however, a good example of how interesting models can be missed when overlooking
operators with derivatives in this model-building framework. The model generates
the following combination of basis operators at the low scale
\begin{equation}
  \mathcal{O}^{[rs]}_{\mathcal{S}_{1} + \Xi} = (L_{r}^{i}L_{s}^{j}) H^{k} H^{l} \Box( H^{m} \tilde{H}^{n}) \epsilon_{ij} \epsilon_{km} \epsilon_{ln} .
\end{equation}
Note that the operator is already symmetric under the interchage
$m \leftrightarrow n$, so another structure need not be added.

\subsection{An algorithm for model building}
\label{sec:algorithm}

With our basic completion recipe established, in the following we outline the
procedures we use to build the UV models that generate the operators listed in
Table~\ref{tab:long}, along with relevant metadata: the tree-level diagrams and
the models' Lagrangians. The methods are presented as they are implemented in
our example code~\cite{neutrinomass2020}.

We use a computational representation for tensors representing fields
transforming as irreducible representations of
$\mathrm{SU}(2)_{+} \otimes \mathrm{SU}(2)_{-} \otimes G_{\text{SM}}$ built on
top of the \textsf{SymPy} package~\cite{10.7717/peerj-cs.103} for symbolic
computation in \textsf{Python}, as well as
\textsf{BasisGen}~\cite{Criado:2019ugp} for group-theory functionality. The code
implements the Butler--Portugal algorithm~\cite{butler1991, MANSSUR_2002} for
obtaining the canonical form of tensorial expressions, which we use to simplify
operators and compare them for equality. Strings of fields and their derivatives
representing gauge- and Lorentz-invariant effective operators are dressed with
$\epsilon$ and $\delta$ tensors to form all possible invariants. In our specific
case, the content of these operators is constructed directly by taking the
product of all field combinations and keeping only those that contain a singlet
part in the decomposition. We checked this against results from the Hilbert
series, projecting out the $\Delta L = 2$ component for the pertinent operators,
and removing the spurions accounting for redundancies from field redefinitions
involving the classical EOM and IBP. For our study of the $\Delta L = 2$
operators, since we are interested in model discovery, we excluded derivative
operators that are non-explosive along with those that contain field-strength
tensors and contracted pairs of derivatives that do not lead to a vanishing
structure upon removal.

In practice we start with a template pattern of contractions corresponding to
the topologies that can accommodate the field content of the operators at
tree-level. These are generated using \textsf{FeynArts}~\cite{Hahn:2000kx}
through \textsf{Mathematica}, and filtered for isomorphism with graph-theory
tools~\cite{igraph2006, the_igraph_core_team_2020_3774399,
  szabolcs_horvat_2020_3739056, SciPyProceedings_11}. These templates provide
the order and pattern of contractions for classes of operators based on the
number of scalars and fermions they contain. Since no distinction is made at
this level between $(\mathbf{2}, \mathbf{1})$- and
$(\mathbf{1}, \mathbf{2})$-fermions, for some operators only a subset of these
templates will be relevant for our purposes, since some contractions may always
imply Proca fields. These templates are used to open up the operator with the
assumptions and methods presented in Sec.~\ref{sec:treelevelmatching}. Every
time a replacement rule is applied, the Feynman graph information is updated and
a Lagrangian term is generated as described in
Sec.~\ref{sec:exploding-operators}. After the procedure is finished, the full
Lagrangian of the model can be generated in the same way as the input effective
operators, described above.

We keep track of the quantum numbers of the heavy fields so as to be economic
with exotic degrees of freedom, while still providing some flexibility in the
model database. Concretely, if a field arises from a contraction whose
corresponding term has already appeared in the Lagrangian, the two associated
exotic fields are identified. If two fields come about from different
contractions but share the same quantum numbers they are distinguished, since it
may be possible that some symmetry would forbid one term but not the other. The
choice to identify fields not only reduces the number of fields in each model,
but may also reduce the total number of completions for a given operator. This
is due to couplings between exotic fields that vanish in the absence of some
exotic generational structure. For example,
$\phi^{i} \phi^{j} \epsilon_{ij} = 0$ for some exotic isodoublet $\phi$, or
$\eta^{a} \eta^{b} \eta^{c} \epsilon_{abc} = 0$ for a colour-triplet $\eta$.

We have attempted to validate our example code against many results in the
literature. We have been able to reproduce the results of
Refs.~\cite{delAguila:2012nu, Cai:2014kra, Herrero-Garcia:2016uab,
  deBlas:2017xtg, deGouvea:2019xzm}, which give systematic listings of models
that generate effective operators at tree-level. Ref.~\cite{deBlas:2017xtg}
provides a UV dictionary for the dimension-six SMEFT. Validation of these
results first required the adaptation of the dimension-six operators to
something analogous to the overcomplete spanning set of type-3 operators used
here. The entire process in this case---including generating the set of
operators, finding the completions and matching examples back onto the Warsaw
basis---is provided as an interactive notebook accompanying our example code. We
note that such matching calculations can also be automated with the help of
automated tools~\cite{Criado:2017khh, Bakshi:2018ics}. For the other studies
mentioned, we provide our validation of their results along with our example
code.

\section{Neutrino mass model building}
\label{sec:modelbuilding}

Up until now we have tried to keep the discussion of exploding operators
general, but in this and following sections we specialise to the case of opening
up operators to build radiative models of Majorana neutrino mass. We discuss the
process of turning $\Delta L = 2$ operators into neutrino self-energy graphs,
the tree-level topologies of the operators, and the methods we use to ensure a
given model's contribution to the neutrino mass is the dominant one.

\subsection{Operator closures and neutrino-mass estimates}
\label{sec:operator-closures}

For operators other than the Weinberg-like ones, neutrino masses are necessarily
generated at loop level. The fields of the $\Delta L = 2$ operator need to be
looped off using SM interactions in such a way that a Weinberg-like operator is
generated after the SM fields are integrated out. We call this the operator
\textit{closure} and it represents the mixing between the $\Delta L = 2$
operator and the Weinberg-like ones. Examples of $\Delta L = 2$ operator
closures are given in Table~\ref{tab:example-closures}, and these are referred
to throughout this section. The closure provides enough information to know the
number of loops in the neutrino self-energy graph (since the $\Delta L = 2$
operator is generated at tree level) and to estimate the scale of the new
physics underlying the operator. We automate the operator-closure process by
applying the methods discussed below through a pattern-matching
algorithm~\cite{krebber2018, krebber2017nonlinear}. The program is a part of our
public example-code repository.

Current neutrino oscillation data provide a lower bound on the mass of the
heaviest neutrino, coming from the measured atmospheric mass-squared difference
$\Delta m^{2}_{\text{atm}} \approx (0.05 \text{ eV})^{2}$~\cite{nufitweb,
  Esteban:2018azc}. We take the neutrino-mass scale $m_{\nu} \approx 0.05 \text{
  eV}$, so that the new-physics scale is bounded above by the implied scale we
estimate for each operator. This is derived by estimating the loop-level
operator closure diagrams. In our case we are interested in estimating the scale
of the neutrino mass in the UV models generating the operator, rather than the
calculable loop-level contributions to the neutrino mass in the EFT. We
associate a factor of $(16\pi^{2})^{-1} \approx 6.3 \cdot 10^{-3}$ with each
loop and assume unit operator coefficients for the non-renormalisable
$\Delta L = 2$ vertices. We take the SM Yukawa couplings to be diagonal and
include factors of $g \approx 0.63$ appropriately for interaction vertices
involving $W$ bosons. Neutrino-mass matrices proportional to Yukawa couplings
will be dominated by the contributions from the third generation of SM fermions
in the absence of any special flavour hierarchy in the new-physics couplings.
For this reason, we consider only the effects of third-generation SM fermions in
our estimates, but mention that our program can be straightforwardly extended to
accommodate the general case where light-fermion Yukawas and off-diagonal CKM
matrix elements appear in the neutrino-mass matrix. For derivative-operator
closures, we can include the $W$ boson from the covariant derivative if it is
present and necessary to correctly close off the diagram. Otherwise, the vertex
should come with an additional factor of momentum. We work in the Feynman gauge
to avoid spurious factors of $\Lambda$ in the neutrino-mass
estimates~\cite{delAguila:2012nu}. The overall scale-suppression of the neutrino
mass is determined by the Weinberg-like operator generated at the low scale. In
most cases, this is the dimension-five operator $\mathcal{O}_{1}$, which implies
$m^{\mathcal{O}_{1}}_{\nu} \sim v^{2} / \Lambda$. Closures leading to the
loop-level generation of $\mathcal{O}_{1}^{\prime}$ and
$\mathcal{O}_{1}^{\prime\prime}$ can also be found, and these naively imply a
significant suppression of the neutrino mass compared to the $\mathcal{O}_{1}$
case: $m^{\mathcal{O}_{1}^{\prime}}_{\nu} \sim v^{4}/ \Lambda^{3}$ and
$m^{\mathcal{O}_{1}^{\prime\prime}}_{\nu} \sim v^{6} / \Lambda^{5}$. However, a
diagram with additional Higgs loops can always be drawn to recover the Weinberg
operator at the low scale. Despite the additional loop suppression, these
diagrams will dominate over those generating $\mathcal{O}_{1}^{\prime}$ and
$\mathcal{O}_{1}^{\prime\prime}$ as long as
$\Lambda \gtrsim 4 \pi v \approx \SI{2.2}{\TeV}$~\cite{Babu:2001ex,
  deGouvea:2007qla}.

It is still true that higher-dimensional operators typically imply smaller
neutrino masses. There are two main reasons for this. First, the number of loops
required for the closure of the operator generally increases with increasing
mass dimension. Second, operators containing more fields imply neutrino
self-energy diagrams containing more couplings. Many of these are SM Yukawas
which (with the exception of $y_{t}$) are small and tend to suppress the
neutrino mass, despite the contributions being dominated by the third
generation. Non-minimal choices such as small exotic Yukawa couplings or
hierarchical flavour structures in the operator coefficients can also lead to
additional suppression of the neutrino mass, and in turn of the implied scale of
the new physics.

\begin{figure}[t]
  \centering
  \includegraphics[width=0.6\linewidth]{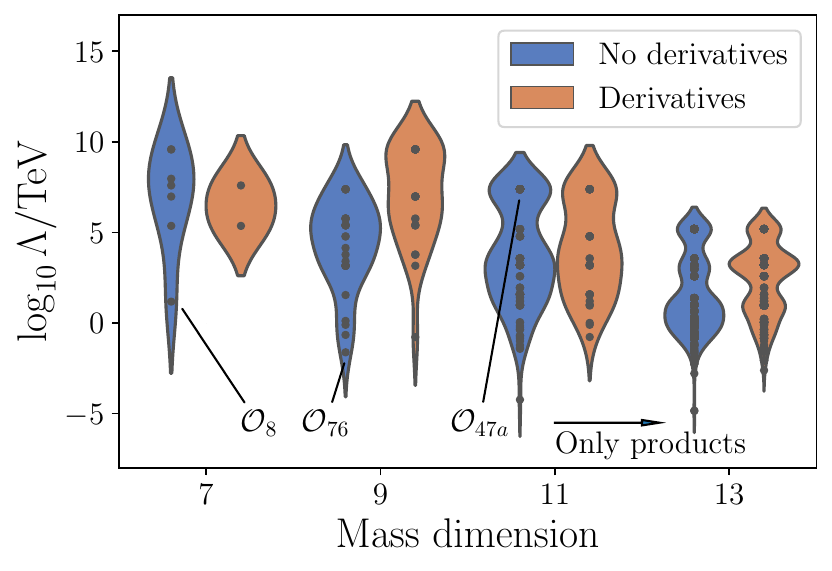}
  \caption{The figure shows smoothed histograms of the number of operators that
    have an estimated upper bound of $\Lambda$ on the new-physics scale. Black
    dots generally represent more than one operator. The strips are broken up by
    mass dimension and whether the operators contain derivatives or not. We
    assume unit operator coefficients and the dominance of third-generation
    SM-fermion contributions in the closure diagrams. Operators containing no
    derivatives (blue) are separated from those containing one derivative
    (orange). Those containing more than one derivative are not included in the
    figure, since in most cases these come about at next-to-leading order in the
    EFT expansion, and therefore imply the same $\Lambda$ values as the
    lower-dimensional operator with two fewer derivatives. The dimension-eleven
    operators containing one derivative and all of the dimension-13 operators
    shown are constructed from the lower-dimensional operators in our listing
    only as products. This means that the set of operators plotted above that do
    not feature in Table~\ref{tab:long} are incomplete. However, we do not find
    that similar product-type operators at dimensions nine and eleven give
    special estimates for the neutrino mass or $\Lambda$, and so we expect these
    results to be representative of the true situation up to mass-dimension 13.
    The general decrease in $\Lambda$ with increasing operator mass dimension is
    evident in the figure. The most suppressed closures tend to be of
    non-derivative operators. By mass-dimension eleven it becomes clear that a
    class of operators, those with the structure shown in
    Eq.~\eqref{eq:unsuppressed-ops}, are less suppressed than the rest.}
  \label{fig:violinplot-closures}
\end{figure}

In Fig.~\ref{fig:violinplot-closures} we show the new-physics scales $\Lambda$
associated with neutrino-mass generation from the $\Delta L = 2$ operators in
the SMEFT up to dimension 13, assuming unit operator coefficients and the
dominance of third-generation couplings. We separate single-derivative operators
from those that contain no derivatives, and choose not to include operators
containing more than one derivative in the figure. This is because these
operators most often arise at next-to-leading order in the EFT expansion, and
therefore usually imply a neutrino-mass scale identical to that of
lower-dimensional operators. The dimension-eleven operators with derivatives as
well as the dimension-13 operators are constructed only as products of
lower-dimensional ones, making the set of operators incomplete. We highlight
that similar kinds of product operators at dimensions eleven and nine do not
imply special values for the estimated neutrino-mass scale or $\Lambda$, and
therefore we expect the results to be representative of the situation up to
dimension 13. From the figure, it is clear that there is a trend towards smaller
values of $\Lambda$ with increasing mass dimension. By dimension 13, the implied
new-physics scale is between $1$ and $100~\text{TeV}$ for most operators. It
seems to be the case that the most constrained closures are generally those of
non-derivative operators.

\begin{longtable}[t]{ccc}
  \caption{The table shows an assortment of $\Delta L = 2$ operator closures, displaying a number of paradigmatic motifs. We represent flavour indices in a sans-serif typeface here to avoid confusion with subscripts labelling the Yukawa couplings. The expressions given for $m_{\nu}^{\textsf{rs}}$ needs to be symmetrised in $\textsf{rs}$, something we do not explicitly indicate in the table. These expressions carry flavour indices in alphabetical order on the fields as they appear in Table~\ref{tab:long}. Here $\kappa$ represents the operator coefficient, $V$ is the CKM matrix and $y^{\textsf{r}}_{e,u,d}$ are the diagonal electron, up-type and down-type Yukawa couplings in the SM. A number of operators require an external electron to be converted into a neutrino. This often necessitates the introduction of a $W$ boson or an unphysical charged Higgs $H^{+}$. Operator $\mathcal{O}_{8}$ generates the dimension-seven analogue of the Weinberg operator with the two-loop diagram shown. (There is a lower order diagram with an $H^{+}$ in place of the $W$ that happens to vanish~\cite{Babu:2010vp}.) A three-loop diagram in which two of the external Higgs lines are looped off leads to mixing with the Weinberg operator. Operator $\mathcal{O}_{76}$ generates the dimension-nine operator $LLHH(H^{\dagger}H)^{2}$, and hence five- and six-loop diagrams are also implied. There is usually more than one choice about where to attach the $W$ boson if one is present in a diagram, and the additional diagrams with the $W$ connecting in other possible ways are left implicit.\label{tab:example-closures}}\\
  \toprule
  Operator & Diagram & $m_{\nu}^{\textsf{rs}}$ \\
  \midrule
  \endfirsthead
  \toprule
  Operator & Diagram & $m_{\nu}^{\textsf{rs}}$ \\
  \midrule
  \endhead
  $4b$ &
\begin{minipage}{0.2\linewidth}
  \centering
  \includegraphics[width=1.2\linewidth]{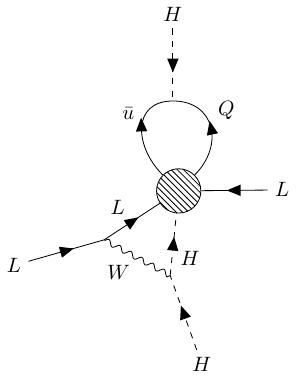}
\end{minipage}
  & $\displaystyle \kappa_{\textsf{[rs]tt}} \frac{g^{2} y_u^{\textsf{t}}}{(16\pi^{2})^{2}} \frac{v^{2}}{\Lambda}$  \\
    $8$ &
    \begin{minipage}{0.2\linewidth}
        \centering
        \includegraphics[width=1.2\linewidth]{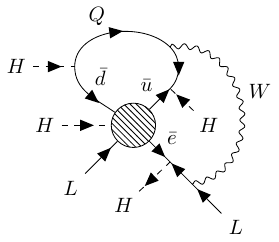}
      \end{minipage} & $\displaystyle \kappa_{\textsf{rstu}} V_{\textsf{tu}} \frac{g^{2} y_{e}^{\textsf{s}} y_{u}^{\textsf{t}} y_{d}^{\textsf{u}}}{(16\pi^{2})^{2}} \frac{v^{2}}{\Lambda} \left( \frac{v^{2}}{\Lambda^{2}} + \frac{1}{16 \pi^{2}}\right)$ \\
    $D3$ &
      \begin{minipage}{0.2\linewidth}
        \centering
        \includegraphics[width=\linewidth]{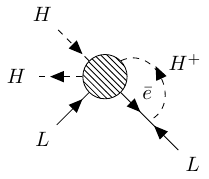}
      \end{minipage} & $\displaystyle \kappa_{\textsf{rs}} \frac{y_{e}^{\textsf{s}}}{16\pi^{2}}  \frac{v^{2}}{\Lambda}$ \\
  $11b$ &
          \begin{minipage}{0.3\linewidth}
            \centering
            \includegraphics[width=\linewidth]{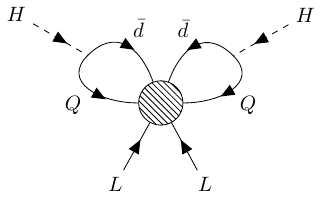}
          \end{minipage}
        & $\displaystyle \kappa_{\textsf{rstutu}} \frac{y_d^{\textsf{t}} y_d^{\textsf{u}}}{(16\pi^{2})^{2}} \frac{v^{2}}{\Lambda}$ \\
    $76$ &
\begin{minipage}{0.3\linewidth}
  \centering
  \includegraphics[width=0.94\linewidth]{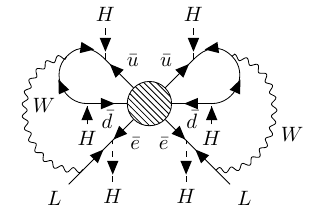}
\end{minipage}
  & $\displaystyle \kappa_{\textsf{rstuvw}} V_{\textsf{tv}} V_{\textsf{uw}} \frac{g^{4} y^{\textsf{r}}_{e} y^{\textsf{s}}_{e} y_{u}^{\textsf{t}} y_{u}^{\textsf{u}} y_{d}^{\textsf{v}} y_{d}^{\textsf{w}}}{(16\pi^{2})^{4}} \frac{v^{2}}{\Lambda} \left(\frac{v^{2}}{\Lambda^{2}} + \frac{1}{16\pi^{2}} \right)^{2} $ \\
    $47a$ &
    \begin{minipage}{0.2\linewidth}
        \centering
        \includegraphics[width=\linewidth]{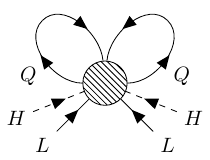}
      \end{minipage} & $\displaystyle \kappa_{\textsf{rstuvw}}\frac{1}{(16\pi^{2})^{2}} \frac{v^{2}}{\Lambda}$ \\
  $56$ &
 \begin{minipage}{0.2\linewidth}
  \centering
  \includegraphics[width=1.1\linewidth]{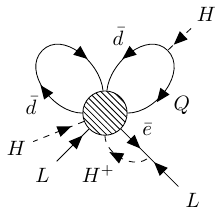}
\end{minipage}
  & $\displaystyle \kappa_{\textsf{rsttvv}}\frac{y_{e}^{\textsf{s}} y_{d}^{\textsf{t}}}{(16\pi^{2})^{3}} \frac{v^{2}}{\Lambda}$ \\
  \bottomrule
\end{longtable}

We note that at dimension eleven it begins to become clear that the
neutrino-mass estimates associated with a category of operators remain large.
These operators include $\mathcal{O}_{47a}$, whose closure is shown in
Table~\ref{tab:example-closures}, and 44 others like it which have loops that
contain no connecting Higgs, and therefore no additional suppression from SM
Yukawa couplings\footnote{A UV example of such a model was presented and studied
  in Ref.~\cite{Gargalionis:2019drk} for $\mathcal{O}_{47j}$. A number of other
  examples were also mentioned in Ref.~\cite{Babu:2019mfe}, including a two-loop
  model generating a dimension-13 operator at tree level.}. These operators have
the form
\begin{equation}
  \label{eq:unsuppressed-ops}
\mathcal{O}_{1} \cdot \prod_{i=1}^{n} \psi_{i}^{\dagger} \psi_{i},
\end{equation}
where $\psi_{i}$ are SM fermion fields, and imply
\begin{equation}
  m_{\nu} \sim \kappa \frac{1}{(16\pi^{2})^{n}} \frac{v^{2}}{\Lambda},
\end{equation}
with $\kappa$ the operator coefficient. The loop suppression becomes too great
to meet the atmospheric bound at $n=6$. Although five loops are viable in the
absence of any other suppression, the operators
$\mathcal{O}_{1} \cdot \prod_{i=1}^{5} (\psi_{i}^{\dagger} \psi_{i})$ cannot
form a Lorentz-singlet without a derivative. This suggests that dimension-21
operators of the form
\begin{equation}
  \label{eq:dim21}
  LLH(\partial^{\mu}H) \cdot (\psi_{0} \sigma_{\mu} \psi_{0}^{\dagger}) \prod_{i=1}^{4} \psi_{i}^{\dagger} \psi_{i}
\end{equation}
are the highest-dimensional operators leading to phenomenologically viable
neutrino masses. They require new physics below $\sim 6~\text{TeV}$. All of the
tree-level topologies associated with the structure in Eq.~\eqref{eq:dim21}
imply that the neutrino mass depends on the product of nine or more
dimensionless couplings. It is clear from Fig.~\ref{fig:violinplot-closures}
that these operators are outliers, and the associated new-physics scale is
already heavily constrained by dimension 13 for most.

\begin{figure}[t]
  \centering
  \includegraphics[width=0.3\linewidth]{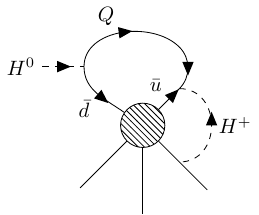}
  \caption{For some operators containing $\bar{d} \bar{u}^{\dagger}$ the
    operator closure involves a motif like that shown in the figure. There is
    always an additional diagram with the roles of the unphysical Higgs and
    $H^{0}$ interchanged. Both diagrams are proportional to $y_{u} y_{d}$ but
    related by a negative sign coming from the couplings of $H^{+}$ to up- and
    down-type quarks as shown in Eq.~\eqref{eq:sm-yukawa-negative}, and
    therefore their sum vanishes.}
  \label{fig:vanishing-loop}
\end{figure}

Estimates for the neutrino mass for the majority of the $\Delta L = 2$ operators
without derivatives have been given previously in Ref.~\cite{deGouvea:2007qla}.
Those that we present here differ in two ways:
\begin{enumerate}
  \item We aim to estimate the contribution to the neutrino mass implied by the
    completions of the operator, not the operator alone. This means, for
    example, that we do not need more loops of gauge bosons to provide
    additional factors of momentum on fermion loops with no mass insertions,
    since it is implicit that the appropriate factors of momentum will arise at
    higher orders in the EFT expansion. Such arrow-preserving loops, as shown in
    the closures of $\mathcal{O}_{47a}$ and $\mathcal{O}_{56}$ in
    Table~\ref{tab:example-closures}, vanish by even--odd parity arguments
    absent these higher-order contributions~\cite{deGouvea:2007qla}. Indeed, in
    UV models built from these operators the additional gauge-boson loops are
    not necessary~\cite{Angel:2012ug, Gargalionis:2019drk}. This means that for
    operators such as $\mathcal{O}_{47a}$ and $\mathcal{O}_{56}$, our
    neutrino-mass estimates are enhanced with respect to those presented in
    Ref.~\cite{deGouvea:2007qla} by $16\pi^{2} / g^{2}$.
  \item In some cases, operators containing a factor of
    $\bar{u}^{\dagger} \bar{d}$ require a closure with $W$ bosons rather than
    $H^{+}$, since the sum of the diagrams with the unphysical Higgs fields
    vanishes~\cite{Babu:2010vp}. The situation is shown in
    Fig.~\ref{fig:vanishing-loop} for a general one-loop case of this
    phenomenon. Ultimately this comes from the relative negative sign in the
    Lagrangian between the up- and down-type Yukawa interactions:
    \begin{equation}
      \label{eq:sm-yukawa-negative}
    \mathscr{L}_{\text{Yuk}} \supset y_{u}^{r} V_{rt} d_{t} \bar{u}_{r} H^{+} - y_{d}^{r} V_{tr} \bar{d}^{\dagger}_{r} u^{\dagger}_{t} H^{+} \ .
  \end{equation}
    As shown in the Fig.~\ref{fig:vanishing-loop}, the fermion loop requires a
    mass insertion on the quark line to which the $H^{+}$ does not connect,
    making both loops proportional to $y_{u} y_{d}$ but with differing signs.
    Care must be taken to ensure that the loop functions are also necessarily
    the same in cases where this property is used.
\end{enumerate}
It might be possible that, in a similar way to (2) above, the sum of diagrams
with different $W$ placements or of the neutrino-flavour-symmetrised diagrams
might also lead to additional cancellations which further decrease the upper
bound on the new-physics scale. This not a possibility we explore in detail
here, but note that similar cancellations have been noted in the
literature~\cite{Gargalionis:2019drk}.

Our estimates for the neutrino mass are provided as symbolic mathematical
expressions in our model database. Where possible these been checked against
more detailed calculations and UV models in the literature generating the
operators to ensure acceptable agreement~\cite{Duerr:2011zd, Babu:2009aq,
  Babu:2010vp, delAguila:2012nu, Cai:2014kra, Zee:1985id, Babu:1988ki,
  Angel:2013hla, Gargalionis:2019drk}. The predictions for the new-physics scale
associated with each operator are provided in Table~\ref{tab:long}, along with
the number of loops in the closure. Operators for which a range is given for the
number of loops are those that generate the dimension-seven or dimension-nine
analogues of the Weinberg operator. As touched on above, the additional Higgs
fields in these closures can always be closed off, adding more loops to the
neutrino self-energy diagram while reducing the overall scale suppression. The
contribution with the highest number of loops will dominate for scales
$\Lambda \gtrsim 4\pi v$.

We note that in some cases, more insights can be made about the structure of the
neutrino-mass matrix from the nature of the operator, even in the general form
with which they appear in our classification. For example, there is only one
independent Lorentz-structure associated with $\mathcal{O}_{4b}$:
$\kappa^{\mathcal{O}_{4b}}_{rstt} (L^{i}_{r} L^{j}_{s} \epsilon_{ij})(Q^{\dagger}_{t k}\bar{u}^{\dagger}_{t})H^{k}$,
from which it can be seen that the operator coefficient must be antisymmetric in
$rs$ from Fermi--Dirac statistics. It is clear from the diagram associated with
the operator in Table~\ref{tab:example-closures} that the loop integral will
depend on an external lepton flavour, and this dependence can only come from
charged-lepton masses, \textit{i.e.} $I(m_{e}^{r})$. Then the complete
expression for the estimated neutrino mass will be something like
\begin{align}
  m_{\nu}^{\{rs\}} &\sim \sum_{t} \frac{g_{2}^{2} y_{u}^{t}}{(16\pi^{2})^{2}} \frac{v^{2}}{\Lambda} [\kappa_{[rs]tt} I(m_{e}^{r}) + \kappa_{[sr]tt} I(m_{e}^{s})] \\
  &= \sum_{t} \frac{g_{2}^{2} y_{u}^{t}}{(16\pi^{2})^{2}} \frac{v^{2}}{\Lambda} \kappa_{[rs]tt}  [I(m_{e}^{r}) - I(m_{e}^{s})],
\end{align}
which implies a neutrino-mass matrix with zeros down the diagonal, similar to
that following from the Lagrangian in Eq.~\eqref{eq:18d-comp-lag}. Such a
texture is disfavoured by neutrino oscillation data. Studying the structure of
the neutrino-mass matrices implied by a complete basis of $\Delta L = 2$
operators would allow more, similar conclusions to be drawn in a
model-independent way. Recently, a complete basis of operators in the SMEFT at
dimension nine has been written down~\cite{Li:2020xlh}, and this could
facilitate such an effort.

\subsection{UV considerations}
\label{sec:uvconsiderations}

We now turn to the UV structure of the operators: their completion topologies,
the associated neutrino self-energy graphs, and the nature of the exotic fields
that feature therein. Central to our study of neutrino mass is the requirement
that a model represent the leading contribution to the neutrino mass, a
condition we impose through a process of model filtering, also discussed in the
present section.

\subsubsection{Tree-level completion topologies}

The tree-level UV topologies depend on the number of fermions and scalars in the
operator, and this is how we choose to label them. Thus, a dimension-eleven
operator with two scalars and six fermions has topologies labelled $2s6f_{i}$.
We do not distinguish between $(\mathbf{2}, \mathbf{1})$- and
$(\mathbf{1}, \mathbf{2})$-fermions in this classification, and some of these
topologies will therefore always imply the existence of heavy vector particles
in the completions. In our analysis these models are not considered, but the
topologies are still presented here in general. Each topology corresponds to a
pattern of contractions in the language of Sec.~\ref{sec:treelevelmatching}, and
sometimes we use this perspective.

We present the different topology types in Table~\ref{tab:topology-data} along
with peripheral information relating to these. The number of propagators in the
diagrams represents an inclusive upper bound on the number of exotic fields
allowed in the completions of the associated operators, counting Dirac fermions
as one exotic field. In many cases, repetition in the operator's field content
can lead to fewer fields furnishing the internal lines of the diagram, since we
identify fields with the same quantum numbers. To avoid clutter we keep the
complete gallery of tree-level diagrams in our online example-code repository,
and instead only show some of the graphs here. For some topology types the
relevant diagrams have already appeared in earlier parts of the paper, and these
figures are referenced in the table. We make more specific comments about the
topology types by operator mass dimension below.

\begin{table}[t]
  \centering
  \begin{tabular}{ccccc}
    \toprule
    Topology type & Operators & Topologies & Propagators & Figure \\
    \midrule
    $0s4f$        & $2$       & 1          & 1           & $\hookrightarrow$         \\
    $0s6f$        & $16$      & 2          & 3           & \ref{fig:6f-topologies}   \\
    $1s4f$        & $16$      & 2          & 2           & \ref{fig:1s4f-topologies} \\
    $2s2f$        & $7$       & 1          & 1           & \ref{fig:seesaw-figs}     \\
    $2s4f$        & $29$      & 8          & 2,3         & \ref{fig:2s4f-topologies} \\
    $2s6f$        & $137$     & 35         & 4,5         & \ref{fig:2s6f-topologies} \\
    $3s2f$        & $3$       & 4          & 1,2         & \ref{fig:3s2f-topologies} \\
    $3s4f$        & $15$      & 23         & 3,4         & \ref{fig:3s4f-topologies} \\
    $4s2f$        & $8$       & 10         & 2,3         & \ref{fig:4s2f-topologies} \\
    $5s2f$        & $1$       & 24         & 2,3,4       & \ref{fig:5s2f-topologies} \\
    $5s4f$        & $15$      & 264        & 4,5,6       & $\hookrightarrow$         \\
    $6s2f$        & $1$       & 66         & 3,4,5       & $\hookrightarrow$         \\
    \bottomrule
  \end{tabular}
  \caption{The table shows the topology classes encountered in our operator
    listing along with related information: the number of pertinent operators,
    the number of tree-level topologies associated with each topology type, the
    number of internal lines featuring in the diagrams (given as a range), and
    the appropriate figure reference in the text. Although there is one $0s4f$
    topology, all of the pertinent operators in our listing are non-explosive
    because they contain derivatives. The symbol $\hookrightarrow$ indicates
    that we do not present these topologies in this paper; instead, we point the
    interested reader to our online database and example code for the relevant
    diagrams. We highlight that although the topologies are labelled only by
    their field content, the pertinent operators may include one or more
    derivatives. We point the reader to the main text for a detailed breakdown
    by mass-dimension of the topologies that are relevant to each operator.}
  \label{tab:topology-data}
\end{table}

\begin{figure}[t]
  \centering
  \subcaptionbox{\label{fig:4s2f-topologies}}{\centering
    \includegraphics[scale=0.13]{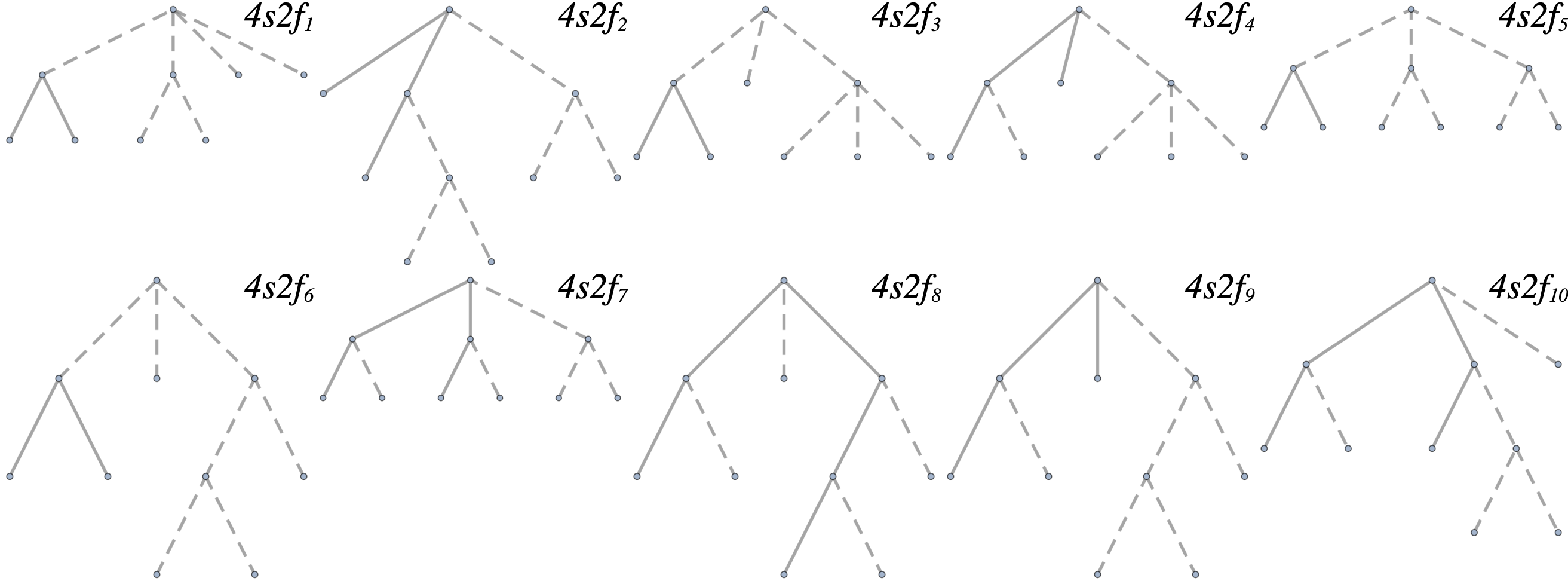}
  }
  \subcaptionbox{\label{fig:1s4f-topologies}}{
    \includegraphics[scale=0.13]{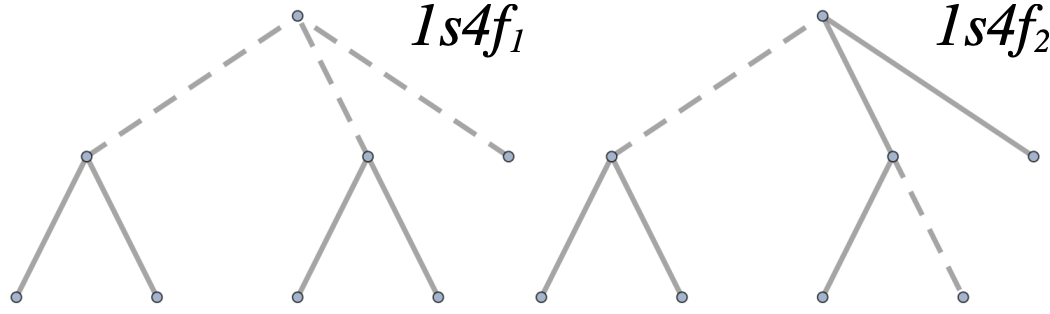}}
  \caption{(a) The tree-level topologies relevant for the completions of the
    four-scalar--two-fermion operator $\mathcal{O}_{1}^{\prime}$. Only topology
    $4s2f_{3}$ leads to a novel completion that does not feature a seesaw field.
    We point out that topology $4s2f_{8}$ permits fermion-only completions,
    which are expected only for the Weinberg-like operators in the absence of
    derivatives. (b) The two tree-level topologies relevant for the completions
    of the one-scalar--four-fermion dimension-seven operators in
    Table~\ref{tab:long}. The internal fermion line on $1s4f_{2}$ must be
    arrow-violating for all of the operators we consider.}
  \label{fig:d7-topologies}
\end{figure}

\paragraph{Dimension seven} At dimension seven there are three broad classes of
$\Delta L = 2$ operators by field-content: $0s4f$, $1s4f$ and $4s2f$ in our
classification scheme. Operator $\mathcal{O}_{D1}$ is one of only two $0s4f$
operators in the entire listing, both of which are non-explosive\footnote{We
  note that although $\mathcal{O}_{D1}$ is non-explosive, one-loop completions
  exist that lead to three-loop neutrino mass models.}. The Weinberg-like
$\mathcal{O}_{1}^{\prime}$ is the only $4s2f$ operator at dimension seven, while
there are six $4s2f$ operators: $\mathcal{O}_{2}$, $\mathcal{O}_{3a,b}$,
$\mathcal{O}_{4a,b}$ and $\mathcal{O}_{8}$. The UV topologies relevant for the
dimension-seven operators are presented in Fig.~\ref{fig:d7-topologies}. There
are only two tree-level topologies associated with the $1s4f$ operators. One
involves two exotic scalars, the other an exotic scalar and a heavy fermion with
an arrow-violating propagator line. There are ten topologies associated with the
$4s2f$ class, for which the only pertinent operator is
$\mathcal{O}_{1}^{\prime}$. Only topology $4s2f_{3}$ is associated with a model
that does not contain seesaw fields. Topology $4s2f_{6}$ accommodates up to
three exotic scalars and $4s2f_{8}$ allows up to three exotic fermions. Such
fermion-only models are expected only for the Weinberg-like operators, in the
absence of derivatives. The remaining topologies allow all other combinations up
to three fields for the number of exotic scalars and fermions introduced.
Radiative neutrino mass from the dimension-seven operators without derivatives
was also studied in Ref.~\cite{Cai:2014kra}.

\begin{figure}[t]
  \centering
  \subcaptionbox{\label{fig:3s4f-topologies}}{
    \includegraphics[scale=0.13]{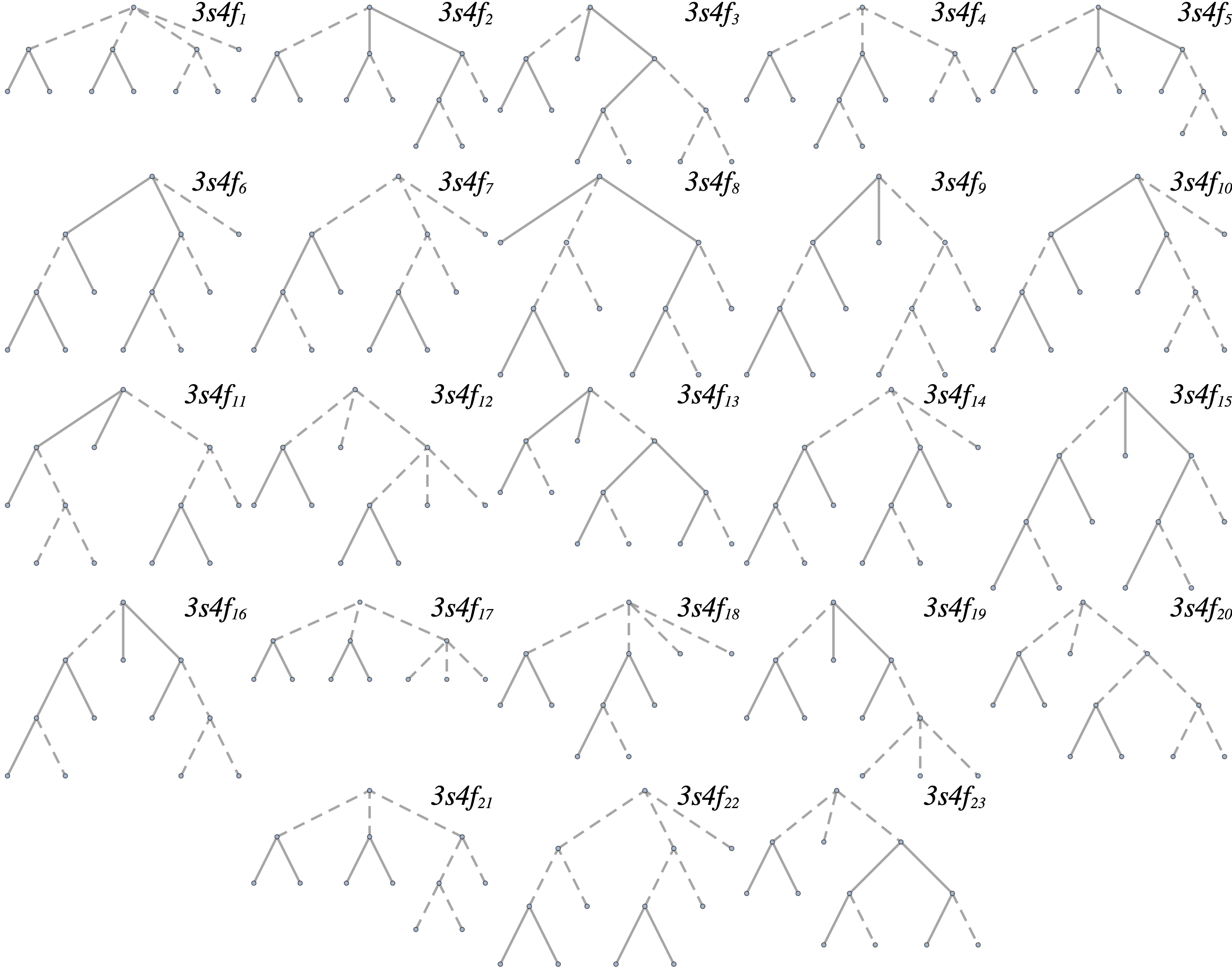}}
  \subcaptionbox{\label{fig:2s4f-topologies}}{
    \includegraphics[scale=0.13]{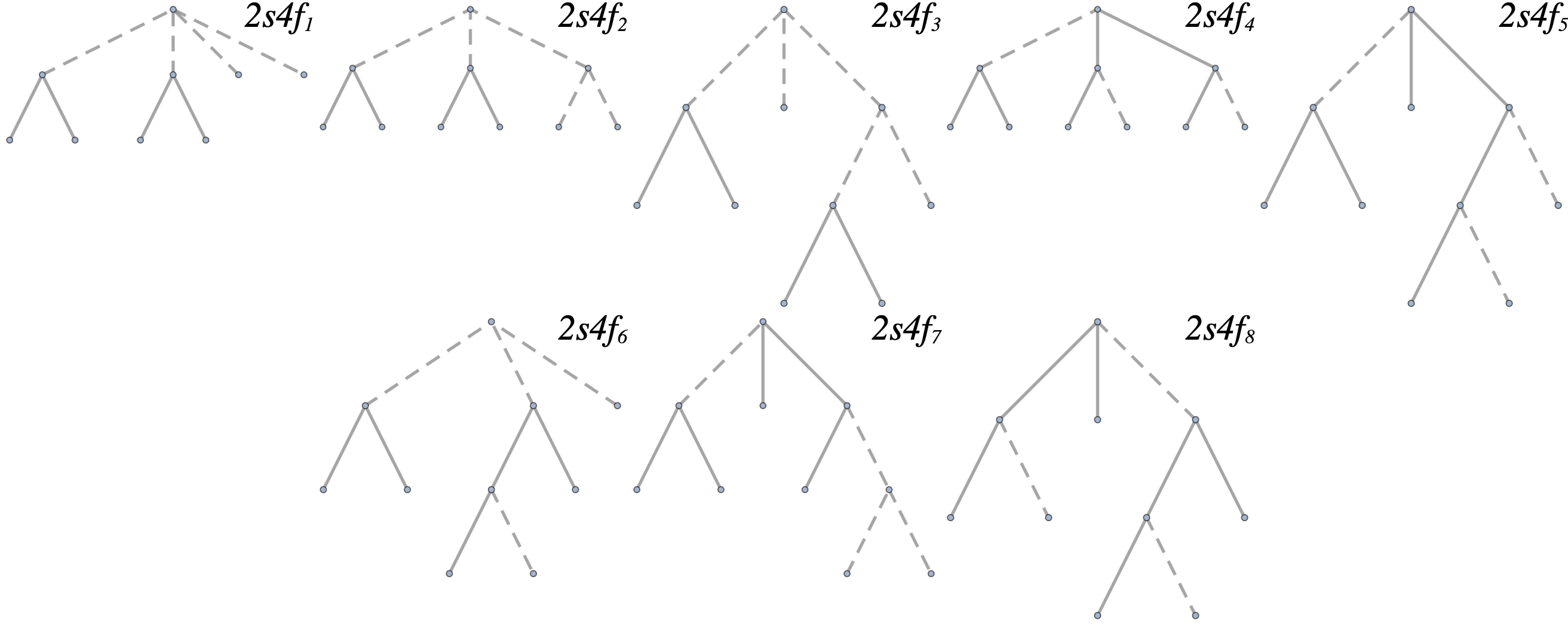}}
  \caption{(a) The figure shows the tree-level topologies relevant to $3s4f$
    operators. Topologies $3s4f_{3}$, $3s4f_{13}$, $3s4f_{20}$ and $3s4f_{23}$
    imply one exotic field that couples only to other exotics in the diagram.
    This topology class is relevant to a large number of dimension-nine
    operators, and these are the lowest-dimensional examples of operators
    containing this property in our listing. (b) The two-scalar--four-fermion
    topologies associated with dimension-nine single-derivative operators in our
    catalogue. Since only single-derivative operators furnish these graphs, only
    those topologies containing at least one arrow-preserving internal fermion
    line are relevant. These are topologies $2s4f_{4}$ -- $2s4f_{8}$; the other
    fermion propagator in $2s4f_{4}$ and $2s4f_{5}$ must be arrow violating.
    Topologies $2s4f_{1}$ -- $2s4f_{3}$ each give rise to completions involving
    exotic Proca fields.}
  \label{fig:d9-topologies}
\end{figure}

\begin{figure}[t]
  \centering
  \subcaptionbox{\label{fig:5s2f-topologies}}{
    \includegraphics[scale=0.13]{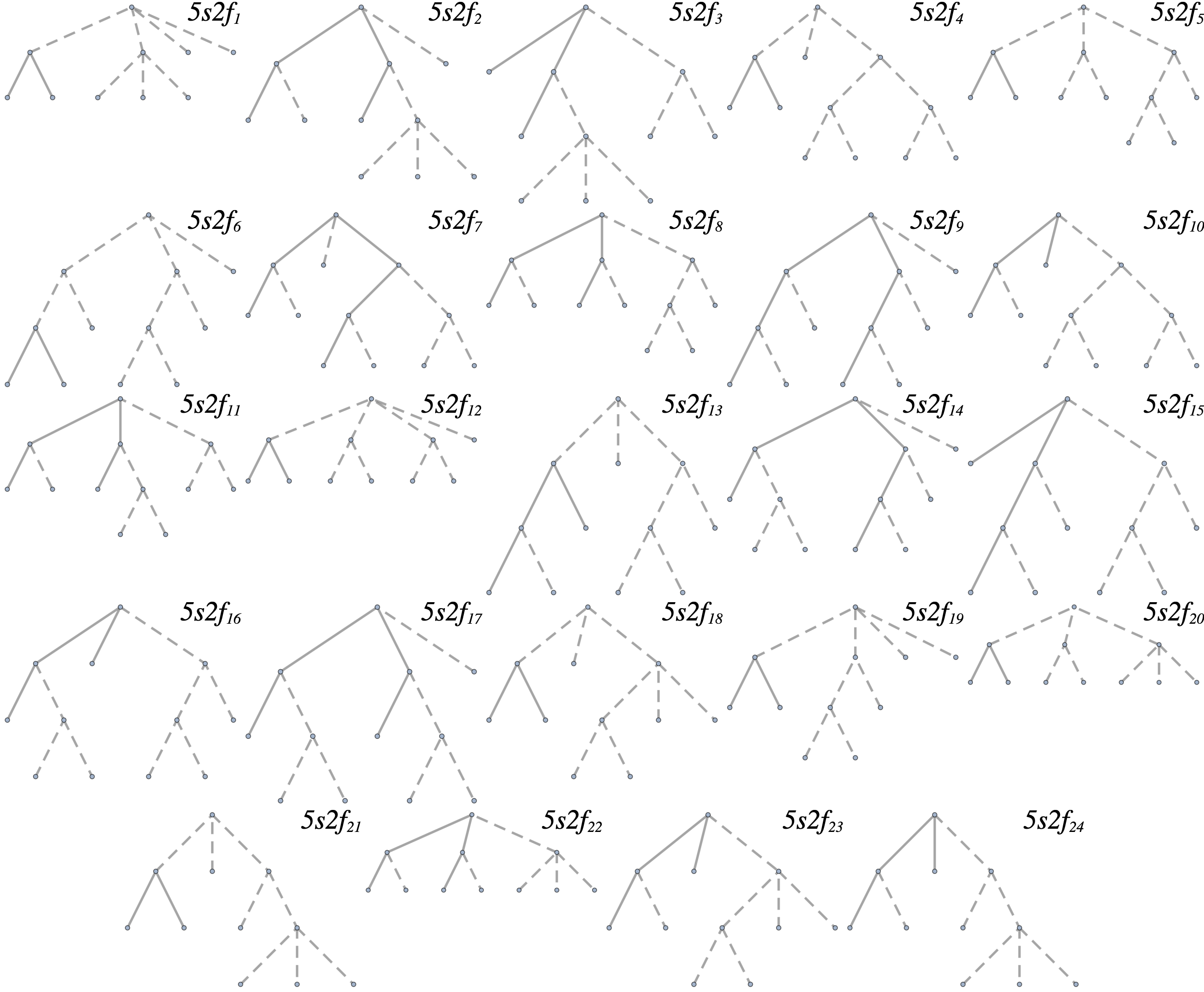}}
  \subcaptionbox{\label{fig:3s2f-topologies}}{
    \includegraphics[scale=0.13]{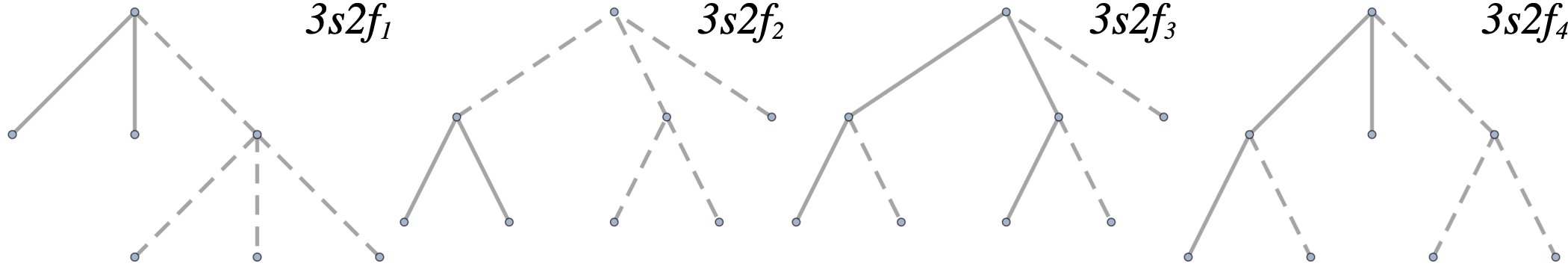}}
  \caption{(a) The $5s2f$ topologies relevant only to the single-derivative
    operator $\mathcal{O}_{D20}$. Only those topologies allowing one
    arrow-preserving internal fermion line give completions allowed in our
    framework. Topologies $5s2f_{4}$, $5s2f_{7}$ and $5s2f_{10}$ contain heavy
    fields that couple only to other exotics in the diagram. (b) The UV diagrams
    associated with the $3s2f$ operator $\mathcal{O}_{D21}$. Only the last two
    diagrams can generate the operator under our model-building assumptions.}
  \label{fig:d9-topologies-derivs}
\end{figure}

\paragraph{Dimension nine} At dimension nine there are 79 operators in our
catalogue. There are 16 operators containing six fermions, these are
$\mathcal{O}_{9}$ through to $\mathcal{O}_{20}$ as well as $\mathcal{O}_{76}$.
The relevant tree-level topologies are presented in
Fig.~\ref{fig:6f-topologies}. There are 15 $3s4f$ operators, most of which have
the form $\mathcal{O}_{1} \cdot \mathcal{O}_{\text{SM Yukawa}}$ or $H^\dagger H$
times a $1s4f$ dimension-seven operator. These are operators $\mathcal{O}_{5}$
through to $\mathcal{O}_{7}$ as well as $\mathcal{O}_{61}$, $\mathcal{O}_{71}$,
$\mathcal{O}_{77}$, $\mathcal{O}_{78}$ and $\mathcal{O}_{8}^{\prime}$. These
topologies are shown in Fig.~\ref{fig:3s4f-topologies}. There is a single $6s2f$
operator: the Weinberg-like $\mathcal{O}_{1}^{\prime\prime}$. The remaining 47
operators contain derivatives. Those that contain an even number share
topologies with dimension-five or dimension-seven operators. These include
$\mathcal{O}_{D19}$, a $2s2f$ operator, $\mathcal{O}_{D18}$ and
$\mathcal{O}_{D22}$ which are $4s2f$ operators with associated topologies shown
in Fig.~\ref{fig:4s2f-topologies}, as well as $\mathcal{O}_{D4}$,
$\mathcal{O}_{D7}$, $\mathcal{O}_{D13}$ and $\mathcal{O}_{D15}$ for which the
$1s4f$ topologies of Fig.~\ref{fig:1s4f-topologies} are relevant. The remaining
operators contain an odd number of derivatives. The operators
$\mathcal{O}_{D5}$, $\mathcal{O}_{D6}$, $\mathcal{O}_{D8}\text{ --
}\mathcal{O}_{D10}$, $\mathcal{O}_{D12}$, $\mathcal{O}_{D14}$,
$\mathcal{O}_{D16}$ and $\mathcal{O}_{D17}$ are of type $2s4f$, implying
entirely new topologies, shown in Fig.~\ref{fig:2s4f-topologies}. To these we
add the $5s2f$ operator $\mathcal{O}_{D20}$ and the $3s2f$ operator
$\mathcal{O}_{D21}$, which also have novel structure.
Fig.~\ref{fig:5s2f-topologies} and Fig.~\ref{fig:3s2f-topologies} are relevant
in this case. For the operators that contain an odd number of derivatives, only
the topologies allowing at least one arrow-preserving fermion propagator do not
contain exotic Proca fields. Some $3s4f$ and $5s2f$ topologies have the
interesting property that they involve exotic fields that couple only to other
exotic fields in the diagram. These are the lowest-dimensional operators in our
listing having this feature, although this becomes more common at dimension
eleven.

We note that the tree-level topologies can also be important in telling which
derivative operators might provide novel completions. As discussed in
Sec.~\ref{sec:exploding-derivative-operators}, many operators containing more
than one derivative have no model-discovery utility. These are operators
generated past the leading order in the expansion of the heavy propagators in
the UV theory, and their completions are always found by exploding the
lower-dimensional operators with an even number of fewer derivatives. One way to
diagnose such a situation is to check how many arrow-preserving fermion lines
are present in the tree-level topologies associated with an operator. If all of
the graphs contain fewer such propagators than the number of derivatives in the
operator, then any model generating this operator will also generate the
corresponding lower-dimensional one. At dimension nine there are seven operator
classes that fall into this category. The four operator families
$\mathcal{O}_{D4}$, $\mathcal{O}_{D7}$, $\mathcal{O}_{D13}$ and
$\mathcal{O}_{D15}$ each contain two derivatives. These operators are identified
above as fitting into the $1s4f$ topology class. It is clear from
Fig.~\ref{fig:1s4f-topologies} that no two-fermion completions are relevant to
this class, and the Lorentz structure of these operators is such that the
internal fermion can only by arrow-violating. This suggests that models
generating these operators at tree-level will always also generate the
derivative-free dimension-seven operators $\mathcal{O}_{2}$, $\mathcal{O}_{3}$,
$\mathcal{O}_{4}$ and $\mathcal{O}_{8}$, respectively. There are two
three-derivative operators: $\mathcal{O}_{D21}$, of topology class $3s2f$, and
$\mathcal{O}_{D11}$, a $0s4f$ operator. The latter is non-explosive and
therefore not relevant to a discussion of tree-level model building. The $3s2f$
class admits completions that contain one and two fermions: those associated
with topologies $3s2f_{4}$ and $3s2f_{2}$, respectively. In both both cases we
find that the operator's structure allows for only a single arrow-preserving
propagator in each diagram. As before, this suggests that $\mathcal{O}_{D21}$ is
not interesting for model discovery, and its completions will be found by
studying $\mathcal{O}_{D3}$. Finally, there is also one four-derivative operator
at dimension nine: the $2s2f$ operator $\mathcal{O}_{D19}$ whose completions
coincide with those of the Weinberg operator $\mathcal{O}_{1}$. This means that
the only two-derivative operators in our listing that could contribute new
completions to the pool of neutrino-mass models are $\mathcal{O}_{D18}$ and
$\mathcal{O}_{D22}$. Operator $\mathcal{O}_{D22}$ has the feature that the
removal of the derivatives causes the operator to vanish, while this is not true
for all of the $\mathrm{SU}(2)_{L}$ structures associated with
$\mathcal{O}_{D18}$.

\begin{figure}[t]
  \centering
    \includegraphics[scale=0.13]{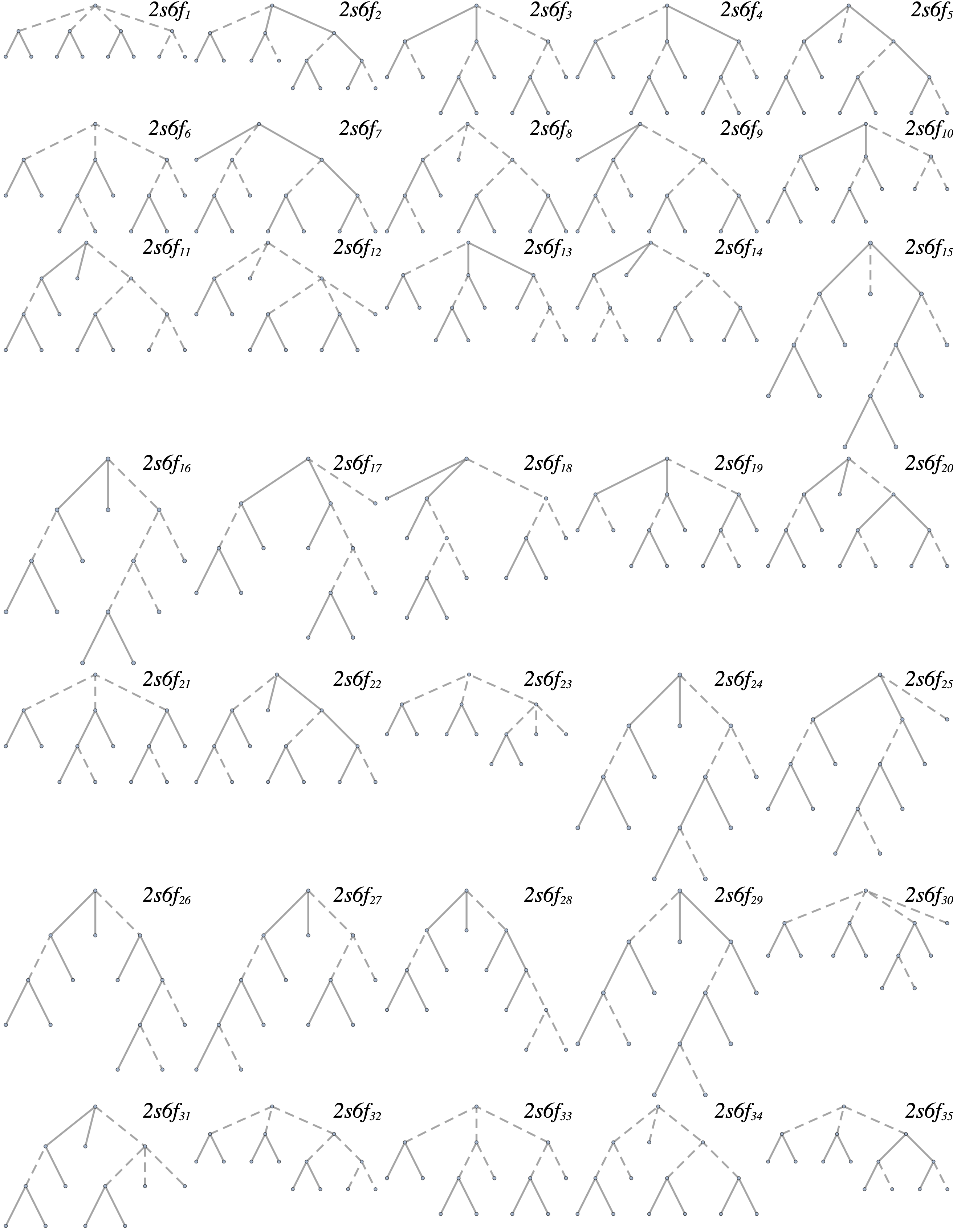}
    \caption{The tree-level topologies associated with the large class of $2s6f$
      dimension-eleven operators in our listing. A number of graphs display the
      feature --- less common at dimension nine --- that an exotic field in the
      diagram couples only to other internal lines.}
  \label{fig:2s6f-topologies}
\end{figure}

\paragraph{Dimension eleven} By far the largest class of operators at dimension
eleven is the $2s6f$ topology type, for which the topologies are presented in
Fig.~\ref{fig:2s6f-topologies}. These operators are mostly formed as products of
$0s6f$ dimension-nine operators with $H^{\dagger} H$, or $1s4f$ dimension-seven
operators with SM Yukawa couplings. They are operators $\mathcal{O}_{21}$
through to $\mathcal{O}_{65}$, excluding the structures associated with
$\mathcal{O}_{61}$, as well as $\mathcal{O}_{75}$, $\mathcal{O}_{76}^{\prime}$
and $\mathcal{O}_{82}$. The only other major class relevant to the
derivative-free dimension-eleven operators is $5s4f$ for which there are 264
tree-level topologies. These are presented with our example code, along with the
topologies relevant to the single $6s2f$ operator
$\mathcal{O}_{1}^{\prime\prime\prime}$. This dimension-eleven generalisation of
the Weinberg operator has already received some attention in the
literature~\cite{Anamiati:2018cuq}.

\subsubsection{Model filtering}
\label{sec:filtering}

The completions constructed by exploding the $\Delta L = 2$ operators are not
all automatically guaranteed to provide the leading-order contribution to the
neutrino mass. The same $\Delta L = 2$ Lagrangian may, for example, inevitably
imply another, larger contribution. Alternatively, the dominant contribution may
come from other LNV combinations of couplings in the model's full Lagrangian.
The relative importance of different mechanisms may also depend on the
assumptions of the model builder. Some neutrino-mass diagrams will dominate over
others only in certain regions of parameter space. Are these regions accessible
without large hierarchies in exotic couplings? Are such hierarchies acceptable,
if necessary to render a mechanism dominant? What about exotic flavours or
additional symmetries? Model filtering is the process of removing those models
that, under some set of assumptions, do not provide the leading-order
contribution to the neutrino mass. Our approach to filtering neutrino mass
models is contrasted against other possible approaches below, and we also make
more general comments about model filtering in other contexts. We mention that
the following discussion of filtering is similar in intent to that of
`genuineness' in the loop-level matching paradigm~\cite{Bonnet:2012kz,
  Sierra:2014rxa, Cepedello:2018rfh, Farzan:2012ev}. We sometimes adopt this
notation as well, and call models `genuine' if they represent the dominant
contribution to the neutrino mass.

\paragraph{Filtering criterion} We begin by noting that model filtering is
ubiquitous when considering tree-level effects. Here, the filtering criterion is
unambiguously the operator dimension, since higher-dimensional operators are
inevitably suppressed compared to lower-dimensional ones. With regard to the
$\Delta L = 2$ EFT, such a dimension-focused criterion is useful for thinking
about LNV scattering events, for example. As discussed in
Sec.~\ref{sec:operator-closures}, the operator dimension is also a rough
indication of the predicted neutrino-mass scale, and therefore has some utility
in anticipating which models will dominate the neutrino mass.

We point out that this approach to model filtering allows for the immediate
rejection of some models, already during the process of opening up the operator.
This can happen, for example, when a contraction introduces an exotic particle
transforming like a SM field. Taking $\mathcal{O}_{2}$ as an example,
contractions like
\begin{equation}
  \overset{\tiny \hspace{-2em} \varphi^{\dagger} \sim (\mathbf{1}, \mathbf{2}, \tfrac{1}{2})}{\wick{L^{i} \c L^{j} L^{k} \c {\bar{e}} H^{l}} \epsilon_{ij} \epsilon_{kl}} \rightarrow L^{i} L^{k} \tilde{\varphi}^{j} H^{l} \epsilon_{ij} \epsilon_{kl}, \label{eq:weinberglike}
\end{equation}
with $\tilde{\varphi}$ a second Higgs doublet, always imply that further
contractions will produce seesaw fields, since the RHS of
Eq.~\eqref{eq:weinberglike} has the same structure as the Weinberg operator. We
note that for fermions the situation is more subtle because of the Lorentz
structure. Specifically, although $H \bar{u}$ transforms like $Q^{\dagger}$
under $G_{\text{SM}}$, the Lorentz transformation properties are different. The
derivative contraction $(D^{\alpha \dot{\alpha}} H) \bar{u}_{\alpha}$ does
transform like $Q$ under
$\mathrm{SU}(2)_{+} \otimes \mathrm{SU}(2)_{-} \otimes G_{\text{SM}}$ and that
makes a number of such contractions forbidden if one is interested in only
dominant contributions according to the mass-dimension criterion. This is the
same phenomenon as that seen in the paradigmatic opening of the derivative
operator $\mathcal{O}_{D3}$ given in Sec.~\ref{sec:deriv-op-examples}, where an
exotic field transforming like $L$ [see Eq.~\eqref{eq:contraction-choices}] lead
to a similar Weinberg-like operator at an intermediate stage in the completion
procedure. We note that this does not completely rule out exotic copies of SM
fields featuring in radiative neutrino mass models. Using $\mathcal{O}_{2}$
again as an example:
\begin{equation}
  \overset{\tiny \varphi^{\dagger} \sim (\mathbf{1}, \mathbf{2}, \tfrac{1}{2})}{\wick{L^{i} L^{j} \c L^{k} \c {\bar{e}} H^{l}} \epsilon_{ij} \epsilon_{kl}} \rightarrow L^{i} L^{j} \tilde{\varphi}^{k} H^{l} \epsilon_{ij} \epsilon_{kl}
\end{equation}
is allowed, since the $\mathrm{SU}(2)_{L}$ structure of this operator differs to
that of the Weinberg operator. Similarly, vector-like quarks and leptons are
extensively found in completions of both derivative and non-derivative operators
after the filtering procedure, but their SM and Lorentz quantum numbers are
interchanged with respect to their SM counterparts. For example, a particular
completion of $\mathcal{O}_{3a}$ is
\begin{equation}
  \overset{\tiny U \sim (\mathbf{3}, \mathbf{1}, \tfrac{2}{3})}{\wick{L^{i} L^{j} \c Q^{k} \bar{d} \c H^{l} \epsilon_{ij} \epsilon_{kl}}} \rightarrow \overset{\tiny \hspace{-3em} \mathcal{S}^{\dagger}_{1} \sim (\mathbf{1}, \mathbf{1}, -1)}{\wick{\c L^{i} \c L^{j} U \bar{d} \epsilon_{ij} \epsilon_{kl}}} \rightarrow \mathcal{S}^{\dagger}_{1} U \bar{d} \ ,
\end{equation}
which contains the vector-like quark $U + \bar{U}^{\dagger}$. Note however that
$U$ transforms like $\bar{u}^{\dagger}$ under $G_{\text{SM}}$, but oppositely
under the Lorentz group. It is true that $\bar{U}$, the vector-like partner of
$U$, does transform like $\bar{u}$, but this plays no role in the operator.

Since we are most interested in radiative neutrino mass, a more direct and
relevant filtering criterion in our case is the neutrino-mass estimate from the
closure graph of the operator. This is the metric we use to compare and filter
models in the results we present in Sec.~\ref{sec:models}. Whichever filtering
criterion is chosen, the conditions for generating the lower-dimensional
operator or the dominant neutrino self-energy graph still depend on the
filtering philosophy.

\paragraph{Filtering philosophy} The filtering criterion defines a hierarchy
among the effective operators. If one is interested in tree-level effects, then
operators of low dimension have a high priority in the sense that their effects
are dominant over those of high-dimensional operators, whose influence is
suppressed by additional powers of $\Lambda$. Similarly, the operators whose
closure graphs imply large contributions to the neutrino mass have a higher
priority than those implying small contributions.

One could take the view that it is sufficient for a subset of the field content
associated with a completion of a high-priority operator to be present in that
of a lower-priority one for it to be filtered out, even if the relevant diagrams
depend on entirely different couplings and interactions. We call this
perspective \textit{democratic}, in the sense that it treats all allowed
couplings and interactions fairly and ignores possible hierarchies in free
parameters. A democratic approach would then filter out all completions of
$\Delta L = 2$ operators of mass dimension larger than five containing one of
the seesaw fields, for example, since these always imply a dominant contribution
from the dimension-five Weinberg operator. Even if the same couplings are not
present in both diagrams, there is no reason, on this view, for one coupling to
be very much larger than another, making the tree-level contribution dominant.

An alternative approach might be to filter out only those completions that
necessarily lead to subdominant contributions to the neutrino mass in all
regions of parameter space. Naively it seems that neutrino-mass mechanisms
involving different couplings would all survive the filtering process in this
case, since the relative ordering of the contributions from each diagram depends
on the chosen values of the coupling constants. This is in general only
guaranteed if a symmetry is recovered in the Lagrangian when one coupling is
turned off, so that the forbidden coupling is not generated at some higher order
in perturbation theory. We call this approach \textit{stringent} filtering,
since the conditions for removing a model are more difficult to satisfy.

For our results in Sec.~\ref{sec:models} we take an intermediate view, leaning
more towards the democratic side. We filter on the basis of particle content,
but always keep track of the baryon-number assignment of the field. We then keep
models with identical SM quantum numbers if the baryon-number assignments of the
fields differ. With a concrete example, we treat
$\zeta^{(\prime)} \sim (\bar{\mathbf{3}}, \mathbf{1}, \tfrac{1}{3})$ in
$x (L^{i} Q^{j}) \zeta^{\{kl\}} \epsilon_{ik} \epsilon_{jl}$ and
$y (Q^{i} Q^{j}) \zeta^{\prime \{kl\}} \epsilon_{ik} \epsilon_{jl}$ as different
fields.

In practice, we enumerate the completions of the operators in order of their
estimated contribution to the neutrino mass. We associate a prime number with
each exotic field encountered, including baryon-number as a distinguishing
property. Models then correspond to products of prime numbers. As we explode
each operator in order, we remove models from the list of completions if their
characteristic number is divisible by that of any models already seen. In this
way, we remove those mechanisms that are subdominant contributions to the
neutrino mass in the democratic sense.

We emphasise that this procedure is not sufficient to fully ensure that the
remaining models are genuinely dominant contributions to the neutrino mass. For
example, it may be the case that the Weinberg operator is generated by loops of
a subset of the exotic particles in one of our models. We are not sensitive to
these models since we are concerned only with tree-level completions of the
operators. One-loop contributions to the neutrino mass from heavy loops can be
diagnosed easily on topological grounds. For example, topology \textsf{T-3} of
Ref.~\cite{Bonnet:2009ej} will come about whenever the neutrinos in the diagram
are connected by a single exotic fermion~\cite{Angel:2012ug}. At two-loops, one
could check the full gauge- and Lorentz-invariant Lagrangian for each model
against Table 1 of Ref.~\cite{Sierra:2014rxa}, for example. We do not include
this in our default filtering procedure, since it would require generating the
full Lagrangian of each model. This is a computationally prohibitive task,
especially since Table~\ref{tab:topology-data} suggests that the completions of
some operators can contain up to six exotic fields. Should any model from our
database be chosen for further study, the full Lagrangian can be generated with
the functions in our example code and studied for the presence such heavy loops.
We note that sometimes the presence of a heavy loop can be diagnosed from the
neutrino self-energy graph, or even the tree-level topology, and we give a
detailed example of such a case in Sec.~\ref{sec:simple-models}. An additional
filter on the models that goes beyond our initial tree-level filtering analysis
is the possibility of exotic fields gaining vacuum expectation values. In this
case, diagrams may exist that imply larger contributions to the neutrino mass
than that suggested by our approach, and we are not sensitive to these since
they generate exotic operators other than the Weinberg operator at the low
scale. Examples are presented in Refs.~\cite{Popov:2019tyc, Babu:2020hun}, where
in both cases a two-loop completion of the Weinberg operator also generates the
exotic operator $LLH^{\dagger}\Theta_{3}$, where
$\Theta_{3} \sim (\mathbf{1}, \mathbf{4}, \tfrac{3}{2})_{S}$.

We note that our model database~\cite{gargalionis_john_2020_4054618} contains
both the unfiltered completions of the operators in Table~\ref{tab:long}, as
well as the models filtered according to the above method. Our example code also
includes functions for filtering on interactions rather than fields, and finding
$\mathrm{U}(1)$ symmetries present in models' Lagrangians. Thus, the results
presented in Sec.~\ref{sec:models} and Table~\ref{tab:long} can be readily
reproduced with alternative filtering criteria, philosophies or approaches.

\section{Models}
\label{sec:models}

In this section we present the radiative models derived by exploding the
$\Delta L = 2$ operators catalogued in Table~\ref{tab:long}. We give an overview
of the models, and explore their particle content and the effects of the partial
model-filtering method we present in Sec.~\ref{sec:filtering}. We do not provide
the entire listing of models here because there are very many, but instead give
some examples. We point the interested reader to our database for the full
searchable listing.

We distinguish the terms `model' and `neutrino-mass mechanism' or
`$\Delta L = 2$ Lagrangian' in this section. By model we mean a collection of
particle content. Those same multiplets may have many combinations of couplings
that violate lepton-number by two units, leading to meany neutrino-mass
mechanisms, or $\Delta L = 2$ Lagrangians. We use the word `completion' here to
mean a neutrino-mass mechanism derived from a particular effective operator.
Used in this way, the same $\Delta L = 2$ Lagrangian may be shared by two
completions, but they correspond to the same model. We also remind the reader
that we use the words `field' and `multiplet' interchangeably.

We note here that the following analysis does not include the dimension-eleven
generalisation of the Weinberg operator $\mathcal{O}_{1}^{\prime\prime\prime}$,
since the operator has an unwieldy number of topologies and the relevant
tree-level completions have already been studied in the
literature~\cite{Anamiati:2018cuq}.

\subsection{Overview}
\label{sec:modelsoverview}

The models are generated by running the algorithm summarised in
Sec.~\ref{sec:algorithm}, as found in our example code, on our catalogue of the
$\Delta L = 2$ operators. The results for the number of completions before and
after filtering are presented in Table~\ref{tab:long}. In the language of
Sec.~\ref{sec:filtering}, we use the democratic filtering procedure with the
neutrino-mass scale as the filtering criterion for these data. We note again
that this leaves us with an overestimate of the actual number of genuine
neutrino-mass models. Even so, one can see that 54 operators end up with no
completions after filtering, ruling them out as possibly playing a dominant role
in generating the neutrino masses, at least according to our model-building
assumptions. The complete list of unfiltered $\Delta L = 2$ Lagrangians and
tree-level completion diagrams is compiled in our database, and the
documentation provides information for how to perform different kinds of
filtering on the models.

The database contains 430,810 inequivalent $\Delta L = 2$ Lagrangians before
filtering. Counted democratically (\textit{i.e.} by particle content) these
correspond to 141,989 unfiltered models. Of the distinct Lagrangians, only
around $3\%$ (11,483) survive democratic filtering with the neutrino-mass
criterion. This corresponds to 11,216 distinct models. In our filtering analysis
we also incorporate information from the one-loop study of the Weinberg
operator\footnote{We anticipate the number of models in our database generating
  the Weinberg operator with exotic loops at higher-loop order to be small. Such
  models would need to contain upwards of four exotic fields, and it becomes
  increasingly less likely that a model will contain a subset of these fields to
  be filtered out.} done in Ref.~\cite{Bonnet:2012kz}. We generate a listing of
the models from Tables~2 and 3 of Ref.~\cite{Bonnet:2012kz} with hypercharges
that are multiples of $1/6$ in the range $[-3,3]$ and ranges for the
$\mathrm{SU}(3)_{c}$ and $\mathrm{SU}(2)_{L}$ representations that cover those
of the exotic fields featuring in our models. We remove the completions in our
listing that contain a subset of these fields and imply neutrino masses
suppressed by more than one loop factor, since the models presented in
Ref.~\cite{Bonnet:2012kz} generate the Weinberg operator at one loop.

\begin{figure}[t]
  \centering
  \includegraphics[width=0.56\textwidth]{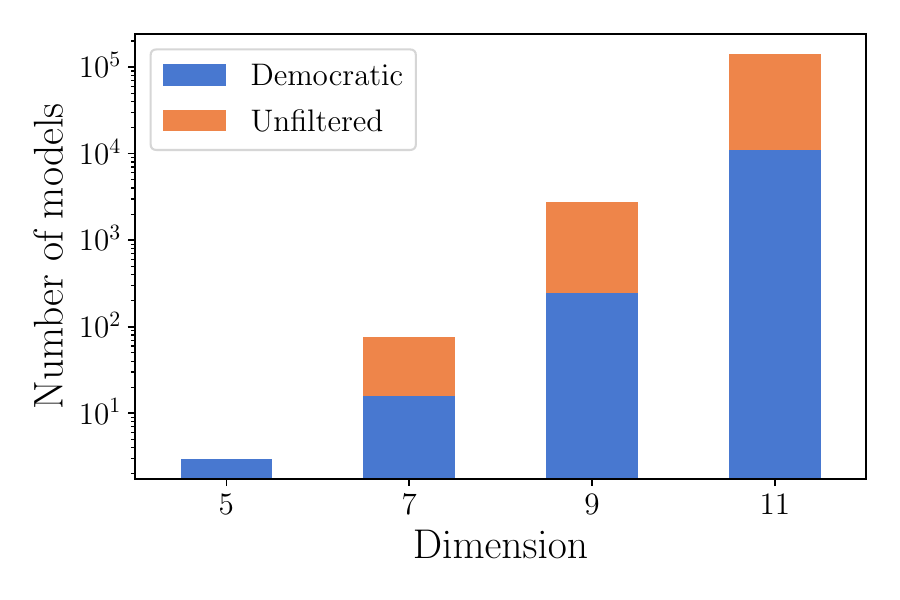}
  \caption{The bar chart shows the number of distinct Lagrangians derived from
    operators of different mass dimension. The orange bars show the number of
    distinct unfiltered models. The blue bars show the number after democratic
    filtering. The number of filtered completions grows with mass dimension.}
  \label{fig:filter-bar-dimension}
\end{figure}

We visualise the number of models with democratic and no filtering in
Fig.~\ref{fig:filter-bar-dimension} broken down by mass dimension. After
filtering there are three models at dimension five, 16 models at dimension
seven, 244 models at dimension nine and 10,969 models at dimension
eleven\footnote{We note that the sum of these numbers is not 11,216 since one
  model can generate multiple operators of different mass dimension in a way
  consistent with our neutrino-mass filtering criterion.}. It is clear that the
number of filtered neutrino-mass models grows with operator dimension, which is
perhaps unintuitive. For any high dimensional operator, there are competing
effects influencing the number of viable completions. First, the large number of
models derived already from lower-dimensional operators means that the chances
some model will be filtered out are larger. Second, high dimensional operators
involve more fields, meaning that there are more combinations of contractions
that can be made, and therefore more completions expected. Despite the increased
filtering odds, evidently the combinatorial explosion of different models wins.

\begin{figure}[t]
  \centering
  \subcaptionbox{\label{fig:scalar-fermion-heatmap}}{
    \includegraphics[width=0.55\linewidth]{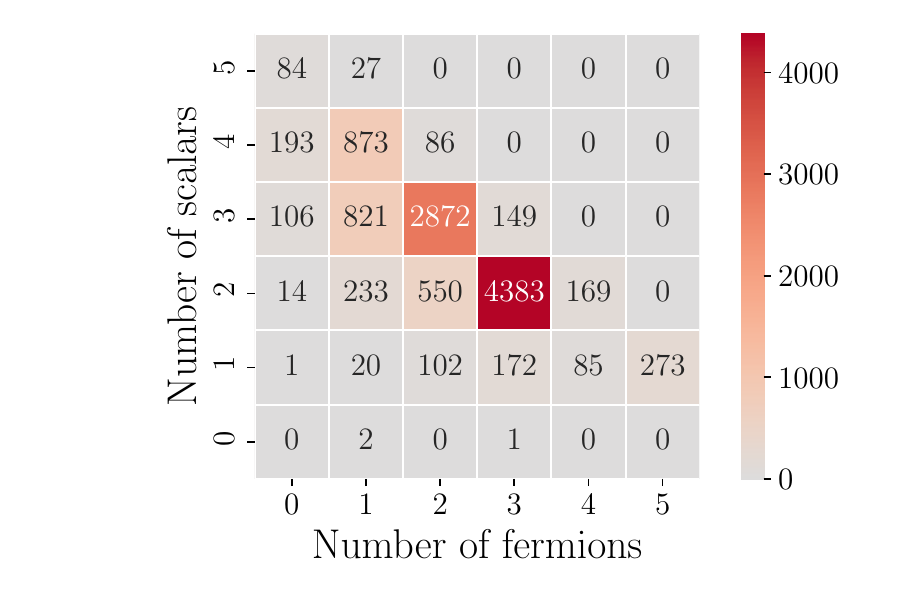}
  }
  \subcaptionbox{\label{fig:lambda-fields}}{
    \includegraphics[width=0.55\linewidth]{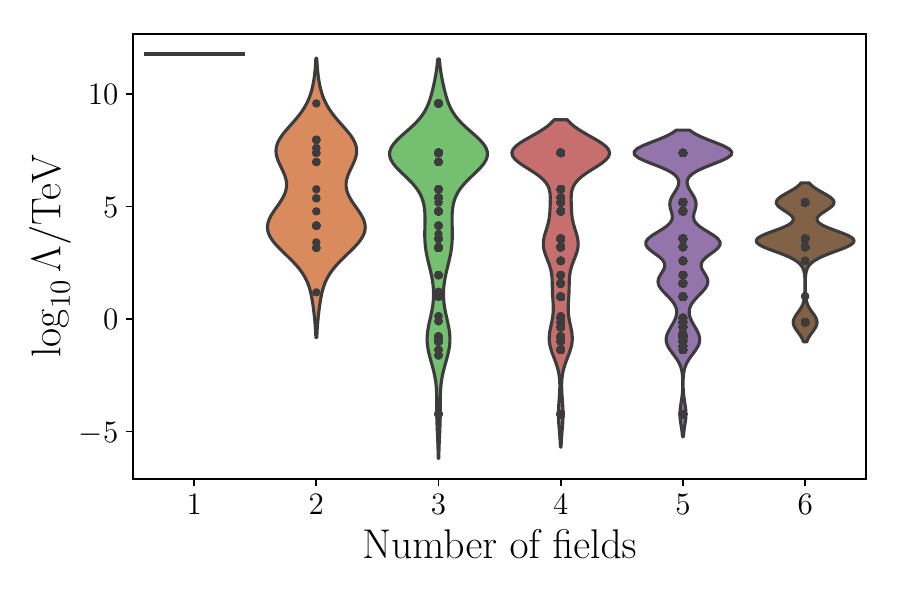}
  }
  \caption{(a) The number of filtered models containing different numbers of
    exotic scalar and fermion fields. Most models contain five fields, with the
    most common combination being three fermions and two scalars. The
    fermion-only models are associated only with Weinberg-like operators. (b)
    The rough upper bound on the new-physics scale $\Lambda$ shown against the
    number of exotic fields introduced in the models. The black dots show the
    upper bound on the scale of the new physics for each model. A given black
    dot generally denotes more than one model. Each strip is a smoothed
    histogram of the number of models having a given $\Lambda$ as the
    new-physics upper bound for the specified number of fields. A sizeable class
    of models are testable at current or future collider experiments.}
  \label{fig:number-of-fields}
\end{figure}

In Fig.~\ref{fig:number-of-fields} we present data relevant to the number of
fields present in the models. Fig.~\ref{fig:scalar-fermion-heatmap} shows the
number of exotic scalars and fermions present in the completions. Despite the
fact that the UV topologies associated with some derivative operators allow
completions containing no scalars, we find that only the Weinberg-like operators
keep their fermion-only models after the democratic filtering procedure. By far
the most common kinds of models contain five heavy fields, especially three
fermions and two scalars, or two fermions and three scalars. This is due to the
fact that, as is clear from Fig.~\ref{fig:filter-bar-dimension}, most of the
models generate dimension-eleven operators. In Fig.~\ref{fig:lambda-fields} we
show the estimated new-physics scale $\Lambda$ against the number of fields
featuring in the models. With the exception of one model with two fields, those
required to lie at collider-accessible energies contain three or more fields.
Models with few fields that imply suppressed neutrino masses, or equivalently a
low new-physics scale, have a kind of selection pressure acting against them:
since there are few fields, it is likely they will arise in the completion of
other operators, that generally will filter out the former and imply a larger
value of $\Lambda$. At dimension-seven, for example, $\mathcal{O}_{8}$ is
generated by models featuring two fields and predicts that these should not be
heavier than about $\SI{15}{\TeV}$. However, of its four tree-level completions,
only one survives the filtering procedure. This is the outlier two-field model
evident in the figure. It was first derived\footnote{We note that the other
  completion of $\mathcal{O}_{8}$ listed in Ref.~\cite{Cai:2014kra} also
  generates $\mathcal{O}_{50a}$ through a diagram which dominates the neutrino
  mass.} in Ref.~\cite{Cai:2014kra} and later in Ref.~\cite{Klein:2019iws}. The
model contains the fields
$\Pi_{1} \sim (\mathbf{3}, \mathbf{1}, \tfrac{1}{6})_{S}$ and
$Q_{7} \sim (\mathbf{3}, \mathbf{2}, \tfrac{7}{6})_{F}$. We list the models
containing three exotic fields that are required to lie below $\SI{100}{\TeV}$
in Sec.~\ref{sec:example-models}.

Of the unfiltered 430,810 models, close to $67\%$ (290,492) contain at least one
of the seesaw fields: $N \sim (\mathbf{1}, \mathbf{1}, 0)_{F}$,
$\Xi_{1} \sim (\mathbf{1}, \mathbf{3}, 1)_{S}$ or
$\Sigma \sim (\mathbf{1}, \mathbf{3}, 0)_{F}$. We present the exact breakdown by
the interactions involved in the models in Table~\ref{tab:models-with-seesaws}.
These are by far the most common fields appearing in the list of unfiltered
models. Since our default filtering philosophy in this analysis is democratic,
all of these are absent from the filtered list of models, and they only appear
in completions of the Weinberg operator and $\mathcal{O}_{D2}$.

\begin{table}[t]
  \centering
    \begin{tabular}{ccll}
    \toprule
    Field & Interactions & $\Delta L = 2$ Lagrangians & Models \\
    \midrule
    \multirow{2}{*}{$N \sim (\mathbf{1}, \mathbf{1}, 0)_{F}$} & $L H N$ & 51,245 (11.9\%) & \multirow{2}{*}{17,139 (17.1\%)} \\
                & Other & 12,433 (2.9\%) &  \\
    \midrule
    \multirow{2}{*}{$\Sigma \sim (\mathbf{1}, \mathbf{3}, 0)_{F}$} & $L H \Sigma$ & 87,535 (20.3\%) & \multirow{2}{*}{31,629 (31.5\%)} \\
                & Other & 28,157 (6.5\%) &  \\
    \midrule
    \multirow{4}{*}{$\Xi_{1} \sim (\mathbf{1}, \mathbf{3}, 1)_{S}$} & $L L \Xi_{1}$ & 59,791 (13.0\%) & \multirow{4}{*}{51,576 (51.4\%)} \\
                & $H H \Xi_{1}^{\dagger}$ & 95,410 (22.1\%) &  \\
                & Both & 10,323 (2.4\%) &  \\
                & Other & 30,761 (7.1\%) &  \\
    \bottomrule
  \end{tabular}
  \caption{The table shows the number of unfiltered models in which the seesaw
    fields appear. The category `other' includes interactions such as
    $L \varphi N$, where one of the SM fields in the interaction has been
    replaced with an exotic copy, as well as couplings involving other exotic
    fields whose quantum numbers are unrelated to those of SM fields.}
  \label{tab:models-with-seesaws}
\end{table}

\begin{figure}
  \centering
  \subcaptionbox{\label{fig:scalar-piechart}}{
    \includegraphics[width=0.8\linewidth]{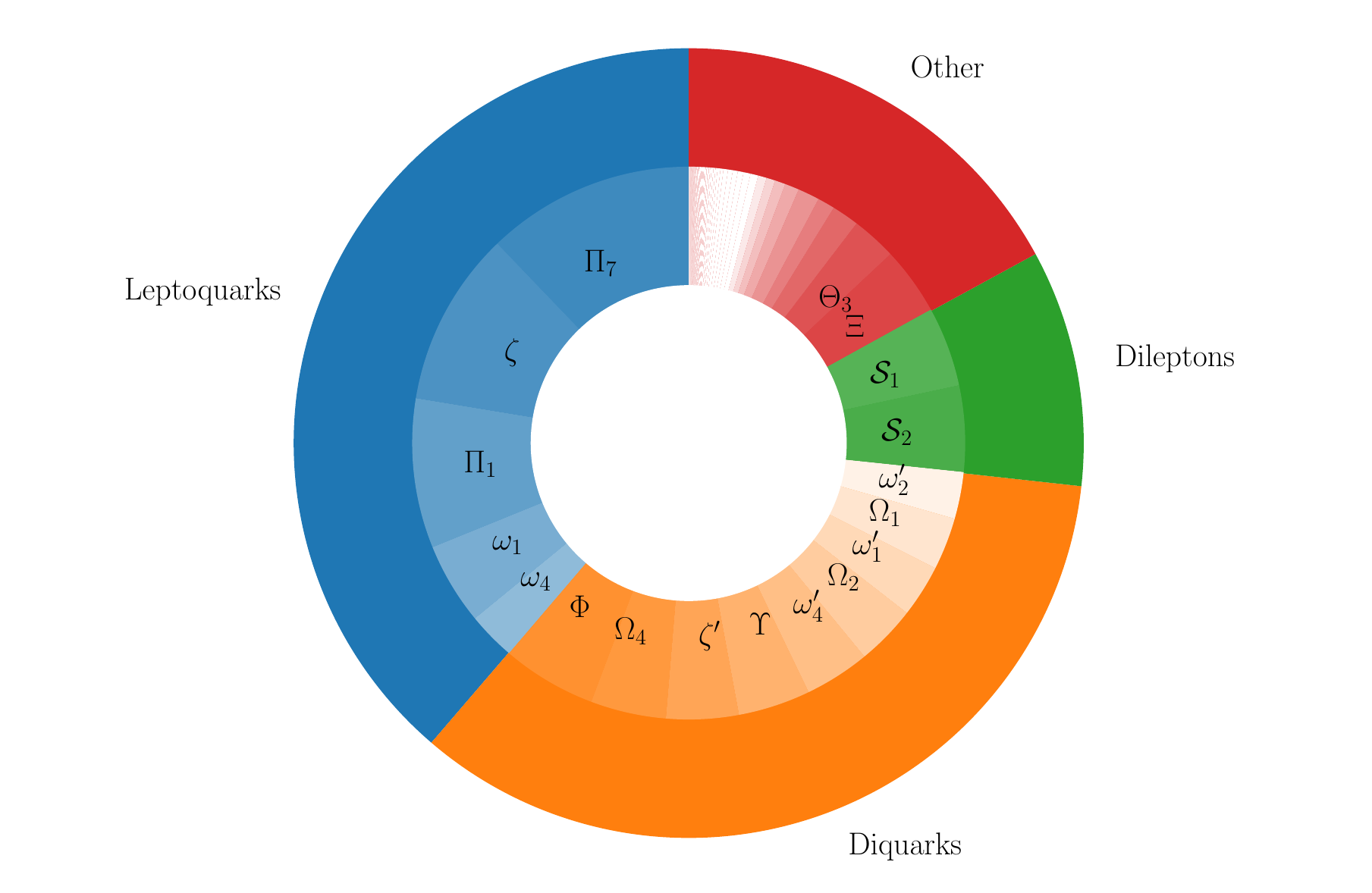}
  }
  \subcaptionbox{\label{fig:fermion-piechart}}{
    \includegraphics[width=0.8\linewidth]{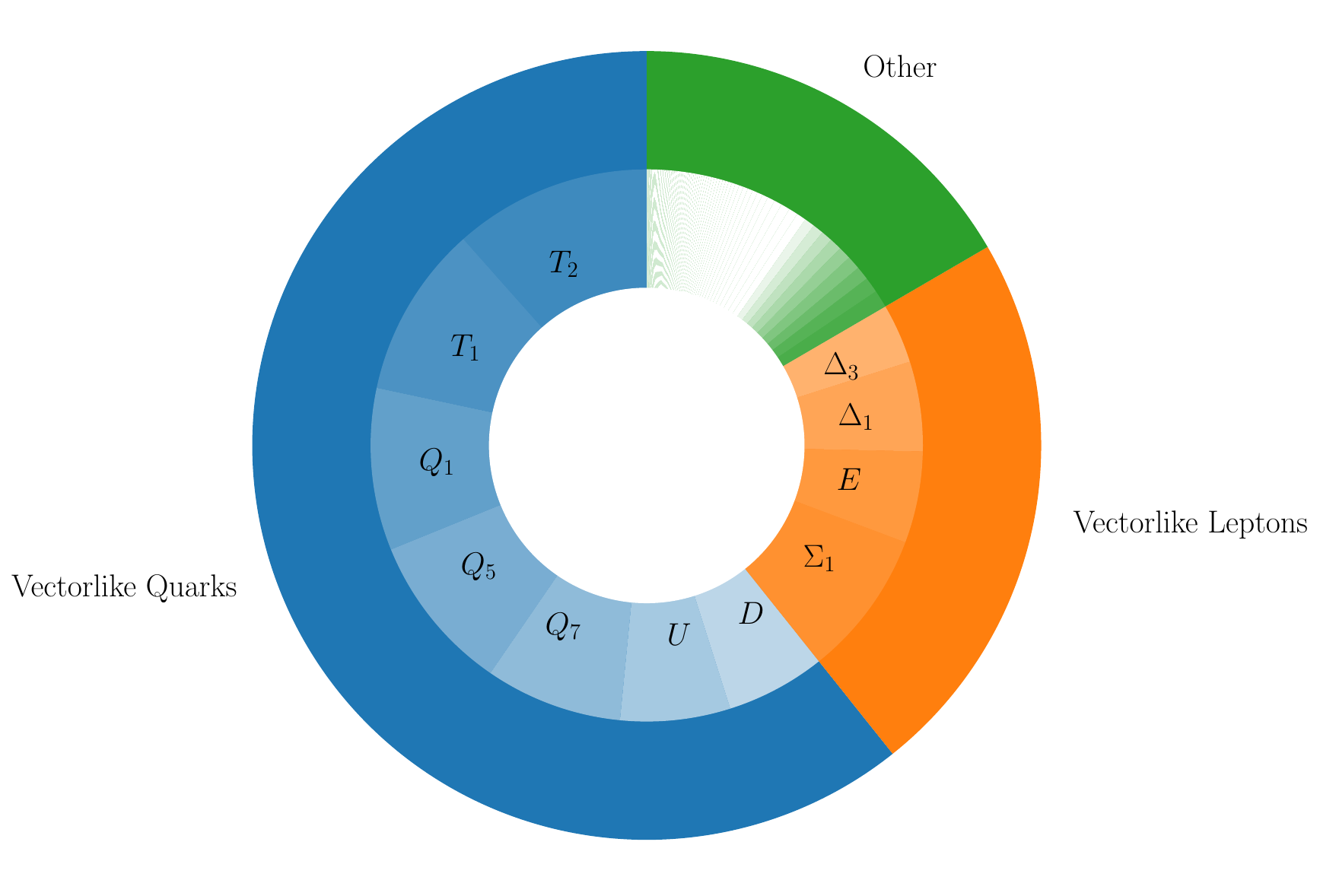}
  }
  \caption{The number of models in which each field appears in the completions
    shown as a pie chart for scalars and fermions separately. The exotics are
    distinguished by their couplings to SM fields. (a) The pie chart of scalar
    fields appearing in the completions. Primed fields represent leptoquarks
    whose baryon-number assignment allows only the diquark couplings. (b) The
    pie chart of fermion fields appearing in the completions. See
    Table~\ref{tab:field-labels} or Ref.~\cite{deBlas:2017xtg} for the
    convention used for the field names.}
  \label{fig:field-piecharts}
\end{figure}

The distinct exotic fields appearing in the completions number 171, although
five fields are completely removed following filtering. These are
$(\mathbf{\bar{6}}, \mathbf{1}, \tfrac{7}{6})_{S}^{1/3}$,
$(\mathbf{\bar{6}}, \mathbf{3}, \tfrac{5}{3})_{S}^{1/3}$,
$(\mathbf{1}, \mathbf{3}, 3)^{0}_{S}$, $(\mathbf{1}, \mathbf{5}, 2)^{0}_{F}$ and
$(\mathbf{\bar{6}}, \mathbf{1}, \tfrac{5}{3})^{1/3}_{S}$, where the superscript
represents the $B$ assignment of the field. There are 83 different scalar fields
and 83 different fermion species. We distinguish three broad classes of scalars
on the basis of their interaction with the SM fermions: leptoquarks, diquarks
and dileptons. For exotic fermions we differentiate between those arising from
contractions between the Higgs and a SM quark (vectorlike quarks), and the Higgs
and a SM lepton (vectorlike leptons). The relative frequencies with which these
field classes appear in the filtered completions are shown in
Fig.~\ref{fig:field-piecharts} as pie charts. The wedges represent the number of
Lagrangians in which the field couples as a leptoquark, diquark, dilepton,
vectorlike quark or vectorlike lepton. We label fields coupling in all other
ways as `other' in the figure. The most represented family of scalars are
leptoquarks, with the most common field being
$\Pi_{7} \sim (\mathbf{3}, \mathbf{2}, \tfrac{7}{6})_{S}$, commonly called
$R_{2}$ in the literature~\cite{Dorsner:2016wpm}. This leptoquark appears in
simplified models of $R_{D^{(*)}}$ and the neutral-current flavour anomalies
like $R_{K^{(*)}}$, see \textit{e.g.} Refs.~\cite{Sakaki:2013bfa,
  Angelescu:2018tyl, Becirevic:2018uab, Popov:2019tyc, Becirevic:2017jtw}. It
was recently shown to be able to reconcile the discrepant measurements in the
anomalous magnetic moments of both the muon and the
electron~\cite{Bigaran:2020jil, Dorsner:2020aaz}. The second most common scalar
appearing in our neutrino-mass models is
$\zeta \sim (\mathbf{\bar{3}}, \mathbf{3}, \tfrac{1}{3})_{S}$, frequently
referred to as $S_{3}$. This leptoquark is a popular explanation of the
neutral-current $b \to s$ anomalies such as $R_{K^{(*)}}$, see
\textit{e.g.}~\cite{Hiller:2014yaa, Gripaios:2014tna, Hiller:2017bzc,
  Dorsner:2017ufx, Angelescu:2018tyl}. The most frequently encountered fermions
are vectorlike quarks, with the most common being
$T_{2}\sim (\mathbf{3}, \mathbf{3}, \tfrac{2}{3})_{F}$. It contains components
that mix with both the up- and down-type SM quarks. We emphasise that the plots
and numbers presented here are directly related to our filtering and
model-counting conventions. In Fig.~\ref{fig:field-piecharts} for example, we do
not count fields just by their quantum numbers, but also include coupling
information as discussed above. Additionally, we count independent Lagrangians
as different models rather than just counting distinct sets of fields, which is
perhaps more in line with our `democratic' approach to filtering. We note that
the qualitative features discussed here are all relatively robust against these
different conventions. We encourage the interested reader to explore our model
database to see how different approaches to filtering and counting can answer
specific questions they may have of the data.

\begin{figure}[t]
  \centering
  \includegraphics[width=0.8\textwidth]{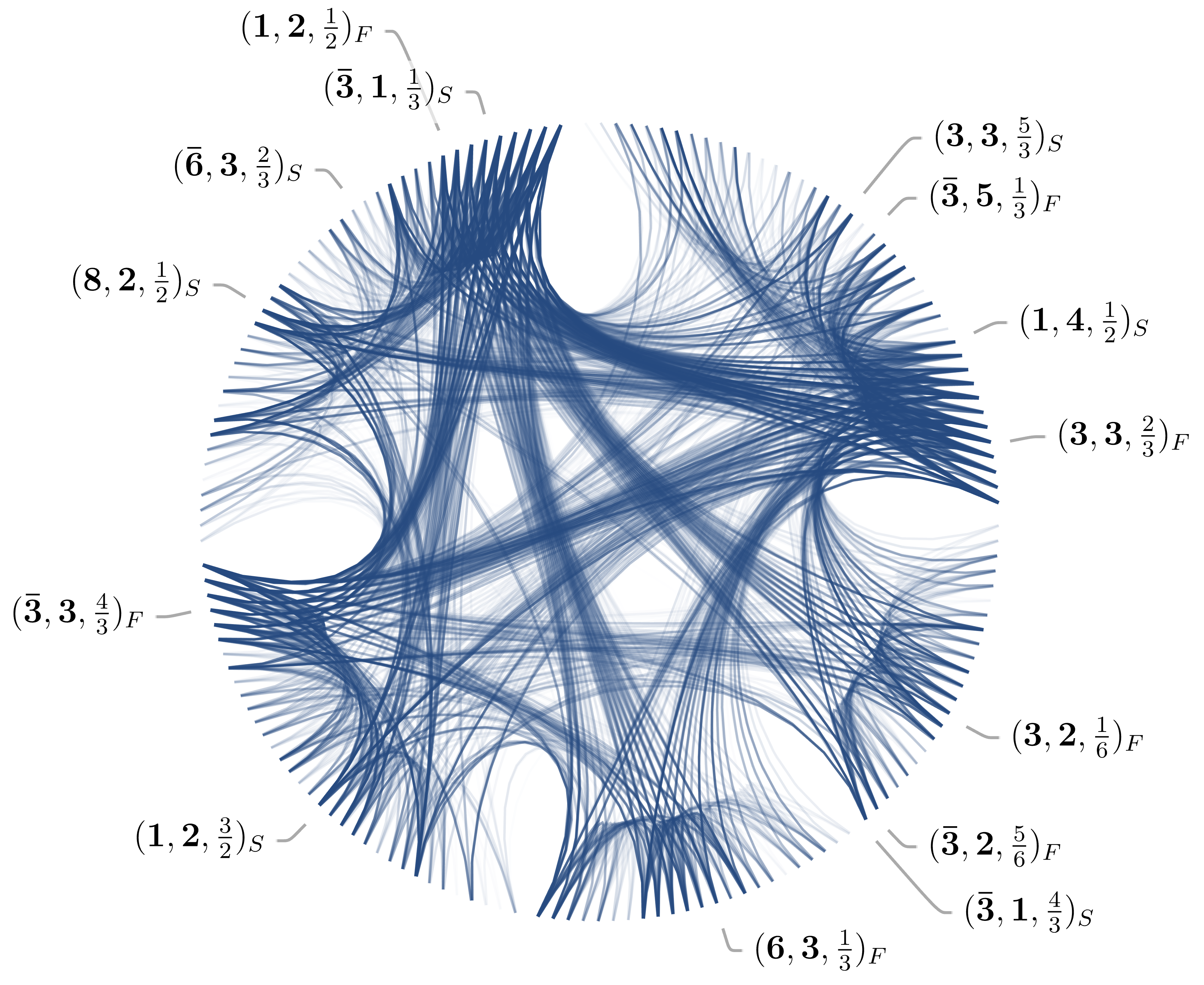}
  \caption{The graph is a representation of the connectivity between exotic
    fields in the neutrino-mass models. Each node represents an exotic field and
    edges connect fields featuring together in a neutrino-mass model. The colour
    is an indication of the weight of the edge, \textit{i.e.} the number of
    times the two nodes appear in models together. The graph is clustered into
    roughly five communities within which there are many mutual connections.
    Only a handful of node labels are shown.}
  \label{fig:model-graph}
\end{figure}

We are also interested in the connectivity between fields as they feature in the
models. To explore this we study a graph in which each vertex represents one of
the 163 exotic fields introduced in the completions that contain at least two
fields, and an edge is drawn between fields featuring together in a model. The
graph is shown shown Fig.~\ref{fig:model-graph}. The exterior sectors at each
node represent the number of degree of the node. The edges in the graph are
weighted by the number of times the corresponding pair of fields appears in the
models; this is shown with a linear colour scaling in the figure. There are 3036
edges in the graph, and the average node degree is approximately 37. About a
fifth of all possible connections in the graph are realised. The ten most
heavily weighted edges, representing the ten most common pairs of fields
appearing in the models, are shown in Table~\ref{tab:graph-info}. Many of these
correlations can be understood on the basis of common contractions in the
derivation of the models, especially those involving $H$ or $L$. There is a
propensity for scalars and fermions with the same gauge quantum numbers to
appear in models together. This seems to come about from the fact that
$H \otimes L$ is a gauge singlet but transforms like $(\mathbf{2}, \mathbf{1})$
under $\mathrm{SU}(2)_{+} \otimes \mathrm{SU}(2)_{-}$. We note that all of the
fields in the table have $|B| = \tfrac{1}{3}$, and so this edge cannot come
about from
$(\mathbf{\bar{3}}, \mathbf{2}, \tfrac{5}{6})_{F} \otimes \bar{d} \sim (\mathbf{3}, \mathbf{2}, \tfrac{7}{6})_{S}$.

\begin{table}[t]
  \centering
  \begin{tabular}[t]{lc}
    \toprule
    Rank & Edge \\
    \midrule
    1 & $(\mathbf{3}, \mathbf{3}, \tfrac{2}{3})_{F}$, $(\mathbf{3}, \mathbf{4}, \tfrac{1}{6})_{S}$ \\
    2 & $(\mathbf{3}, \mathbf{2}, \tfrac{1}{6})_{S}$, $(\mathbf{3}, \mathbf{2}, \tfrac{1}{6})_{F}$ \\
    3 & $(\mathbf{3}, \mathbf{3}, \tfrac{2}{3})_{S}$, $(\mathbf{3}, \mathbf{2}, \tfrac{7}{6})_{S}$ \\
    4 & $(\mathbf{3}, \mathbf{2}, \tfrac{7}{6})_{F}$, $(\mathbf{3}, \mathbf{2}, \tfrac{1}{6})_{S}$ \\
    5 & $(\mathbf{3}, \mathbf{3}, \tfrac{2}{3})_{F}$, $(\mathbf{3}, \mathbf{4}, \tfrac{7}{6})_{F}$ \\
    6 & $(\mathbf{\bar{3}}, \mathbf{3}, \tfrac{1}{3})_{S}$, $(\mathbf{3}, \mathbf{4}, \tfrac{1}{6})_{S}$ \\
    7 & $(\mathbf{3}, \mathbf{2}, \tfrac{1}{6})_{F}$, $(\mathbf{3}, \mathbf{3}, \tfrac{2}{3})_{S}$ \\
    8 & $(\mathbf{\bar{3}}, \mathbf{3}, \tfrac{4}{3})_{F}$, $(\mathbf{\bar{3}}, \mathbf{2}, \tfrac{5}{6})_{F}$ \\
    9 & $(\mathbf{3}, \mathbf{2}, \tfrac{1}{6})_{S}$, $(\mathbf{3}, \mathbf{3}, \tfrac{2}{3})_{S}$ \\
    10 & $(\mathbf{3}, \mathbf{2}, \tfrac{7}{6})_{S}$, $(\mathbf{\bar{3}}, \mathbf{2}, \tfrac{5}{6})_{F}$ \\
    \bottomrule
  \end{tabular}
  \caption{The table shows the pairs of fields that most often appear together
    in the filtered completions of the $\Delta L = 2$ operators we consider. In
    the context of the graph of field connections introduced in the main text,
    these are the top ten edges by edge weight. Many of the connections can be
    understood on the basis of common couplings to SM fields, especially $L$ and
    $H$. For example,
    $(\mathbf{3}, \mathbf{3}, \tfrac{2}{3})_{F} \otimes L \sim (\mathbf{3}, \mathbf{4}, \tfrac{1}{6})_{S}$
    and
    $(\mathbf{3}, \mathbf{3}, \tfrac{2}{3})_{S} \otimes H \sim (\mathbf{3}, \mathbf{2}, \tfrac{7}{6})_{S}$.
    All of the fields in the table have $|B| = \tfrac{1}{3}$.}
  \label{tab:graph-info}
\end{table}

\subsection{Example models}
\label{sec:example-models}

In this section we present some example neutrino-mass models, illustrating use
cases of the model database, aspects to be careful of in its use, and
representative features of the novel models derived from dimension-eleven and
dimension-nine operators.

\subsubsection{Simple models at the TeV scale}
\label{sec:simple-models}

We are particularly interested in models that are simple, in the sense that they
involve few exotic fields, and testable, in that they predict new physics at
currently or nearly accessible energy ranges. We query our model database to
return models featuring three fields or fewer with the estimated upper bound on
the new-physics scale required to be between \SI{700}{\GeV} and \SI{100}{\TeV}.
The results of the query are presented in Table~\ref{tab:simple-models}. There
are twelve\footnote{We note that there are technically more models: those for
  which the colour-sextet fields in Table~\ref{tab:simple-models} are replaced
  with colour triplets, with a corresponding baryon-number assignment such that
  the same interactions as the sextet are picked out.} models listed, only one
of which has explicitly appeared before in the literature to our knowledge: the
completion of $\mathcal{O}_{8}$ discussed in Sec.~\ref{sec:modelsoverview}. It
is interesting to note that the scalar leptoquark\footnote{We mention
  parenthetically that although this leptoquark does not possess diquark
  couplings, baryon-number violation does occur through a term in the scalar
  potential. The leading-order contribution is through a dimension-ten operator
  mediating $p \to \pi^{+}\pi^{+} e^{-} \nu \nu$~\cite{Arnold:2012sd}.}
$\Pi_{1} \sim (\mathbf{3}, \mathbf{2}, \tfrac{1}{6})_{S}$ appears in almost
every model listed in the table. This suggests that our general analysis of the
frequency of fields appearing in the completions in
Sec.~\ref{sec:modelsoverview} may look different if specific selection criteria
are placed on the data. We have checked the full Lagrangians implied by the
field content of each model and found that seven of the models listed in the
table imply the generation of the Weinberg operator through heavy loops. We
emphasise that these non-genuine completions are potentially interesting and new
radiative models, although the neutrino self-energy diagram will look different
to that implied by the closure of the tree-level graph from which the model was
derived. This means that the bound on the implied new-physics scale is in
general higher than that suggested by the closure of the original operator. In
this class are all of the models for which the upper bound on the new-physics
scale is larger than \SI{15}{\TeV}. This means that there are only five models
in our database with fewer than four fields for which the upper bound on
$\Lambda$ is between \SI{700}{\GeV} and \SI{100}{\TeV}, and they all predict new
physics below \SI{15}{\TeV}. In the following we present two example models from
the table:
\begin{enumerate}
  \item We look at one of the models---the one derived from
    $\mathcal{O}_{62b}$---that generates the Weinberg operator through a heavy
    loop. We intend this to be an example of how this phenomenon can appear and
    how it is easy to diagnose in some cases.
  \item We present a brief study of the implications for neutrino mass implied
    by the model given in the last row.
\end{enumerate}

\begin{table}[t]
  \centering
  \begin{tabular}[t]{cclc}
    \toprule
    Field content & Operators & $\Lambda~[\TeV]$ & Dominant? \\
    \midrule
    $(\mathbf{3}, \mathbf{2}, \tfrac{1}{6})_{S}$, $(\mathbf{3}, \mathbf{2}, \tfrac{7}{6})_{F}$ & $8$, $D15$ & 15 & Y \\
    $(\mathbf{1}, \mathbf{2}, \tfrac{1}{2})_{F}$, $(\mathbf{1}, \mathbf{1}, 1)_{S}$, $(\mathbf{1}, \mathbf{2}, \tfrac{3}{2})_{S}$ & $62b$ & 16 & N \\
    $(\mathbf{\bar{3}}, \mathbf{2}, \tfrac{5}{6})_{S}$, $(\mathbf{3}, \mathbf{2}, \tfrac{1}{6})_{F}$, $(\mathbf{3}, \mathbf{2}, \tfrac{1}{6})_{S}$ & $8^{\prime}$ & 1 & N \\
    $(\mathbf{\bar{3}}, \mathbf{1}, \tfrac{1}{3})_{S}$, $(\mathbf{\bar{6}}, \mathbf{2}, \tfrac{1}{6})_{S}$, $(\mathbf{3}, \mathbf{2}, \tfrac{1}{6})_{F}$ & $24f$ & 89 & N \\
    $(\mathbf{\bar{3}}, \mathbf{3}, \tfrac{1}{3})_{F}$, $(\mathbf{\bar{6}}, \mathbf{2}, \tfrac{1}{6})_{S}$, $(\mathbf{3}, \mathbf{2}, \tfrac{1}{6})_{S}$ & $24d$ & 89 & N \\
    $(\mathbf{\bar{3}}, \mathbf{2}, \tfrac{5}{6})_{S}$, $(\mathbf{1}, \mathbf{2}, \tfrac{3}{2})_{F}$, $(\mathbf{3}, \mathbf{2}, \tfrac{1}{6})_{S}$ & $8^{\prime}$ & 1 & N \\
    $(\mathbf{\bar{3}}, \mathbf{3}, \tfrac{1}{3})_{F}$, $(\mathbf{\bar{6}}, \mathbf{4}, \tfrac{1}{6})_{S}$, $(\mathbf{3}, \mathbf{2}, \tfrac{1}{6})_{S}$ & $24f$ & 89 & N \\
    $(\mathbf{\bar{3}}, \mathbf{1}, \tfrac{1}{3})_{F}$, $(\mathbf{\bar{6}}, \mathbf{2}, \tfrac{1}{6})_{S}$, $(\mathbf{3}, \mathbf{2}, \tfrac{1}{6})_{S}$ & $24d$ & 89 & N \\
    $(\mathbf{\bar{6}}, \mathbf{2}, \tfrac{7}{6})_{F}$, $(\mathbf{8}, \mathbf{2}, \tfrac{1}{2})_{S}$, $(\mathbf{3}, \mathbf{2}, \tfrac{1}{6})_{S}$ & $20$ & 0.8 & Y \\
    $(\mathbf{6}, \mathbf{1}, \tfrac{4}{3})_{S}$, $(\mathbf{6}, \mathbf{1}, \tfrac{1}{3})_{F}$, $(\mathbf{3}, \mathbf{2}, \tfrac{1}{6})_{S}$ & $20$ & 0.8 & Y \\
    $(\mathbf{6}, \mathbf{2}, \tfrac{5}{6})_{S}$, $(\mathbf{3}, \mathbf{2}, \tfrac{1}{6})_{F}$, $(\mathbf{3}, \mathbf{2}, \tfrac{1}{6})_{S}$ & $50a,b$ & 10 & Y \\
    $(\mathbf{\bar{6}}, \mathbf{2}, \tfrac{1}{6})_{S}$, $(\mathbf{\bar{3}}, \mathbf{2}, \tfrac{5}{6})_{F}$, $(\mathbf{3}, \mathbf{2}, \tfrac{1}{6})_{S}$ & $50a,b$ & 10 & Y \\
    \bottomrule
  \end{tabular}
  \caption{The table shows the models in our filtered list that contain fewer
    than four fields with the estimate of the upper-bound on the new-physics
    scale $\Lambda$ in the range $\SI{700}{\GeV} < \Lambda < \SI{100}{\TeV}$.
    Models containing colour sextet fields can be replaced with the
    corresponding colour-triplet fields with a different baryon-number
    assignment. The fields and models are listed in no special order. The scalar
    leptoquark $\Pi_{1} \sim (\mathbf{3}, \mathbf{2}, \tfrac{1}{6})$ appears in
    almost all of the models listed. Completions marked as non-dominant may be
    viable and interesting neutrino-mass models, but the main contribution to
    the neutrino mass does not come from the closure of the tree-level diagram
    from which the particle content was derived. This means, among other things,
    that the upper bound on the scale of the new physics associated with the
    model will differ to that presented here.}
  \label{tab:simple-models}
\end{table}

\paragraph{Model derived from $\mathcal{O}_{62b}$} The model derived from
$\mathcal{O}_{62b}$ is especially simple since it does not require the
imposition of $\mathrm{U}(1)_{B}$. The exotic fields introduced are
$\Delta_{1} \sim (\mathbf{1}, \mathbf{2}, \tfrac{1}{2})_{F}$,
$\mathcal{S}_{1} \sim (\mathbf{1}, \mathbf{1}, 1)_{S}$ and
$\chi \sim (\mathbf{1}, \mathbf{2}, \tfrac{3}{2})_{S}$. The additional
interaction Lagrangian necessary to generate $\mathcal{O}_{62b}$ at tree level
is $\Delta \mathscr{L} = \mathscr{L}_{Y} - \mathcal{V}$, with
\begin{align}
  - \mathscr{L}_{Y} &= m_{\Delta_{1}} \bar{\Delta}_{1} \Delta_{1} + x_{[rs]} L^{i}_{r} L^{j}_{s} \mathcal{S}_{1}\epsilon_{ij} + y_{r} \bar{e}_{r} \bar{\Delta}_{1}^{i} \tilde{H}^{j} \epsilon_{ij} + z_{r}\bar{e}_{r} \tilde{\chi}^{i} \Delta_{1}^{j} \epsilon_{ij} \ , \\
  \mathcal{V} &= m_{\mathcal{S}_{1}}^{2} \mathcal{S}_{1}^{\dagger} \mathcal{S}_{1} + m_{\chi}^{2} \chi^{\dagger} \chi + w H^{i} \chi^{j} \mathcal{S}^{\dagger}_{1} \mathcal{S}^{\dagger}_{1} \epsilon_{ij} \ .
\end{align}
This implies that the neutrino-mass mechanism depends on 13 new parameters: nine
Yukawa couplings, $w$ and the three masses; although there are a much larger
number of terms present in the full Lagrangian of the model. Importantly, one of
these is
$x^{\prime}_{r} L_{r}^{i} \bar{\Delta}^{j}_{1} \mathcal{S}_{1} \epsilon_{ij}$,
which we now show is sufficient to generate the Weinberg operator through a
two-loop diagram containing one heavy loop.

\begin{figure}[t]
  \centering
  \subcaptionbox{\label{fig:leptonic-model-treelevel}}{
    \includegraphics[width=0.35\linewidth]{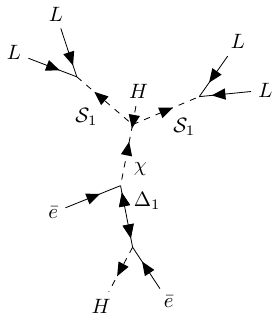}
  }
  \subcaptionbox{\label{fig:leptonic-model-neutrinomass}}{
    \includegraphics[width=0.6\linewidth]{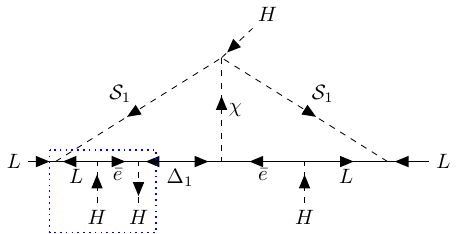}
  }
  \caption{(a) The furnishing of the tree-level topology, labelled $2s6f_4$ in
    our scheme, that generates $\mathcal{O}_{62b}$ at tree level. The
    interactions allowed in the theory are such that the $\Delta_{1}$ line can
    be connected straight into one of the $LL\mathcal{S}_{1}$ vertices in place
    of an $L$, leading to a loop-level completion of $\mathcal{O}_{2}$. (b) The
    neutrino self-energy diagram relevant to the non-genuine completion of
    $\mathcal{O}_{62b}$. It is clear that this diagram does not represent the
    dominant contribution to the neutrino mass, since the the highlighted
    collection of fields can be replaced with the interaction
    $\bar{\Delta}_{1} L \mathcal{S}_{1}$. This leads to a diagram with heavy
    loop involving $\Delta_{1}$, $\chi$ and $\mathcal{S}_{1}$, which dominates
    the neutrino masses. In both cases, the relevant topology is CLBZ-7 in the
    classification of Ref.~\cite{Sierra:2014rxa}.}
  \label{fig:leptonic-model-diagrams}
\end{figure}

The tree-level completion diagram and the neutrino-mass diagram relevant to the
model are shown in Fig.~\ref{fig:leptonic-model-diagrams}. There are two- and
three-loop neutrino self-energies, where the three-loop models arise by
connecting the $H$ and $H^{\dagger}$ lines in
Fig.~\ref{fig:leptonic-model-neutrinomass} in all possible ways. In this case,
the first part of the fermion line (highlighted in
Fig.~\ref{fig:leptonic-model-neutrinomass}) can be replaced with the
aforementioned $L\bar{\Delta}_{1}\mathcal{S}_{1}$ vertex so that the left loop
contains only $\mathcal{S}_{1}$, $\chi$ and the Dirac fermion
$\Delta_{1} + \bar{\Delta}_{1}^{\dagger}$. (It can also be noticed from the
tree-level opening in Fig.~\ref{fig:leptonic-model-treelevel} that the
$\Delta_{1}$ line can be connected directly to one of the $LL\mathcal{S}_{1}$
vertices, giving rise to a loop-level completion of $\mathcal{O}_{2}$.) This
heavy-loop neutrino-mass diagram, although interesting in its own right,
predicts a different mass-scale for the exotic fields (roughly \SI{e6}{\TeV}),
and a different structure for the neutrino-mass matrix.

\begin{figure}[t]
  \centering
  \subcaptionbox{\label{fig:lowscale-model-treelevel}}{
    \includegraphics[width=0.35\linewidth]{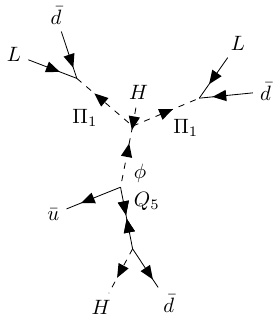}
  }
  \subcaptionbox{\label{fig:lowscale-model-neutrinomass}}{
    \includegraphics[width=0.6\linewidth]{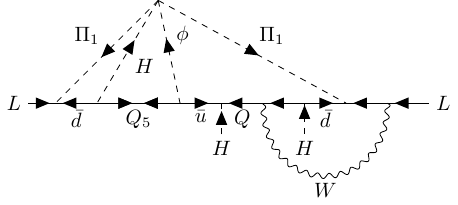}
  }
  \caption{(a) The tree-level completion diagram for the model derived from
    exploding $\mathcal{O}_{50}$ and discussed in the main text. The topology
    is $2s6f_{4}$ in our classification scheme. The closure involves an
    arrow-preserving loop connecting the $\bar{d}^{\dagger}$ to one of the
    $\bar{d}$ lines, and the $W$-boson closure motif discussed in
    Sec.~\ref{sec:operator-closures}. (b) One of the neutrino-mass diagrams
    relevant to the model derived from $\mathcal{O}_{50}$. There is a
    three-loop diagram with the $H$ line broken into an $H^{\dagger}, H$ pair
    that generates the dimension-seven generalised Weinberg operator. The
    four-loop diagrams all involve connecting the $H^{\dagger}$ to each of the
    three $H$ legs in the diagram. There are also multiple places the $W$ could
    end in the diagram, although it must couple to the $Q$ line. The four-loop
    diagrams will give larger contributions to the neutrino mass than the
    three-loop diagrams for $\Lambda \gtrsim \SI{2}{\TeV}$.}
  \label{fig:lowscale-example}
\end{figure}

\paragraph{A genuine low-scale model} Below we present a brief exploration of
the model derived from $\mathcal{O}_{50}$ that contains the exotic fields
$\phi \sim (\mathbf{6}, \mathbf{2}, -\tfrac{1}{6})_{F}$,
$\Pi_{1} \sim (\mathbf{3}, \mathbf{2}, \tfrac{1}{6})_{S}$ and
$Q_{5} \sim (\mathbf{3}, \mathbf{2}, -\tfrac{5}{6})_{F}$. The estimate for the
neutrino mass derived from the operator closure suggests this model's exotic
particle content should live roughly below \SI{10}{\TeV}. The corresponding
$\Delta L = 2$ Lagrangian we write again as
$\Delta \mathscr{L} = \mathscr{L}_{Y} - \mathcal{V}$, with
\begin{align}
  -\mathscr{L}_{Y} &= x_{rs} L^{i}_{r} \bar{d}_{sa} \Pi_{1}^{aj} \epsilon_{ij} + y_{r} \phi^{\{ab\}i} \bar{u}_{ra} \bar{Q}^{j}_{5 b} \epsilon_{ij} + z_{r} \bar{d}_{ra} H^{i} Q^{aj}_{5} \epsilon_{ij} + \text{h.c.} \label{eq:low-scale-yuks} \\
  \mathcal{V} &= \lambda \tilde{\Pi}_{1a}^{i} \tilde{\Pi}_{1b}^{j} \phi^{\{ab\}k} H^{l} (\epsilon_{ik}\epsilon_{jl} + \epsilon_{il}\epsilon_{jk}) \ .
\end{align}
We note that $\mathrm{U}(1)_{B}$ must be imposed on the Lagrangian to prevent
terms like $\Pi^{3}_{1} H^{\dagger}$, $|\phi|^{2}\phi^{\dagger}\Pi_{1}$ and
$|\Pi_{1}|^{2} \Pi_{1} \phi$ that destabilise the proton in the presence of the
Yukawa interactions of Eq.~\eqref{eq:low-scale-yuks}. The field $\phi$ only
couples to SM fermions together with $Q_{5}$ in this model, and so it generates
no dimension-six operators at tree level. The completion graph and one of the
neutrino self-energy diagrams are shown in Fig.~\ref{fig:lowscale-example}. The
tree-level topology is again $2s6f_{4}$, and the neutrino masses are realised at
three and four loops, with the additional loop arising from the connection of an
$H$ and $H^{\dagger}$. One of the loops involves a $W$ boson, and so the diagram
does not fit into existing topological classifications. The three-loop diagram
is similar to the topology $D_{9}^{M}$ of Ref.~\cite{Cepedello:2018rfh}, with
one of the scalar lines replaced with a vector boson. The $W$ boson line must
connect to $Q$ in the diagram, but could end on any field with non-trivial
$\mathrm{SU}(2)_{L}$ charge. The connection to the $L$ line is shown, since the
loop integral then depends on leptonic flavour indices, which can change the
structure of the neutrino-mass matrix. There are also several ways of connecting
the Higgs lines and only one combination is shown in the figure. The four-loop
diagrams will be the dominant contribution to the neutrino masses for exotic
fields above $4 \pi v \approx \SI{2}{\TeV}$.

The neutrino-mass matrix in this model can be estimated as
\begin{equation}
  [\mathbf{m}_{\nu}]_{rs} = \frac{\lambda g^{2}}{(16\pi^{2})^{3}} \left( \frac{v^{2}}{\Lambda^{2}} + \frac{1}{16\pi^{2}}\right) \frac{1}{\Lambda}\sum_{t,u,v} x_{rt} z^{*}_{t} y^{*}_{u} [\mathbf{m}_{u}]_{u} V_{uv} [\mathbf{m}_{d}]_{v} x_{sv} I_{rstuv} + (r \leftrightarrow s) \ ,
\end{equation}
where the $V_{rs}$ are CKM matrix elements, $\Lambda$ is the generic UV scale,
and $I_{rstuv}$ is the loop function. The dependence on the masses of the up-
and down-type quarks implies that the largest contributions to the neutrino
masses will come from loops containing top and bottom quarks. If the parameters
$y_{1,2}$, $x_{r1}$ and $x_{r2}$ play no significant role in the physics of
neutrino mass, then the matrix will have rank 1 if the loop function carries no
leptonic flavour indices. It may be the case that an additional generation of
$Q_{5}$, $\phi$ or $\Pi_{7}$ is therefore required for the model to successfully
reproduce the measured pattern of neutrino masses and mixings.

\subsubsection{A model derived from a derivative operator}

We move on to discuss a model generating the single-derivative dimension-nine
operators $\mathcal{O}_{D10a,b,c}$. The estimated upper-bound on the exotic
scale is close to \SI{1.5e3}{\TeV} in this case. The model contains the fields
$\rho \sim (\mathbf{1}, \mathbf{2}, \tfrac{3}{2})_{S}$,
$Q_{5} \sim (\mathbf{3}, \mathbf{2}, \tfrac{5}{6})_{F}$ and
$\Sigma_{1} \sim (\mathbf{1}, \mathbf{3}, 1)_{F}$. Such two-fermion--one-scalar
models are unique to completions of single-derivative operators at dimension
nine.

The part of the Lagrangian relevant to lepton-number violation is
\begin{equation}
  - \Delta \mathscr{L} = x_{r} L_{r}^{i} \Sigma_{1}^{\{jk\}} H^{\dagger}_{k} \epsilon_{ij} + y_{r} L_{r}^{i} \rho^{j} \bar{\Sigma}_{1}^{\{kl\}}\epsilon_{ik}\epsilon_{jl} + z_{r} \bar{d}_{ra} H^{i} Q^{aj}_{5} \epsilon_{ij} + w_{r} \bar{u}_{ra} \rho^{i} Q_{5}^{aj} \epsilon_{ij} + \text{h.c.}
\end{equation}
The only additions to the scalar potential are the expected $|\rho|^{2}|H|^{2}$
and $|\rho|^{4}$ terms, and these play no role in the lepton-number violation.
Notably, there are no Yukawa couplings involving $\bar{Q}_{5}$, and the field
$\rho$ generates no dimension-six operators at tree-level, since the naively
expected coupling $H^{i}H^{j}H^{k}\rho_{k} \epsilon_{ij}$ vanishes. The model
also has the nice feature that no baryon-number violating interactions are
present.

\begin{figure}[t]
  \centering
  \subcaptionbox{\label{fig:derivative-example-treelevel}}{
    \includegraphics[width=0.5\linewidth]{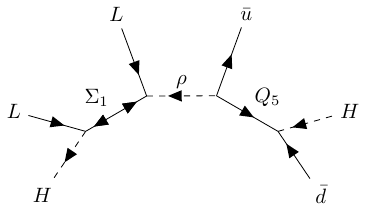}
  }
  \subcaptionbox{\label{fig:derivative-example-neutrinomass}}{
    \includegraphics[width=0.4\linewidth]{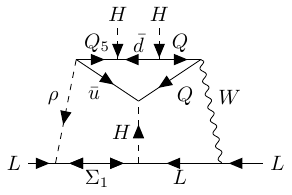}
  }
  \caption{(a) The tree-level completion diagram for the model that generates
    the single-derivative operators $\mathcal{O}_{D10,a,b,c}$ and discussed in
    the main text. The topology is $2s4f_{8}$ in our classification scheme. This
    class of topologies is only relevant to single-derivative operators, and
    contains an arrow-preserving fermion propagator, that of $Q_{5}$ in the
    diagram. The closure of the diagram involves a $W$-boson loop, similar to
    that required in Fig.~\ref{fig:lowscale-example}. (b) One of the
    neutrino-mass diagrams relevant to the model generating
    $\mathcal{O}_{D10a,b,c}$. The diagram generates the Weinberg operator as
    drawn, but additional diagrams exist with the central $H$ line cut into an
    $H,H^{\dagger}$ pair that generate $\mathcal{O}^{\prime}_{1}$ instead. These
    diagrams will only be relevant for exotic masses less than about
    $\SI{2}{\TeV}$. Additional three-loop diagrams exist in which the Higgs
    coming from the $\Sigma_{1}LH^{\dagger}$ interaction loops into any of the
    other external $H$ fields. The $W$ boson must connect to the $Q$ line, but
    could end on any other field with non-trivial $\mathrm{SU}(2)_{L}$ charge.
    The topology is cocktail-like~\cite{Gustafsson:2012vj}, and resembles
    $D^{M}_{15}$ in Ref.~\cite{Cepedello:2018rfh}.}
  \label{fig:derivative-example-diagrams}
\end{figure}

The tree-level completion diagram and one of the neutrino-mass diagrams are
shown in Fig.~\ref{fig:derivative-example-diagrams}. The completion diagram has
topology $2s4f_{8}$, which requires one of the heavy fermions to have an
arrow-preserving propagator. The neutrino-mass diagram shown is
cocktail-like~\cite{Gustafsson:2012vj}, although there are also two-loop diagrams generating
$\mathcal{O}_{1}^{\prime}$ at the low scale, as well as other diagrams with the
$W$ and $H$ lines in different places. The topology of the neutrino self-energy
diagram is similar to $D^{M}_{15}$ in Ref.~\cite{Cepedello:2018rfh}.

The flavour structure of the neutrino-mass matrix has the approximate form
\begin{equation}
  [\mathbf{m}_{\nu}]_{rs} = \frac{g^{2}}{(16\pi^{2})^{3}} \frac{1}{\Lambda} \sum_{t,u} y_{r} x_{s} w^{*}_{t} z_{t} [\mathbf{m}_{d}]_{t} V_{ut} [\mathbf{m}_{u}]_{u} I_{rstu} + (r \leftrightarrow s) \ .
\end{equation}
The dependence on the up- and down-type mass matrices, as in the example
presented in Sec.~\ref{sec:simple-models}, means that the couplings $w_{1,2}$
and $z_{1,2}$ will not play an important role in generating the observed pattern
of neutrino masses and mixings. In this case the matrix has at least rank 2,
even if the leptonic-flavour structure of the loop integrals $I_{rstu}$ is flat.
Thus, the structure of the neutrino masses and mixing parameters emerges mostly
from the six parameters $x_{r}$ and $y_{r}$.

\subsubsection{A model of neutrino mass and the flavour anomalies}

Here we present a model designed specifically to generate a particular set of
dimension-six operators. The example is motivated by the recent flavour
anomalies: deviations from the SM seen in charged- and neutral-current $B$-meson
decays. Key examples are the lepton-flavour-universality (LFU)
ratios~\cite{Lees:2012xj, Lees:2013uzd, Huschle:2015rga, Hirose:2016wfn,
  Abdesselam:2016cgx, Aaij:2017tyk, Aaij:2017uff}
\begin{equation}
  \label{eq:rd}
  R_{D^{(*)}} = \frac{\Gamma(B \to D^{(*)} \tau \nu)}{\Gamma(B \to D^{(*)} \ell \nu)}\quad \text{ with } \ell \in \{e, \mu\}  \ ,
\end{equation}
for which the combined significance of the deviation from the SM is
$3.1\sigma$~\cite{Amhis:2019ckw}, and
\begin{equation}
  \label{eq:rk}
  R_{K^{(*)}} = \frac{\Gamma(B \to K^{(*)} \mu \mu)}{\Gamma(B \to K^{(*)} e e)} \ ,
\end{equation}
both measured to be about $2.5\sigma$ away from
$R_{K^{(*)}} \approx 1$~\cite{Aaij:2019wad, Aaij:2017vbb}. Along with the ratios
$R_{K^{(*)}}$ sit a large class of discrepant measurements in
$b \to s \ell \ell$ processes. These include differences from the SM expectation
of angular observables in $B \to K^{*} \mu \mu$~\cite{Aaij:2015oid,
  ATLAS-CONF-2017-023, CMS-PAS-BPH-15-008, Khachatryan:2015isa} and suppressed
branching ratios measured for $B \to K^{(*)} \mu\mu$~\cite{Aaij:2014pli} and
$B_{s} \to \phi\mu \mu$~\cite{Aaij:2015esa}. The LFU ratios given in
Eqs.~\eqref{eq:rd} and \eqref{eq:rk} are theoretically very clean, since a large
part of the theory uncertainty cancels in the ratio.

In the following we adhere to the conventions\footnote{These can be accessed
  easily at \url{https://flav-io.github.io/docs/operators.html}.} of
Ref.~\cite{Aebischer:2017ugx} relevant to the Warsaw basis for the SMEFT and the
\textsf{flavio} basis~\cite{Straub:2018kue} for the Weak Effective Theory (WET).
The leptoquarks
$\omega_{1} \sim (\mathbf{\bar{3}}, \mathbf{1}, \tfrac{1}{3})_{S}$ and
$\Pi_{7} \sim (\mathbf{3}, \mathbf{2}, \tfrac{7}{6})_{S}$ can provide an
explanation of the anomalies in $R_{D^{(*)}}$ with contributions to the SMEFT
operators
\begin{align}
  \label{eq:rdrdstar-relation}
  [C_{lequ}^{(1)}]_{3332} = \begin{cases}
    - 4 [C_{lequ}^{(3)}]_{3332} &\text{for } \omega_{1} \\
    4 [C_{lequ}^{(3)}]_{3332}  &\text{for } \Pi_{7} \\
  \end{cases} \ ,
\end{align}
since they have Yukawa couplings to left- and right-handed SM fields. [We note
that Eq.~\eqref{eq:rdrdstar-relation} holds at the high scale, and the relation
between the operators is altered by running.] The Yukawa terms are
\begin{align}
  - \mathcal{L}_{\omega_{1}} &= f_{rs} L_{r} Q_{s} \omega_{1} + g_{rs} \bar{e}^{\dagger}_{r} \bar{u}^{\dagger}_{s} \omega_{1} + \text{h.c.} \label{eq:omega1-coup} \\
  - \mathcal{L}_{\Pi_{7}} &= x_{rs} L_{r} \bar{u}_{s} \Pi_{7} + y_{rs} \bar{e}^{\dagger}_{r} Q^{\dagger}_{s} \Pi_{7} + \text{h.c.} \label{eq:pi7-coup}
\end{align}
and these imply
\begin{align}
  \label{eq:omega1-pi7-clequ}
  [C_{lequ}^{(1)}]_{3332} = \begin{cases}
    \dfrac{f_{33} g_{32}^{*}}{2 m_{\omega_{1}}^{2}} &\text{for } \omega_{1} \\
    \dfrac{x_{32}^{*}y_{33}}{2 m_{\Pi_{7}}^{2}} &\text{for } \Pi_{7} \\
  \end{cases}
\end{align}
at tree level. A satisfactory explanation of $R_{D^{(*)}}$ requires
$\mathcal{O}(1)$ couplings, \textit{e.g.}~\cite{Cai:2017wry, Popov:2019tyc}, and
for $\Pi_{7}$ fits are consistent with the operator coefficient being purely
imaginary, \textit{e.g.}~\cite{Angelescu:2018tyl}.

The $b \to s$ data can be explained by the tree-level exchange of the leptoquark
$\zeta \sim (\mathbf{\bar{3}}, \mathbf{3}, \tfrac{1}{3})_{S}$, which generates
\begin{equation}
  [C_{lq}^{(1)}]_{2223} = 3 [C_{lq}^{(3)}]_{2223} \ ,
\end{equation}
relevant for the neutral-current anomalies. Fits are usually performed to
four-fermion operators in the WET, defined below the electroweak scale. For the
$b \to s \ell \ell$ data, a good fit is given for~\cite{Aebischer:2019mlg}
\begin{equation}
  C_{9}^{bs\mu\mu} = - C_{10}^{bs\mu\mu} = \frac{1}{2} \left(  V_{tb} V_{ts}^{*} \frac{e^{2}}{16\pi^{2}} \frac{4 G_{F}}{\sqrt{2}} \right)^{-1} \left[ C_{lq}^{(1)} + C_{lq}^{(3)} \right]_{2232} \approx -0.5 \ .
\end{equation}

It was pointed out in Ref.~\cite{Aebischer:2019mlg} that there exists a mild
tension between the fit to $R_{K^{(*)}}$ and the other anomalous $b \to s$ data,
which can be reconciled with an additional LFU contribution to
$C_{9}^{bs\ell\ell}$ such that
\begin{equation}
  C_{9}^{bs\mu\mu} \approx -0.44\quad \text{ and }\quad C_{9}^{bs\ell\ell} \approx -0.5 \ ,
\end{equation}
for $\ell \in \{e, \mu, \tau\}$. A potential source of this universal
contribution to $C_{9}$ is new physics in four-quark operators
like~\cite{Aebischer:2019mlg}
\begin{equation}
  \label{eq:ouq}
  [\mathcal{O}_{qu}^{(1)}]_{2322} = (\bar{Q}_{2} \gamma_{\mu} Q_{3}) (\bar{u}_{2} \gamma^{\mu} \bar{u}_{2}) \ ,
\end{equation}
which can be generated, for example, by
$\Phi \sim (\mathbf{8}, \mathbf{2}, \tfrac{1}{2})_{S}$. The relevant Yukawa terms are
\begin{equation}
  \label{eq:phi-yuks}
  - \mathscr{L}_{\Phi} = w_{rs} Q_{r}^{ai} \bar{u}_{sb} \Phi^{bj}_{\ a} \epsilon_{ij} + \text{h.c.}
\end{equation}
and a contribution of about the right size to $C_{9}^{bs\ell\ell}$ can be
generated while avoiding dijet exclusion bounds from the LHC for
$m_{\Phi} \sim \SI{2}{\TeV}$ and
$|w_{22}|, |w_{32}| \sim 1$~\cite{Aebischer:2019mlg}.

We construct a UV model that contains $\zeta$ and $\Phi$ as well as one of
$\omega_{1}$ or $\Pi_{7}$ in an attempt to incorporate this explanation into a
model of neutrino mass. We emphasise that our goal here is not to present the
most elegant or motivated model of neutrino mass and the flavour anomalies, but
rather to show that our database can be used to motivate complex models with a
specific structure.

We query the filtered model database for neutrino-mass models that contain the
interactions $Q \bar{u} \Phi$, needed to generate $\mathcal{O}_{qu}^{(1)}$;
$LQ\zeta$, needed to generate $C_{9}^{bs\mu\mu} = -C_{10}^{bs\mu\mu}$; and one
of $\omega_{1}$ or $\Pi_{7}$, required to explain $R_{D^{(*)}}$. Our query
returns a number of models, and we choose one to study briefly below. We note
that none of the models involve the leptoquark $\omega_{1}$, and none feature
the interaction $\bar{e}^{\dagger} Q^{\dagger} \Pi_{7}$, implying some freedom
in the explanation of $R_{D^{(*)}}$ since the couplings $y_{rs}$ of
Eqs.~\eqref{eq:pi7-coup} and \eqref{eq:omega1-pi7-clequ} will be
unrelated\footnote{Expanding our search criteria, we find no viable models in
  the database in which both sets of couplings presented in
  Eqs.~\eqref{eq:omega1-coup} and \eqref{eq:pi7-coup} feature. This can be
  understood in the following way. Any neutrino self-energy diagram containing
  both couplings will also imply another where
  $\bar{e}^{\dagger} \bar{u}^\dagger \omega_{1}$ or
  $\bar{e}^{\dagger} Q^{\dagger} \Pi_{7}$ is replaced with the corresponding
  coupling to $L$, which contains a neutrino field. This generally gives a
  larger contribution to the neutrino mass, since the closure of the diagram
  containing the $\bar{e}$ will involve an additional loop with a $W$ boson.
  Thus, diagrams with both sets of Yukawa interactions to SM fermions relevant
  to $\omega_{1}$ and $\Pi_{7}$ are likely to be removed by our filtering
  procedure. We note that, after studying the unfiltered list of models, we find
  that some models can be engineered so that a sizeable (but not dominant)
  contribution to the neutrino masses does come from such diagrams involving
  both sets of leptoquark--fermion Yukawa couplings.} to the neutrino mass.

The model contains the additional fields $\Phi$, $\zeta$, $\Pi_{7}$ and
$\eta \sim (\mathbf{8}, \mathbf{1}, 1)_{S}$, necessary for lepton-number
violation. It generates $\mathcal{O}_{29b}$, which implies an upper bound on the
new-physics scale of roughly $\SI{e7}{\TeV}$. The additional piece of the
Lagrangian is $\Delta \mathscr{L} = \mathscr{L}_{Y} - \mathcal{V}$, with
\begin{align}
  -\mathscr{L}_{Y} &= x_{rs} L^{i}_{r}\bar{u}_{sa} \Pi^{ja}_{7} \epsilon_{ij} +  y_{rs} \bar{e}^{\dagger}_{r} Q_{sai}^{\dagger} \Pi^{ai}_{7} + z_{rs}L^{i}_{r} Q^{ja}_{s} \zeta_{a}^{\{kl\}} \epsilon_{ik} \epsilon_{jl} + w_{rs} Q^{ai}_{r} \bar{u}_{sb} \Phi^{bj}_{\ a} \epsilon_{ij}  + \text{h.c.} \label{eq:flav-anom-yuks} \\
  \mathcal{V} &= \kappa H^{i} \Phi^{a j}_{\ b} \eta^{\dagger b}_{\ \ a} \epsilon_{ij} + \lambda H^{i} \eta^{a}_{\ b} \tilde{\Pi}^{j}_{7a} \tilde{\zeta}^{b \{kl\}} \epsilon_{ik}\epsilon_{jl} + \text{h.c.} + \cdots \ ,
\end{align}
where we have only shown the part of the scalar potential relevant to
lepton-number violation in this model, since the full expression contains a
large number of terms. The leptoquark $\zeta$ has a diquark coupling which we
forbid by imposing $\mathrm{U}(1)_{B}$ on the Lagrangian, assigning baryon
numbers of $-\tfrac{1}{3}$ and $\tfrac{1}{3}$ to $\zeta$ and $\Pi_{7}$,
respectively. (All other exotic fields have $B = 0$.) The model contains 33 free
parameters, although not all of them are necessary to address the flavour
anomalies and generate viable neutrino masses.

\begin{figure}[t]
  \centering
  \subcaptionbox{\label{fig:flavour-anomalies-model-treelevel}}{
    \includegraphics[width=0.38\linewidth]{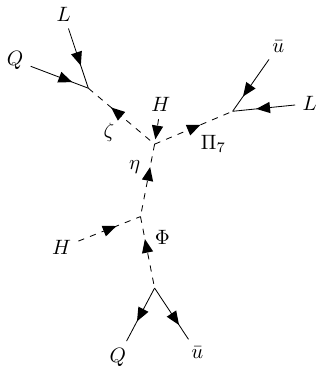}
  }
  \subcaptionbox{\label{fig:flavour-anomalies-model-neutrinomass}}{
    \includegraphics[width=0.53\linewidth]{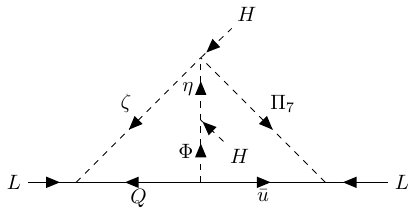}
  }
  \caption{(a) The figure shows the tree-level completion diagram for the model
    constructed to address the flavour anomalies and neutrino masses. The
    topology is labelled $2s6f_{2}$ in our scheme. The closure contains two
    arrow-preserving loops, which arise by looping the $\bar{u}$ into the
    $\bar{u}^{\dagger}$ and the $Q$ into the $Q^{\dagger}$. (b) The self-energy
    diagram for the same model. The diagram has a CLBZ-10 topology in the
    language of Ref.~\cite{Sierra:2014rxa}. The neutrino masses are not
    suppressed by SM-fermion masses on account of the arrow-preserving fermion
    lines. This feature raises the bound on the new-physics scale relevant to
    the model, but also allows couplings to the second generation of fermions to
    play a role in the physics of neutrino mass. This is beneficial in our case
    since many of these couplings are involved in generating the pattern of
    dimension-six operators that motivates this example, and so provides for a
    more intimate connection between the flavour anomalies and neutrino masses.}
  \label{fig:flavour-anomalies-model-diagrams}
\end{figure}

The tree-level completion diagram and the neutrino self-energy diagram are shown
in Fig.~\ref{fig:flavour-anomalies-model-diagrams}. The neutrino mass arises at
two loops, and the topology has the feature that no fermion propagators are
arrow-violating. This implies that the neutrino masses are not proportional to
any SM-fermion masses. This feature has been studied before in the context of a
specific UV model in Ref.~\cite{Gargalionis:2019drk}. The phenomenon is
particular to models derived from operators whose closures feature
arrow-preserving loops, as discussed in Sec.~\ref{sec:operator-closures}. From a
model-building perspective, one consequence is that the neutrino masses need not
be dominated by Yukawa couplings to SM fermions of the third generation. Indeed,
motivated by the pattern of operators required to explain the flavour anomalies,
we adopt textures for the Yukawa couplings of Eq.~\eqref{eq:flav-anom-yuks} that
imply dominance of the bottom-quark couplings for $\zeta$, but the charm-quark
couplings for $\Pi_{7}$:
\begin{equation}
  \mathbf{x} = \begin{pmatrix}
    0 & x_{12} & 0 \\
    0 & x_{22} & 0 \\
    0 & x_{32} & 0 \\
  \end{pmatrix},\quad \mathbf{z} = \begin{pmatrix}
    0 & 0 & z_{13} \\
    0 & z_{22} & z_{23} \\
    0 & 0 & z_{33} \\
  \end{pmatrix} \ ,
\end{equation}
where the additional coupling $z_{22}$ is required to generate the relevant
dimension-six operators $[\mathcal{O}_{lq}^{(1,3)}]_{2232}$. Interestingly, the
minimal set of couplings $w_{rs}$ that gives viable neutrino masses while
incorporating the key ingredients required to generate both
$[\mathcal{O}_{lq}^{(1,3)}]_{2232}$ and $[\mathcal{O}_{lequ}^{(1,3)}]_{3332}$ is
\begin{equation}
  \mathbf{w} = \begin{pmatrix}
    0 & 0 & 0 \\
    0 & w_{22} & 0 \\
    0 & w_{32} & 0 \\
    \end{pmatrix} \ ,
  \end{equation}
  which is exactly the correct set required to also generate the operator given
  in Eq.~\eqref{eq:ouq}. Thus, there is a natural connection in this model
  between the explanation of the charged- and neutral-current anomalies through
  the neutrino masses. With the exception of $y_{33}$, all of the couplings
  featuring in the explanation of the flavour anomalies also play a role in the
  generation of the neutrino masses. The structure of the neutrino-mass matrix
  is
  \begin{equation}
    \label{eq:flav-anom-mv}
    \begin{aligned}
      [\mathbf{m}_{\nu}]_{rs} &\simeq \frac{\lambda \kappa}{(16 \pi^{2})^{2}} \frac{v^{2}}{\Lambda^{2}} \sum_{t,u} [z_{rt} w_{tu} x_{su} + (r \leftrightarrow s)] \\
      &= \frac{\lambda \kappa}{(16 \pi^{2})^{2}} \frac{v^{2}}{\Lambda^{2}} [z_{r2} w_{22} x_{s2} + z_{r3}w_{32}x_{s2} + (r \leftrightarrow s)] \ .
    \end{aligned}
  \end{equation}
  The matrix is rank 2, and so implies an almost massless neutrino. Since there is no suppression of the neutrino-mass scale by SM Yukawa
  couplings, we distinguish the UV scales $\Lambda$ and $\kappa$ so that
  \begin{equation}
    \Lambda \simeq \max\left(m_{\zeta}, m_{\Phi}, m_{\eta}, m_{\Pi_{7}}\right)
  \end{equation}
  and consider the region of parameter space in which
  $\lambda \kappa \ll \Lambda$.

  An explanation of the flavour anomalies in this picture can be achieved with
  $\mathcal{O}(1)$ couplings for $\Pi_{7}$ and $\Phi$ at a few \TeV, and $\zeta$
  at tens of \TeV. We take $\eta$ slightly heavier at $\sim \SI{50}{\TeV}$ to
  decouple its phenomenology and aid in suppressing the neutrino mass. This
  implies $\lambda \kappa \sim \SI{0.05}{\GeV}$ for neutrino masses saturating
  the atmospheric bound. This choice is technically natural, since in the limit
  of vanishing $\lambda$ or $\kappa$ the Lagrangian regains $\mathrm{U}(1)_{L}$.
  We rewrite Eq.~\eqref{eq:flav-anom-mv} as
  \begin{equation}
    [\mathbf{m}_{\nu}]_{rs} = m_{0} [ x_{s2}(z_{r2}w_{22} + z_{r3}w_{32}) + (r \leftrightarrow s) ] \ ,
  \end{equation}
  where $m_{0} \approx \lambda \kappa v^{2} (16\pi^{2})^{-2} m_{\eta}^{-2}$.
  This allows for the adoption of a Casas--Ibarra-like parametrisation of the
  vectors $x_{s2}$ and
  \begin{equation}
    \mathbf{Z} = \begin{pmatrix} z_{13}w_{32} \\ z_{22}w_{22} + z_{23}w_{32} \\ z_{33} w_{32} \end{pmatrix} \ ,
  \end{equation}
  so that~\cite{Cai:2014kra}
  \begin{align}
    x_{r2} &= \frac{\xi}{\sqrt{2m_{0}}} \left(\sqrt{m_{2}} u_{2}^{*} + i \sqrt{m_{3}} u_{3}^{*} \right) \ , \\
    Z_{r} &= \frac{1}{\xi \sqrt{2m_{0}}} \left(\sqrt{m_{2}} u_{2}^{*} - i \sqrt{m_{3}} u_{3}^{*} \right) \ ,
  \end{align}
  where the $u_{i}$ are the $i$th columns of the PMNS matrix; $m_{i}$ are the
  neutrino masses, fixed by the measured squared mass differences and the choice
  of normal ordering; and $\xi$ is a free complex parameter. We find, for
  example, that the choices $m_{\Phi} = \SI{2}{\TeV}$,
  $m_{\Pi_{7}} = \SI{1}{\TeV}$, $m_{\zeta} = \SI{15}{\TeV}$,
  $m_{\eta} = \SI{50}{\TeV}$, $\lambda \kappa = \SI{0.05}{\GeV}$,
  $\xi = e^{3i/2}$, $z_{23} = 1$, $w_{22} = - w_{32} = 1$ and $y_{33} = 2e^{2i}$
  give approximately the right values to generate the pattern of dimension-six
  operators discussed and explain the flavour anomalies. This includes the
  additional lepton-flavour universal contribution to $C_{9}^{bs\ell\ell}$,
  discussed in Ref.~\cite{Aebischer:2019mlg}. Although a more detailed study of
  the phenomenological implications of the model is beyond the scope of this
  simple example, we have shown how a specific UV scenario can be embedded into
  a radiative model in a way consistent with the measured neutrino masses and
  mixing parameters.

\section{Conclusions}
\label{sec:conclusions}

We have described a procedure for building UV-complete models from effective
operators in a way amenable to automation. We have applied the algorithm, as
found in our publicly available example code~\cite{neutrinomass2020}, to the
$\Delta L = 2$ operators in the SMEFT up to and including dimension eleven,
producing just over 11,000 minimal and predictive models of radiative Majorana
neutrino mass. We share our complete listing of models, as well as the set
reduced by model filtering, in our searchable model
database~\cite{gargalionis_john_2020_4054618}.

Our analysis includes new operators that have not appeared in previous
catalogues, along with updated estimates for the upper bounds on the new-physics
scales associated with these. We performed a preliminary study of the UV models,
showing that the most represented exotic fields featuring in the completions are
leptoquarks. We find that a number of simple models predict new physics that
must live below \SI{100}{\TeV}. Adding the additional requirements that the
models contain fewer than four exotic fields and that the new-physics scale
should be larger than \SI{700}{\GeV} gives at most five models fitting this
description, all of which predict new fields below \SI{15}{\TeV}. One of these
models was studied briefly, along with a model derived from a derivative
operator, and one that addresses the flavour anomalies.

Our model database is perhaps a good laboratory for experiments in automated
phenomenological analysis. Now that the models have been written down and
compiled into this computationally accessible format, our hope is that a large
number of them can be ruled out in a systematic way through improved model
filtering, neutrino oscillation data, or collider constraints. Our results also
pave the way for more detailed studies of the models that are currently
accessible to experiments. As each model is tested, we will either get very
lucky and discover the origin of neutrino masses at low energies, or else
falsify these scenarios and build a stronger circumstantial case for those that
cannot be tested at collider experiments.

\appendix

\section{Table of operators}

Below we present the catalogue of $\Delta L = 2$ operators we use in our study.
The operators are listed and labelled in a way consistent with the previous
catalogues~\cite{Babu:2001ex,deGouvea:2007qla}, although we enforce that
operators with the same field content carry the same numerical labels. This
means that our listing may contain more $\mathrm{SU}(2)_{L}$ structures for any
numbered family of operators. Product operators as presented in the table must
be read with care. This is just a convenient shorthand to represent the
field-content of an operator and illustrate that isospin indices are internally
contracted. For example, by writing
$\mathcal{O}_{5b} = \mathcal{O}_{1} Q^{i} \bar{d} \tilde{H}^{j} \epsilon_{ij}$,
we do not mean to suggest that Lorentz indices must be contracted internally to
$\mathcal{O}_{1}$ and the down-type Yukawa.

In each row we also provide information relevant to the number of completions.
The number of unfiltered models (sets of field content) derived from the
operator using our techniques is presented, along with the number that survive
the democratic filtering procedure with the neutrino-mass filtering criterion. A
sizeable number of operators end up with no completions that can play a dominant
role in the physics of neutrino mass.

Other information relevant to the operators is also shown, including the number
of loops required for the operator closure (the same as the number of loops
appearing in the associated neutrino self-energy diagram) and the upper-bound on
the scale of the new physics generating the operator at tree level, derived from
the atmospheric lower bound on the mass of the heaviest neutrino. Operators for
which a range is given for the number of loops are those that generate the
dimension-seven or dimension-nine analogues of the Weinberg operator. The
additional Higgs fields in these diagrams can always be closed off, adding more
loops to the neutrino self-energy while reducing the overall scale suppression.
The contribution with the highest number of loops will dominate for scales
$\Lambda \gtrsim 4\pi v$.

We remind the reader that our analysis does not include the number of unfiltered
completions of $\mathcal{O}_{1}^{\prime\prime\prime}$. In this case, the number
of filtered models comes from Ref.~\cite{Anamiati:2018cuq}. Other operators
featuring a `---' are non-explosive, \textit{i.e.} they do not support
tree-level topologies containing only scalars and fermions.

\begin{longtable}[c]{ | l | l | c | c | c | c |}
  \caption{The table displays our listing of the $\Delta L = 2$ operators along with the number of completions before and after our model-filtering procedure, the number of loops in the neutrino self-energy diagram, and the upper bound on the new-physics scale associated with each operator. See the main text of the appendix for more information.\label{tab:long}}\\
  \hline
  Labels & Operator & Models & Filtered & Loops & $\Lambda~[\text{TeV}]$ \\
  \endfirsthead \hline
  Labels & Operator & Models & Filtered & Loops & $\Lambda~[\text{TeV}]$ \\
  \hline \endhead \hline
$1$ & $L^{i} L^{j} H^{k} H^{l}  \cdot  \epsilon_{i k} \epsilon_{j l}$ & 3 & 3 & 0 & \mynum{605520000000.000} \\
$2$ & $L^{i} L^{j} L^{k} \bar{e} H^{l}  \cdot  \epsilon_{i k} \epsilon_{j l}$ & 8 & 2 & 1 & \mynum{39226496.2471310} \\
$3a$ & $L^{i} L^{j} Q^{k} \bar{d} H^{l}  \cdot  \epsilon_{i j} \epsilon_{k l}$ & 9 & 2 & 2 & \mynum{231157.260299850} \\
$3b$ & $L^{i} L^{j} Q^{k} \bar{d} H^{l}  \cdot  \epsilon_{i k} \epsilon_{j l}$ & 14 & 5 & 1 & \mynum{92116154.1084314} \\
$4a$ & $L^{i} L^{j} \tilde{Q}^{k} \bar{u}^{\dagger} H^{l}  \cdot  \epsilon_{i k} \epsilon_{j l}$ & 5 & 0 & 1 & \mynum{3807173871.71594} \\
$4b$ & $L^{i} L^{j} \tilde{Q}^{k} \bar{u}^{\dagger} H^{l}  \cdot  \epsilon_{i j} \epsilon_{k l}$ & 4 & 2 & 2 & \mynum{9553762.74866082} \\
$5a$ & $L^{i} L^{j} Q^{k} \bar{d} H^{l} H^{m} \tilde{H}^{n}  \cdot \epsilon_{i l} \epsilon_{j n} \epsilon_{k m}$ & 790 & 36 & 2 & \mynum{583332.360427892} \\
$5b$ & $\mathcal{O}_1 \cdot Q^{i} \bar{d} \tilde{H}^{j}  \cdot \epsilon_{i j}$ & 492 & 14 & 1,2 & \mynum{583332.360436087} \\
$5c$ & $\mathcal{O}_{3a} \cdot H^{i} \tilde{H}^{j}  \cdot \epsilon_{i j}$ & 509 & 0 & 2,3 & \mynum{1463.82371741127} \\
$5d$ & $\mathcal{O}_{3b} \cdot H^{i} \tilde{H}^{j}  \cdot \epsilon_{i j}$ & 799 & 16 & 1,2 & \mynum{583332.360436087} \\
$6a$ & $L^{i} L^{j} \tilde{Q}^{k} \bar{u}^{\dagger} H^{l} H^{m} \tilde{H}^{n}  \cdot  \epsilon_{i l} \epsilon_{j n} \epsilon_{k m}$ & 289 & 14 & 2 & \mynum{24109210.1884026} \\
$6b$ & $\mathcal{O}_1 \cdot \tilde{Q}^{i} \bar{u}^{\dagger} \tilde{H}^{j}  \cdot  \epsilon_{i j}$ & 177 & 0 & 1,2 & \mynum{24109210.1884027} \\
$6c$ & $\mathcal{O}_{4a} \cdot H^{i} \tilde{H}^{j}  \cdot  \epsilon_{i j}$ & 262 & 0 & 1,2 & \mynum{24109210.1884027} \\
$6d$ & $\mathcal{O}_{4b} \cdot H^{i} \tilde{H}^{j}  \cdot  \epsilon_{i j}$ & 208 & 0 & 2,3 & \mynum{60499.9094497991} \\
$7$ & $L^{i} \bar{e}^{\dagger} Q^{j} \tilde{Q}^{k} H^{l} H^{m} H^{n}  \cdot  \epsilon_{i l} \epsilon_{j m} \epsilon_{k n}$ & 240 & 15 & 2 & \mynum{248404.689368816} \\
$8$ & $L^{i} \bar{e}^{\dagger} \bar{u}^{\dagger} \bar{d} H^{j}  \cdot  \epsilon_{i j}$ & 5 & 1 & 2,3 & \mynum{15.1766163003309} \\
$9$ & $L^{i} L^{j} L^{k} L^{l} \bar{e} \bar{e}  \cdot  \epsilon_{i k} \epsilon_{j l}$ & 14 & 1 & 2 & \mynum{2541.15141997984} \\
$10$ & $L^{i} L^{j} L^{k} \bar{e} Q^{l} \bar{d}  \cdot  \epsilon_{i k} \epsilon_{j l}$ & 50 & 1 & 2 & \mynum{5967.42299748072} \\
$11a$ & $L^{i} L^{j} Q^{k} Q^{l} \bar{d} \bar{d}  \cdot  \epsilon_{i j} \epsilon_{k l}$ & 48 & 0 & 3 & \mynum{35.1653418765092} \\
$11b$ & $L^{i} L^{j} Q^{k} Q^{l} \bar{d} \bar{d}  \cdot  \epsilon_{i k} \epsilon_{j l}$ & 72 & 16 & 2 & \mynum{14013.3865895896} \\
$12a$ & $L^{i} L^{j} \tilde{Q}^{k} \tilde{Q}^{l} \bar{u}^{\dagger} \bar{u}^{\dagger}  \cdot  \epsilon_{i k} \epsilon_{j l}$ & 19 & 0 & 2 & \mynum{23937397.4261404} \\
$12b$ & $L^{i} L^{j} \tilde{Q}^{k} \tilde{Q}^{l} \bar{u}^{\dagger} \bar{u}^{\dagger}  \cdot  \epsilon_{i j} \epsilon_{k l}$ & 17 & 4 & 3 & \mynum{60068.7605913504} \\
$13$ & $L^{i} L^{j} L^{k} \bar{e} \tilde{Q}^{l} \bar{u}^{\dagger}  \cdot  \epsilon_{i k} \epsilon_{j l}$ & 12 & 0 & 2 & \mynum{246634.449053772} \\
$14a$ & $L^{i} L^{j} Q^{k} \tilde{Q}^{l} \bar{u}^{\dagger} \bar{d}  \cdot  \epsilon_{i j} \epsilon_{k l}$ & 29 & 1 & 3 & \mynum{1453.38862741285} \\
$14b$ & $L^{i} L^{j} Q^{k} \tilde{Q}^{l} \bar{u}^{\dagger} \bar{d}  \cdot  \epsilon_{i k} \epsilon_{j l}$ & 43 & 1 & 2 & \mynum{579175.279238636} \\
$15$ & $L^{i} L^{j} L^{k} \tilde{L}^{l} \bar{u}^{\dagger} \bar{d}  \cdot \epsilon_{i k} \epsilon_{j l}$ & 12 & 1 & 3 & \mynum{1453.38862741285} \\
$16$ & $L^{i} L^{j} \bar{e} \bar{e}^{\dagger} \bar{u}^{\dagger} \bar{d}  \cdot  \epsilon_{i j}$ & 13 & 1 & 3 & \mynum{1453.38862741285} \\
$17$ & $L^{i} L^{j} \bar{u}^{\dagger} \bar{d} \bar{d} \bar{d}^{\dagger}  \cdot  \epsilon_{i j}$ & 18 & 12 & 3 & \mynum{1453.38862741285} \\
$18$ & $L^{i} L^{j} \bar{u} \bar{u}^{\dagger} \bar{u}^{\dagger} \bar{d}  \cdot  \epsilon_{i j}$ & 22 & 8 & 3 & \mynum{1453.38862741285} \\
$19$ & $L^{i} \bar{e}^{\dagger} Q^{j} \bar{u}^{\dagger} \bar{d} \bar{d}  \cdot  \epsilon_{i j}$ & 27 & 0 & 3,4 & \mynum{0.221892467282772} \\
$20$ & $L^{i} \bar{e}^{\dagger} \tilde{Q}^{j} \bar{u}^{\dagger} \bar{u}^{\dagger} \bar{d}  \cdot  \epsilon_{i j}$ & 27 & 3 & 3,4 & \mynum{0.797031006138724} \\
$21a$ & $L^{i} L^{j} L^{k} \bar{e} Q^{l} \bar{u} H^{m} H^{n}  \cdot  \epsilon_{i l} \epsilon_{j m} \epsilon_{k n}$ & 3943 & 1 & 2,3 & \mynum{1561.83395520421} \\
$21b$ & $L^{i} L^{j} L^{k} \bar{e} Q^{l} \bar{u} H^{m} H^{n}  \cdot  \epsilon_{i k} \epsilon_{j m} \epsilon_{l n}$ & 4080 & 4 & 3 & \mynum{1561.83089406901} \\
$22a$ & $L^{i} L^{j} L^{k} \tilde{L}^{l} \bar{e} \bar{e}^{\dagger} H^{m} H^{n}  \cdot  \epsilon_{i l} \epsilon_{j m} \epsilon_{k n}$ & 726 & 0 & 2 & \mynum{24282256.1517830} \\
$22b$ & $\mathcal{O}_2 \cdot \tilde{L}^i \bar{e}^\dagger H^j \epsilon_{ij}$ & 931 & 0 & 2 & \mynum{24282256.1517830} \\
$23a$ & $L^{i} L^{j} L^{k} \bar{e} \tilde{Q}^{l} \bar{d}^{\dagger} H^{m} H^{n}  \cdot  \epsilon_{i l} \epsilon_{j m} \epsilon_{k n}$ & 780 & 0 & 2,3 & \mynum{37.9148278684193} \\
$23b$ & $\mathcal{O}_2 \cdot \tilde{Q}^i \bar{d}^\dagger H^j \cdot \epsilon_{ij}$ & 969 & 0 & 2,3 & \mynum{37.9148278684193} \\
$24a$ & $L^{i} L^{j} Q^{k} Q^{l} \bar{d} \bar{d} H^{m} \tilde{H}^{n}  \cdot  \epsilon_{i l} \epsilon_{j n} \epsilon_{k m}$ & 9613 & 193 & 3 & \mynum{88.7408072559298} \\
$24b$ & $L^{i} L^{j} Q^{k} Q^{l} \bar{d} \bar{d} H^{m} \tilde{H}^{n}  \cdot  \epsilon_{i m} \epsilon_{j n} \epsilon_{k l}$ & 6058 & 110 & 3 & \mynum{88.7408072559298} \\
$24c$ & $\mathcal{O}_{3a} \cdot Q^i \bar{d} \tilde{H}^j \cdot \epsilon_{ij}$ & 6022 & 34 & 3,4 & \mynum{1.10099178378389} \\
$24d$ & $\mathcal{O}_{3b} \cdot Q^i \bar{d} \tilde{H}^j \cdot \epsilon_{ij}$ & 9616 & 211 & 2,3 & \mynum{88.7946179153465} \\
$24e$ & $\mathcal{O}_{11a} \cdot H^i \tilde{H}^j \cdot \epsilon_{ij}$ & 3834 & 18 & 3,4 & \mynum{1.10099178378389} \\
$24f$ & $\mathcal{O}_{11b} \cdot H^i \tilde{H}^j \cdot \epsilon_{ij}$ & 5915 & 131 & 2,3 & \mynum{88.7946179153465} \\
$25a$ & $L^{i} L^{j} Q^{k} Q^{l} \bar{u} \bar{d} H^{m} H^{n}  \cdot  \epsilon_{i m} \epsilon_{j n} \epsilon_{k l}$ & 5960 & 151 & 2,3 & \mynum{3667.67160535231} \\
$25b$ & $\mathcal{O}_{3a} \cdot Q^i \bar{u} H^j \cdot \epsilon_{ij}$ & 5913 & 9 & 3,4 & \mynum{9.67388631414653} \\
$25c$ & $\mathcal{O}_{3b} \cdot Q^i \bar{u} H^j \cdot \epsilon_{ij}$ & 14036 & 470 & 2,3 & \mynum{3667.67160535231} \\
$26a$ & $L^{i} L^{j} \tilde{L}^{k} \bar{e}^{\dagger} Q^{l} \bar{d} H^{m} H^{n}  \cdot  \epsilon_{i k} \epsilon_{j m} \epsilon_{l n}$ & 1600 & 0 & 3 & \mynum{37.7891475874534} \\
$26b$ & $L^{i} L^{j} \tilde{L}^{k} \bar{e}^{\dagger} Q^{l} \bar{d} H^{m} H^{n}  \cdot  \epsilon_{i m} \epsilon_{j n} \epsilon_{k l}$ & 1040 & 0 & 2,3 & \mynum{37.9148278684193} \\
$26c$ & $\mathcal{O}_{3a} \cdot \tilde{L}^i \bar{e}^\dagger H^j \cdot \epsilon_{ij}$ & 1149 & 0 & 3 & \mynum{37.7891475874534} \\
$26d$ & $\mathcal{O}_{3b} \cdot \tilde{L}^i \bar{e}^\dagger H^j \cdot \epsilon_{ij}$ & 1797 & 0 & 2,3 & \mynum{37.9148278684193} \\
$27a$ & $L^{i} L^{j} Q^{k} \tilde{Q}^{l} \bar{d} \bar{d}^{\dagger} H^{m} H^{n}  \cdot  \epsilon_{i k} \epsilon_{j m} \epsilon_{l n}$ & 3851 & 164 & 2 & \mynum{24282256.1517830} \\
$27b$ & $L^{i} L^{j} Q^{k} \tilde{Q}^{l} \bar{d} \bar{d}^{\dagger} H^{m} H^{n}  \cdot  \epsilon_{i m} \epsilon_{j n} \epsilon_{k l}$ & 2226 & 74 & 2 & \mynum{24282256.1517830} \\
$27c$ & $\mathcal{O}_{3a} \cdot \tilde{Q}^i \bar{d}^\dagger H^j \cdot \epsilon_{ij}$ & 2469 & 33 & 3 & \mynum{60934.1527582468} \\
$27d$ & $\mathcal{O}_{3b} \cdot \tilde{Q}^i \bar{d}^\dagger H^j \cdot \epsilon_{ij}$ & 3443 & 165 & 2 & \mynum{24282256.1517830} \\
$28a$ & $L^{i} L^{j} Q^{k} \tilde{Q}^{l} \bar{u}^{\dagger} \bar{d} H^{m} \tilde{H}^{n}  \cdot  \epsilon_{i l} \epsilon_{j n} \epsilon_{k m}$ & 4038 & 64 & 3 & \mynum{3667.67030180250} \\
$28b$ & $L^{i} L^{j} Q^{k} \tilde{Q}^{l} \bar{u}^{\dagger} \bar{d} H^{m} \tilde{H}^{n}  \cdot  \epsilon_{i m} \epsilon_{j n} \epsilon_{k l}$ & 4103 & 0 & 3,4 & \mynum{9.67388631414653} \\
$28c$ & $L^{i} L^{j} Q^{k} \tilde{Q}^{l} \bar{u}^{\dagger} \bar{d} H^{m} \tilde{H}^{n}  \cdot  \epsilon_{i k} \epsilon_{j n} \epsilon_{l m}$ & 4305 & 123 & 3 & \mynum{3667.67030180250} \\
$28d$ & $\mathcal{O}_{3a} \cdot \tilde{Q}^i \bar{u}^\dagger \tilde{H}^j \cdot \epsilon_{ij}$ & 2749 & 7 & 3,4 & \mynum{9.67388631414653} \\
$28e$ & $\mathcal{O}_{3b} \cdot \tilde{Q}^i \bar{u}^\dagger \tilde{H}^j \cdot \epsilon_{ij}$ & 4304 & 90 & 2,3 & \mynum{3667.67160535231} \\
$28f$ & $\mathcal{O}_{4a} \cdot Q^i \bar{d} \tilde{H}^j \cdot \epsilon_{ij}$ & 4039 & 74 & 2,3 & \mynum{3667.67160535231} \\
$28g$ & $\mathcal{O}_{4b} \cdot Q^i \bar{d} \tilde{H}^j \cdot \epsilon_{ij}$ & 2748 & 14 & 3,4 & \mynum{9.67388631414653} \\
$28h$ & $\mathcal{O}_{14a} \cdot H^i \tilde{H}^j \cdot \epsilon_{ij}$ & 2701 & 10 & 3,4 & \mynum{9.67388631414653} \\
$28i$ & $\mathcal{O}_{14b} \cdot H^i \tilde{H}^j \cdot \epsilon_{ij}$ & 4177 & 90 & 3 & \mynum{3667.67030180250} \\
$29a$ & $L^{i} L^{j} Q^{k} \tilde{Q}^{l} \bar{u} \bar{u}^{\dagger} H^{m} H^{n}  \cdot  \epsilon_{i m} \epsilon_{j n} \epsilon_{k l}$ & 2226 & 267 & 2 & \mynum{24282256.1517830} \\
$29b$ & $L^{i} L^{j} Q^{k} \tilde{Q}^{l} \bar{u} \bar{u}^{\dagger} H^{m} H^{n}  \cdot  \epsilon_{i k} \epsilon_{j m} \epsilon_{l n}$ & 3846 & 498 & 2 & \mynum{24282256.1517830} \\
$29c$ & $\mathcal{O}_{4a} \cdot Q^i \bar{u} H^j \cdot \epsilon_{ij}$ & 3444 & 422 & 2 & \mynum{24282256.1517830} \\
$29d$ & $\mathcal{O}_{4b} \cdot Q^i \bar{u} H^j \cdot \epsilon_{ij}$ & 2468 & 64 & 3 & \mynum{60934.1527582468} \\
$30a$ & $L^{i} L^{j} \tilde{L}^{k} \bar{e}^{\dagger} \tilde{Q}^{l} \bar{u}^{\dagger} H^{m} H^{n}  \cdot  \epsilon_{i k} \epsilon_{j m} \epsilon_{l n}$ & 1772 & 0 & 3 & \mynum{1561.83089406901} \\
$30b$ & $L^{i} L^{j} \tilde{L}^{k} \bar{e}^{\dagger} \tilde{Q}^{l} \bar{u}^{\dagger} H^{m} H^{n}  \cdot  \epsilon_{i m} \epsilon_{j n} \epsilon_{k l}$ & 1140 & 2 & 3 & \mynum{1561.83089406901} \\
$30c$ & $\mathcal{O}_{4a} \cdot \tilde{L}^i \bar{e}^\dagger H^j \cdot \epsilon_{ij}$ & 1776 & 2 & 2,3 & \mynum{1561.83395520421} \\
$30d$ & $\mathcal{O}_{4b} \cdot \tilde{L}^i \bar{e}^\dagger H^j \cdot \epsilon_{ij}$ & 1398 & 11 & 3 & \mynum{1561.83089406901} \\
$31a$ & $\mathcal{O}_{4a} \cdot \tilde{Q}^i \bar{d}^\dagger H^j \cdot \epsilon_{ij}$ & 3107 & 10 & 2,3 & \mynum{3667.67160535231} \\
$31b$ & $L^{i} L^{j} \tilde{Q}^{k} \tilde{Q}^{l} \bar{u}^{\dagger} \bar{d}^{\dagger} H^{m} H^{n}  \cdot  \epsilon_{i m} \epsilon_{j n} \epsilon_{k l}$ & 1404 & 4 & 2,3 & \mynum{3667.67160535231} \\
$31c$ & $\mathcal{O}_{4b} \cdot \tilde{Q}^i \bar{d}^\dagger H^j \cdot \epsilon_{ij}$ & 1654 & 8 & 3,4 & \mynum{9.67388631414653} \\
$32a$ & $L^{i} L^{j} \tilde{Q}^{k} \tilde{Q}^{l} \bar{u}^{\dagger} \bar{u}^{\dagger} H^{m} \tilde{H}^{n}  \cdot  \epsilon_{i l} \epsilon_{j n} \epsilon_{k m}$ & 2103 & 157 & 3 & \mynum{151585.340033349} \\
$32b$ & $L^{i} L^{j} \tilde{Q}^{k} \tilde{Q}^{l} \bar{u}^{\dagger} \bar{u}^{\dagger} H^{m} \tilde{H}^{n}  \cdot  \epsilon_{i m} \epsilon_{j n} \epsilon_{k l}$ & 1493 & 151 & 3 & \mynum{151585.340033349} \\
$32c$ & $\mathcal{O}_{4a} \cdot \tilde{Q}^i \bar{u}^\dagger \tilde{H}^j \cdot \epsilon_{ij}$ & 2100 & 56 & 3 & \mynum{151585.340033349} \\
$32d$ & $\mathcal{O}_{4b} \cdot \tilde{Q}^i \bar{u}^\dagger \tilde{H}^j \cdot \epsilon_{ij}$ & 1747 & 26 & 3,4 & \mynum{380.402438028539} \\
$32e$ & $\mathcal{O}_{12a} \cdot H^i \tilde{H}^j$ & 1250 & 36 & 3 & \mynum{151585.340033349} \\
$32f$ & $\mathcal{O}_{12b} \cdot H^i \tilde{H}^j$ & 1143 & 24 & 3,4 & \mynum{380.402438028539} \\
$33$ & $\mathcal{O}_1 \cdot \bar{e} \bar{e} \bar{e}^{\dagger} \bar{e}^{\dagger}$ & 451 & 5 & 2 & \mynum{24282256.1517830} \\
$34$ & $L^{i} \bar{e} \bar{e}^{\dagger} \bar{e}^{\dagger} Q^{j} \bar{d} H^{k} H^{l}  \cdot  \epsilon_{i k} \epsilon_{j l}$ & 1377 & 231 & 3 & \mynum{37.7891475874534} \\
$35$ & $L^{i} \bar{e} \bar{e}^{\dagger} \bar{e}^{\dagger} \tilde{Q}^{j} \bar{u}^{\dagger} H^{k} H^{l}  \cdot  \epsilon_{i k} \epsilon_{j l}$ & 1126 & 15 & 3 & \mynum{1561.83089406901} \\
$36$ & $\bar{e}^{\dagger} \bar{e}^{\dagger} Q^{i} Q^{j} \bar{d} \bar{d} H^{k} H^{l}  \cdot  \epsilon_{i k} \epsilon_{j l}$ & 970 & 208 & 4 & \mynum{0.0000588091842232492} \\
$37$ & $\bar{e}^{\dagger} \bar{e}^{\dagger} Q^{i} \tilde{Q}^{j} \bar{u}^{\dagger} \bar{d} H^{k} H^{l}  \cdot  \epsilon_{i k} \epsilon_{j l}$ & 2470 & 58 & 4,5,6,7 & \mynum{0.0425599112941507} \\
$38$ & $\bar{e}^{\dagger} \bar{e}^{\dagger} \tilde{Q}^{i} \tilde{Q}^{j} \bar{u}^{\dagger} \bar{u}^{\dagger} H^{k} H^{l}  \cdot  \epsilon_{i k} \epsilon_{j l}$ & 3358 & 451 & 4 & \mynum{0.100456717300928} \\
$39a$ & $\mathcal{O}_1 \cdot L^{i} L^{j} \tilde{L}^{k} \tilde{L}^{l} \cdot  \epsilon_{i k} \epsilon_{j l}$ & 296 & 0 & 2 & \mynum{24282256.1517830} \\
$39b$ & $L^{i} L^{j} L^{k} L^{l} \tilde{L}^{m} \tilde{L}^{n} H^{p} H^{q}  \cdot  \epsilon_{i k} \epsilon_{j l} \epsilon_{m p} \epsilon_{n q}$ & 220 & 6 & 2 & \mynum{24282256.1517830} \\
$39c$ & $L^{i} L^{j} L^{k} L^{l} \tilde{L}^{m} \tilde{L}^{n} H^{p} H^{q}  \cdot  \epsilon_{i l} \epsilon_{j n} \epsilon_{k p} \epsilon_{m q}$ & 588 & 0 & 2 & \mynum{24282256.1517830} \\
$39d$ & $\mathcal{O}_1 \cdot L^{i} L^{j} \tilde{L}^{k} \tilde{L}^{l} \cdot  \epsilon_{i j} \epsilon_{k l}$ & 324 & 0 & 2 & \mynum{24282256.1517830} \\
$40a$ & $L^{i} L^{j} L^{k} \tilde{L}^{l} Q^{m} \tilde{Q}^{n} H^{p} H^{q}  \cdot  \epsilon_{i l} \epsilon_{j n} \epsilon_{k p} \epsilon_{m q}$ & 963 & 22 & 2 & \mynum{24282256.1517830} \\
$40b$ & $L^{i} L^{j} L^{k} \tilde{L}^{l} Q^{m} \tilde{Q}^{n} H^{p} H^{q}  \cdot  \epsilon_{i l} \epsilon_{j p} \epsilon_{k q} \epsilon_{m n}$ & 729 & 25 & 2 & \mynum{24282256.1517830} \\
$40c$ & $L^{i} L^{j} L^{k} \tilde{L}^{l} Q^{m} \tilde{Q}^{n} H^{p} H^{q}  \cdot  \epsilon_{i n} \epsilon_{j p} \epsilon_{k q} \epsilon_{l m}$ & 759 & 25 & 2 & \mynum{24282256.1517830} \\
$40d$ & $L^{i} L^{j} L^{k} \tilde{L}^{l} Q^{m} \tilde{Q}^{n} H^{p} H^{q}  \cdot  \epsilon_{i k} \epsilon_{j l} \epsilon_{m p} \epsilon_{n q}$ & 953 & 0 & 3 & \mynum{60934.1527582468} \\
$40e$ & $L^{i} L^{j} L^{k} \tilde{L}^{l} Q^{m} \tilde{Q}^{n} H^{p} H^{q}  \cdot  \epsilon_{i l} \epsilon_{j m} \epsilon_{k p} \epsilon_{n q}$ & 1321 & 31 & 2 & \mynum{24282256.1517830} \\
$40f$ & $L^{i} L^{j} L^{k} \tilde{L}^{l} Q^{m} \tilde{Q}^{n} H^{p} H^{q}  \cdot  \epsilon_{i k} \epsilon_{j n} \epsilon_{l p} \epsilon_{m q}$ & 963 & 100 & 2 & \mynum{24282256.1517830} \\
$40g$ & $L^{i} L^{j} L^{k} \tilde{L}^{l} Q^{m} \tilde{Q}^{n} H^{p} H^{q}  \cdot  \epsilon_{i m} \epsilon_{j n} \epsilon_{k p} \epsilon_{l q}$ & 1339 & 30 & 2 & \mynum{24282256.1517830} \\
$40h$ & $L^{i} L^{j} L^{k} \tilde{L}^{l} Q^{m} \tilde{Q}^{n} H^{p} H^{q}  \cdot  \epsilon_{i k} \epsilon_{j m} \epsilon_{l p} \epsilon_{n q}$ & 820 & 56 & 2 & \mynum{24282256.1517830} \\
$40i$ & $L^{i} L^{j} L^{k} \tilde{L}^{l} Q^{m} \tilde{Q}^{n} H^{p} H^{q}  \cdot  \epsilon_{i m} \epsilon_{j p} \epsilon_{k q} \epsilon_{l n}$ & 844 & 9 & 2 & \mynum{24282256.1517830} \\
$40j$ & $L^{i} L^{j} L^{k} \tilde{L}^{l} Q^{m} \tilde{Q}^{n} H^{p} H^{q}  \cdot  \epsilon_{i k} \epsilon_{j p} \epsilon_{l n} \epsilon_{m q}$ & 908 & 60 & 2 & \mynum{24282256.1517830} \\
$40k$ & $L^{i} L^{j} L^{k} \tilde{L}^{l} Q^{m} \tilde{Q}^{n} H^{p} H^{q}  \cdot  \epsilon_{i k} \epsilon_{j p} \epsilon_{l m} \epsilon_{n q}$ & 970 & 98 & 2 & \mynum{24282256.1517830} \\
$40l$ & $L^{i} L^{j} L^{k} \tilde{L}^{l} Q^{m} \tilde{Q}^{n} H^{p} H^{q} \cdot  \epsilon_{i k} \epsilon_{j p} \epsilon_{l q} \epsilon_{m n}$ & 933 & 87 & 2 & \mynum{24282256.1517830} \\
$41a$ & $L^{i} L^{j} L^{k} \tilde{L}^{l} \bar{d} \bar{d}^{\dagger} H^{m} H^{n}  \cdot  \epsilon_{i l} \epsilon_{j m} \epsilon_{k n}$ & 729 & 6 & 2 & \mynum{24282256.1517830} \\
$41b$ & $L^{i} L^{j} L^{k} \tilde{L}^{l} \bar{d} \bar{d}^{\dagger} H^{m} H^{n}  \cdot  \epsilon_{i k} \epsilon_{j m} \epsilon_{l n}$ & 933 & 71 & 2 & \mynum{24282256.1517830} \\
$42a$ & $L^{i} L^{j} L^{k} \tilde{L}^{l} \bar{u} \bar{u}^{\dagger} H^{m} H^{n}  \cdot  \epsilon_{i l} \epsilon_{j m} \epsilon_{k n}$ & 729 & 21 & 2 & \mynum{24282256.1517830} \\
$42b$ & $L^{i} L^{j} L^{k} \tilde{L}^{l} \bar{u} \bar{u}^{\dagger} H^{m} H^{n}  \cdot  \epsilon_{i k} \epsilon_{j m} \epsilon_{l n}$ & 933 & 120 & 2 & \mynum{24282256.1517830} \\
$43a$ & $L^{i} L^{j} L^{k} \tilde{L}^{l} \bar{u}^{\dagger} \bar{d} H^{m} \tilde{H}^{n}  \cdot  \epsilon_{i k} \epsilon_{j n} \epsilon_{l m}$ & 1068 & 7 & 3,4 & \mynum{9.67388631414653} \\
$43b$ & $L^{i} L^{j} L^{k} \tilde{L}^{l} \bar{u}^{\dagger} \bar{d} H^{m} \tilde{H}^{n}  \cdot  \epsilon_{i l} \epsilon_{j m} \epsilon_{k n}$ & 1438 & 7 & 3,4 & \mynum{9.67388631414653} \\
$43c$ & $L^{i} L^{j} L^{k} \tilde{L}^{l} \bar{u}^{\dagger} \bar{d} H^{m} \tilde{H}^{n}  \cdot  \epsilon_{i k} \epsilon_{j m} \epsilon_{l n}$ & 1068 & 8 & 3,4 & \mynum{9.67388631414653} \\
$43d$ & $L^{i} L^{j} L^{k} \tilde{L}^{l} \bar{u}^{\dagger} \bar{d} H^{m} \tilde{H}^{n}  \cdot  \epsilon_{i k} \epsilon_{j l} \epsilon_{m n}$ & 1068 & 8 & 3,4 & \mynum{9.67388631414653} \\
$44a$ & $L^{i} L^{j} \bar{e} \bar{e}^{\dagger} Q^{k} \tilde{Q}^{l} H^{m} H^{n}  \cdot  \epsilon_{i l} \epsilon_{j m} \epsilon_{k n}$ & 1571 & 155 & 2 & \mynum{24282256.1517830} \\
$44b$ & $L^{i} L^{j} \bar{e} \bar{e}^{\dagger} Q^{k} \tilde{Q}^{l} H^{m} H^{n}  \cdot  \epsilon_{i m} \epsilon_{j n} \epsilon_{k l}$ & 1016 & 91 & 2 & \mynum{24282256.1517830} \\
$44c$ & $L^{i} L^{j} \bar{e} \bar{e}^{\dagger} Q^{k} \tilde{Q}^{l} H^{m} H^{n}  \cdot  \epsilon_{i j} \epsilon_{k m} \epsilon_{l n}$ & 1137 & 2 & 3 & \mynum{60934.1527582468} \\
$44d$ & $L^{i} L^{j} \bar{e} \bar{e}^{\dagger} Q^{k} \tilde{Q}^{l} H^{m} H^{n}  \cdot  \epsilon_{i k} \epsilon_{j m} \epsilon_{l n}$ & 1765 & 133 & 2 & \mynum{24282256.1517830} \\
$45$ & $L^{i} L^{j} \bar{e} \bar{e}^{\dagger} \bar{d} \bar{d}^{\dagger} H^{k} H^{l}  \cdot  \epsilon_{i k} \epsilon_{j l}$ & 1016 & 81 & 2 & \mynum{24282256.1517830} \\
$46$ & $L^{i} L^{j} \bar{e} \bar{e}^{\dagger} \bar{u} \bar{u}^{\dagger} H^{k} H^{l}  \cdot  \epsilon_{i k} \epsilon_{j l}$ & 1016 & 49 & 2 & \mynum{24282256.1517830} \\
$47a$ & $L^{i} L^{j} Q^{k} Q^{l} \tilde{Q}^{m} \tilde{Q}^{n} H^{p} H^{q}  \cdot  \epsilon_{i m} \epsilon_{j n} \epsilon_{k p} \epsilon_{l q}$ & 1013 & 236 & 2 & \mynum{24282256.1517830} \\
$47b$ & $L^{i} L^{j} Q^{k} Q^{l} \tilde{Q}^{m} \tilde{Q}^{n} H^{p} H^{q}  \cdot  \epsilon_{i m} \epsilon_{j p} \epsilon_{k n} \epsilon_{l q}$ & 2253 & 423 & 2 & \mynum{24282256.1517830} \\
$47c$ & $L^{i} L^{j} Q^{k} Q^{l} \tilde{Q}^{m} \tilde{Q}^{n} H^{p} H^{q}  \cdot  \epsilon_{i p} \epsilon_{j q} \epsilon_{k m} \epsilon_{l n}$ & 1007 & 200 & 2 & \mynum{24282256.1517830} \\
$47d$ & $L^{i} L^{j} Q^{k} Q^{l} \tilde{Q}^{m} \tilde{Q}^{n} H^{p} H^{q}  \cdot  \epsilon_{i l} \epsilon_{j n} \epsilon_{k p} \epsilon_{m q}$ & 2838 & 690 & 2 & \mynum{24282256.1517830} \\
$47e$ & $L^{i} L^{j} Q^{k} Q^{l} \tilde{Q}^{m} \tilde{Q}^{n} H^{p} H^{q}  \cdot  \epsilon_{i n} \epsilon_{j p} \epsilon_{k l} \epsilon_{m q}$ & 1730 & 387 & 2 & \mynum{24282256.1517830} \\
$47f$ & $L^{i} L^{j} Q^{k} Q^{l} \tilde{Q}^{m} \tilde{Q}^{n} H^{p} H^{q}  \cdot  \epsilon_{i j} \epsilon_{k n} \epsilon_{l p} \epsilon_{m q}$ & 1702 & 60 & 3 & \mynum{60934.1527582468} \\
$47g$ & $L^{i} L^{j} Q^{k} Q^{l} \tilde{Q}^{m} \tilde{Q}^{n} H^{p} H^{q}  \cdot  \epsilon_{i l} \epsilon_{j p} \epsilon_{k n} \epsilon_{m q}$ & 2796 & 530 & 2 & \mynum{24282256.1517830} \\
$47h$ & $L^{i} L^{j} Q^{k} Q^{l} \tilde{Q}^{m} \tilde{Q}^{n} H^{p} H^{q}  \cdot  \epsilon_{i j} \epsilon_{k p} \epsilon_{l q} \epsilon_{m n}$ & 924 & 46 & 3 & \mynum{60934.1527582468} \\
$47i$ & $L^{i} L^{j} Q^{k} Q^{l} \tilde{Q}^{m} \tilde{Q}^{n} H^{p} H^{q}  \cdot  \epsilon_{i l} \epsilon_{j p} \epsilon_{k q} \epsilon_{m n}$ & 2078 & 369 & 2 & \mynum{24282256.1517830} \\
$47j$ & $L^{i} L^{j} Q^{k} Q^{l} \tilde{Q}^{m} \tilde{Q}^{n} H^{p} H^{q}  \cdot  \epsilon_{i p} \epsilon_{j q} \epsilon_{k l} \epsilon_{m n}$ & 902 & 183 & 2 & \mynum{24282256.1517830} \\
$47k$ & $L^{i} L^{j} Q^{k} Q^{l} \tilde{Q}^{m} \tilde{Q}^{n} H^{p} H^{q}  \cdot  \epsilon_{i k} \epsilon_{j l} \epsilon_{m p} \epsilon_{n q}$ & 1203 & 258 & 2 & \mynum{24282256.1517830} \\
$47l$ & $L^{i} L^{j} Q^{k} Q^{l} \tilde{Q}^{m} \tilde{Q}^{n} H^{p} H^{q}  \cdot  \epsilon_{i j} \epsilon_{k l} \epsilon_{m p} \epsilon_{n q}$ & 814 & 46 & 3 & \mynum{60934.1527582468} \\
$48$ & $L^{i} L^{j} \bar{d} \bar{d} \bar{d}^{\dagger} \bar{d}^{\dagger} H^{k} H^{l}  \cdot  \epsilon_{i k} \epsilon_{j l}$ & 921 & 125 & 2 & \mynum{24282256.1517830} \\
$49$ & $L^{i} L^{j} \bar{u} \bar{u}^{\dagger} \bar{d} \bar{d}^{\dagger} H^{k} H^{l}  \cdot  \epsilon_{i k} \epsilon_{j l}$ & 2086 & 384 & 2 & \mynum{24282256.1517830} \\
$50a$ & $L^{i} L^{j} \bar{u}^{\dagger} \bar{d} \bar{d} \bar{d}^{\dagger} H^{k} \tilde{H}^{l}  \cdot  \epsilon_{i k} \epsilon_{j l}$ & 2285 & 68 & 3,4 & \mynum{9.67388631414653} \\
$50b$ & $\mathcal{O}_{17} \cdot H^i \tilde{H}^j \cdot \epsilon_{ij}$ & 1523 & 52 & 3,4 & \mynum{9.67388631414653} \\
$51$ & $L^{i} L^{j} \bar{u} \bar{u} \bar{u}^{\dagger} \bar{u}^{\dagger} H^{k} H^{l}  \cdot  \epsilon_{i k} \epsilon_{j l}$ & 921 & 225 & 2 & \mynum{24282256.1517830} \\
$52a$ & $L^{i} L^{j} \bar{u} \bar{u}^{\dagger} \bar{u}^{\dagger} \bar{d} H^{k} \tilde{H}^{l}  \cdot  \epsilon_{i k} \epsilon_{j l}$ & 2896 & 170 & 3,4 & \mynum{9.67388631414653} \\
$52b$ & $\mathcal{O}_{18} \cdot H^i \tilde{H}^j \cdot \epsilon_{ij}$ & 1872 & 94 & 3,4 & \mynum{9.67388631414653} \\
$53$ & $L^{i} L^{j} \bar{u}^{\dagger} \bar{u}^{\dagger} \bar{d} \bar{d} \tilde{H}^{k} \tilde{H}^{l}  \cdot  \epsilon_{i k} \epsilon_{j l}$ & 939 & 162 & 4,5,6 & \mynum{0.151764140756919} \\
$54a$ & $L^{i} \bar{e}^{\dagger} Q^{j} Q^{k} \tilde{Q}^{l} \bar{d} H^{m} H^{n}  \cdot  \epsilon_{i l} \epsilon_{j m} \epsilon_{k n}$ & 2203 & 92 & 3 & \mynum{37.7891475874534} \\
$54b$ & $L^{i} \bar{e}^{\dagger} Q^{j} Q^{k} \tilde{Q}^{l} \bar{d} H^{m} H^{n}  \cdot  \epsilon_{i m} \epsilon_{j l} \epsilon_{k n}$ & 3393 & 89 & 3 & \mynum{37.7891475874534} \\
$54c$ & $L^{i} \bar{e}^{\dagger} Q^{j} Q^{k} \tilde{Q}^{l} \bar{d} H^{m} H^{n}  \cdot  \epsilon_{i m} \epsilon_{j k} \epsilon_{l n}$ & 2456 & 30 & 3 & \mynum{37.7891475874534} \\
$54d$ & $L^{i} \bar{e}^{\dagger} Q^{j} Q^{k} \tilde{Q}^{l} \bar{d} H^{m} H^{n}  \cdot  \epsilon_{i k} \epsilon_{j m} \epsilon_{l n}$ & 3835 & 100 & 3 & \mynum{37.7891475874534} \\
$55a$ & $L^{i} \bar{e}^{\dagger} Q^{j} \tilde{Q}^{k} \tilde{Q}^{l} \bar{u}^{\dagger} H^{m} H^{n}  \cdot  \epsilon_{i l} \epsilon_{j m} \epsilon_{k n}$ & 3478 & 143 & 3 & \mynum{1561.83089406901} \\
$55b$ & $L^{i} \bar{e}^{\dagger} Q^{j} \tilde{Q}^{k} \tilde{Q}^{l} \bar{u}^{\dagger} H^{m} H^{n}  \cdot  \epsilon_{i m} \epsilon_{j l} \epsilon_{k n}$ & 3493 & 144 & 3 & \mynum{1561.83089406901} \\
$55c$ & $L^{i} \bar{e}^{\dagger} Q^{j} \tilde{Q}^{k} \tilde{Q}^{l} \bar{u}^{\dagger} H^{m} H^{n}  \cdot  \epsilon_{i m} \epsilon_{j n} \epsilon_{k l}$ & 2049 & 86 & 3 & \mynum{1561.83089406901} \\
$55d$ & $L^{i} \bar{e}^{\dagger} Q^{j} \tilde{Q}^{k} \tilde{Q}^{l} \bar{u}^{\dagger} H^{m} H^{n}  \cdot  \epsilon_{i j} \epsilon_{k m} \epsilon_{l n}$ & 2156 & 106 & 3 & \mynum{1561.83089406901} \\
$56$ & $L^{i} \bar{e}^{\dagger} Q^{j} \bar{d} \bar{d} \bar{d}^{\dagger} H^{k} H^{l}  \cdot  \epsilon_{i k} \epsilon_{j l}$ & 2273 & 252 & 3 & \mynum{37.7891475874534} \\
$57$ & $L^{i} \bar{e}^{\dagger} \tilde{Q}^{j} \bar{u}^{\dagger} \bar{d} \bar{d}^{\dagger} H^{k} H^{l}  \cdot  \epsilon_{i k} \epsilon_{j l}$ & 4251 & 481 & 3 & \mynum{1561.83089406901} \\
$58$ & $L^{i} \bar{e}^{\dagger} \tilde{Q}^{j} \bar{u} \bar{u}^{\dagger} \bar{u}^{\dagger} H^{k} H^{l}  \cdot  \epsilon_{i k} \epsilon_{j l}$ & 2408 & 183 & 3 & \mynum{1561.83089406901} \\
$59a$ & $L^{i} \bar{e}^{\dagger} Q^{j} \bar{u}^{\dagger} \bar{d} \bar{d} H^{k} \tilde{H}^{l}  \cdot  \epsilon_{i k} \epsilon_{j l}$ & 2638 & 65 & 3,4,5 & \mynum{0.201597798274807} \\
$59b$ & $\mathcal{O}_{19} \cdot H^i \tilde{H}^j \cdot \epsilon_{ij}$ & 2583 & 65 & 3,4,5 & \mynum{0.201597798274807} \\
$59c$ & $\mathcal{O}_{8} \cdot Q^i \bar{d} \tilde{H}^j \cdot \epsilon_{ij}$ & 2639 & 42 & 4,5,6 & \mynum{0.0606824914422098} \\
$60a$ & $L^{i} \bar{e}^{\dagger} \tilde{Q}^{j} \bar{u}^{\dagger} \bar{u}^{\dagger} \bar{d} H^{k} \tilde{H}^{l}  \cdot  \epsilon_{i l} \epsilon_{j k}$ & 2687 & 35 & 4,5 & \mynum{0.141649379630778} \\
$60b$ & $\mathcal{O}_8 \cdot \tilde{Q}^i \bar{u}^\dagger \tilde{H}^j \cdot \epsilon_{ij}$ & 2687 & 121 & 3,4,5 & \mynum{0.429379935617762} \\
$60c$ & $\mathcal{O}_{20} \cdot H^i \tilde{H}^j \cdot \epsilon_{ij}$ & 2687 & 104 & 3,4,5 & \mynum{0.429379935617762} \\
$61a$ & $\mathcal{O}_1 \cdot L^i \bar{e} \tilde{H}^j \cdot \epsilon_{ij}$ & 382 & 0 & 1,2 & \mynum{248404.689388061} \\
$61b$ & $\mathcal{O}_2 \cdot H^i \tilde{H}^j \cdot \epsilon_{ij}$ & 408 & 0 & 1,2 & \mynum{248404.689388061} \\
$62a$ & $L^{i} L^{j} L^{k} L^{l} \bar{e} \bar{e} H^{m} \tilde{H}^{n}  \cdot  \epsilon_{i l} \epsilon_{j m} \epsilon_{k n}$ & 1820 & 0 & 2,3 & \mynum{16.3788192601811} \\
$62b$ & $\mathcal{O}_9 \cdot H^i \tilde{H}^j \cdot \epsilon_{ij}$ & 830 & 1 & 2,3 & \mynum{16.3788192601811} \\
$63a$ & $L^{i} L^{j} L^{k} \bar{e} Q^{l} \bar{d} H^{m} \tilde{H}^{n}  \cdot  \epsilon_{i k} \epsilon_{j n} \epsilon_{l m}$ & 4619 & 12 & 3 & \mynum{37.7891475874534} \\
$63b$ & $L^{i} L^{j} L^{k} \bar{e} Q^{l} \bar{d} H^{m} \tilde{H}^{n}  \cdot  \epsilon_{i l} \epsilon_{j m} \epsilon_{k n}$ & 7216 & 77 & 2,3 & \mynum{37.9148278684193} \\
$63c$ & $\mathcal{O}_2 \cdot Q^i \bar{d} \tilde{H}^j \cdot \epsilon_{ij}$ & 4621 & 49 & 2,3 & \mynum{37.9148278684193} \\
$63d$ & $\mathcal{O}_{10} \cdot H^i \tilde{H}^j \cdot \epsilon_{ij}$ & 4590 & 45 & 2,3 & \mynum{37.9148278684193} \\
$64a$ & $L^{i} L^{j} L^{k} \bar{e} \tilde{Q}^{l} \bar{u}^{\dagger} H^{m} \tilde{H}^{n}  \cdot  \epsilon_{i l} \epsilon_{j m} \epsilon_{k n}$ & 1370 & 0 & 3 & \mynum{1561.83089406901} \\
$64b$ & $L^{i} L^{j} L^{k} \bar{e} \tilde{Q}^{l} \bar{u}^{\dagger} H^{m} \tilde{H}^{n}  \cdot  \epsilon_{i k} \epsilon_{j n} \epsilon_{l m}$ & 1050 & 0 & 3 & \mynum{1561.83089406901} \\
$64c$ & $\mathcal{O}_2 \cdot \tilde{Q}^i \bar{u}^\dagger \tilde{H}^j \cdot \epsilon_{ij}$ & 1049 & 0 & 2,3 & \mynum{1561.83395520421} \\
$64d$ & $\mathcal{O}_{13} \cdot H^i \tilde{H}^j \cdot \epsilon_{ij}$ & 1008 & 0 & 3 & \mynum{1561.83089406901} \\
$65a$ & $L^{i} L^{j} \bar{e} \bar{e}^{\dagger} \bar{u}^{\dagger} \bar{d} H^{k} \tilde{H}^{l}  \cdot  \epsilon_{i k} \epsilon_{j l}$ & 1925 & 17 & 3,4 & \mynum{9.67388631414653} \\
$65b$ & $\mathcal{O}_{16} \cdot H^i \tilde{H}^j \cdot \epsilon_{ij}$ & 1259 & 11 & 3,4 & \mynum{9.67388631414653} \\
$71$ & $\mathcal{O}_1 \cdot Q^{i} \bar{u} H^{j} \cdot  \epsilon_{i j}$ & 396 & 9 & 2 & \mynum{24109210.1884026} \\
$75$ & $\mathcal{O}_8 \cdot Q^{i} \bar{u} H^j \cdot \epsilon_{i j}$ & 3951 & 84 & 3 & \mynum{37.7891475874534} \\
$76$ & $\bar{e}^\dagger \bar{e}^\dagger \bar{u}^\dagger \bar{u}^\dagger \bar{d} \bar{d}$ & 16 & 4 & 4,5,6 & \mynum{0.0242272389061306} \\
$77$ & $\mathcal{O}_{1} \cdot \tilde{L}^{i} \bar{e}^{\dagger} H^{j} \cdot \epsilon_{ij}$ & 156 & 0 & 2 & \mynum{248404.689368816} \\
$78$ & $\mathcal{O}_{1} \cdot \tilde{Q}^{i} \bar{d}^{\dagger} H^{j} \cdot \epsilon_{ij}$ & 156 & 0 & 2 & \mynum{583332.360427892} \\
$1^{\prime}$ & $\mathcal{O}_{1} \cdot \tilde{H}^{i} H^{j} \cdot \epsilon_{ij}$ & 53 & 1 & 0,1 & \mynum{3834500194.94427} \\
$8^{\prime}$ & $\mathcal{O}_{8} \cdot \tilde{H}^{i} H^{j} \cdot \epsilon_{ij}$ & 301 & 4 & 2,3,4 & \mynum{1.31993072379150} \\
$1^{\prime\prime}$ & $\mathcal{O}_{1} \cdot\tilde{H}^{i} H^{j} \tilde{H}^{k} H^{l} \cdot \epsilon_{ij} \epsilon_{kl}$ & 1893 & 6 & 0,1,2 & \mynum{24282256.1517834} \\
$1^{\prime\prime\prime}$ & $\mathcal{O}_{1} \cdot \tilde{H}^{i} H^{j} \tilde{H}^{k} H^{l} \tilde{H}^{m} H^{n} \cdot \epsilon_{ij} \epsilon_{kl} \epsilon_{mn}$ & --- & 2 & 0,1,2 & \mynum{24282256.1517834} \\
$7^{\prime}$ & $\mathcal{O}_{7} \cdot \tilde{H}^{i} H^{j} \cdot \epsilon_{ij}$ & 24951 & 374 & 2,3 & \mynum{1573.04411114262} \\
$8^{\prime\prime}$ & $\mathcal{O}_{8} \cdot \tilde{H}^{i} H^{j} \tilde{H}^{k} H^{l} \cdot \epsilon_{ij} \epsilon_{kl}$ & 19229 & 197 & 2,3,4,5 & \mynum{0.706389586862170} \\
$71^{\prime}$ & $\mathcal{O}_{71} \cdot \tilde{H}^{i} H^{j}  \cdot \epsilon_{ij}$ & 39331 & 446 & 2,3 & \mynum{152673.357091994} \\
$76^{\prime}$ & $\mathcal{O}_{76} \cdot \tilde{H}^{i} H^{j}  \cdot \epsilon_{ij}$ & 679 & 209 & 4,5,6,7 & \mynum{0.0425599112941507} \\
$77^{\prime}$ & $\mathcal{O}_{77} \cdot \tilde{H}^{i} H^{j}  \cdot \epsilon_{ij}$ & 14598 & 0 & 1,2,3 & \mynum{1573.04715044539} \\
$78^{\prime}$ & $\mathcal{O}_{78} \cdot \tilde{H}^{i} H^{j}  \cdot \epsilon_{ij}$ & 14644 & 1 & 2,3 & \mynum{3693.99662022470} \\
$79a$ & $\mathcal{O}_{61a} \cdot \tilde{H}^{i} H^{j}  \cdot \epsilon_{ij}$ & 31791 & 14 & 1,2,3 & \mynum{1573.04715044539} \\
$79b$ & $\mathcal{O}_{2} \cdot \tilde{H}^{i} H^{j}  \tilde{H}^{k} H^{l} \cdot \epsilon_{ij} \epsilon_{kl}$ & 23931 & 14 & 1,2,3 & \mynum{1573.04715044539} \\
$80a$ & $\mathcal{O}_{5a} \cdot \tilde{H}^{i} H^{j}  \cdot \epsilon_{ij}$ & 72694 & 154 & 2,3 & \mynum{3693.99662022470} \\
$80b$ & $\mathcal{O}_{5b} \cdot \tilde{H}^{i} H^{j}  \cdot \epsilon_{ij}$ & 49108 & 371 & 1,2,3 & \mynum{3693.99791448348} \\
$80c$ & $\mathcal{O}_{3a} \cdot \tilde{H}^{i} H^{j}  \tilde{H}^{k} H^{l} \cdot \epsilon_{ij} \epsilon_{kl}$ & 31569 & 16 & 2,3,4 & \mynum{10.1500704145279} \\
$80d$ & $\mathcal{O}_{3b} \cdot \tilde{H}^{i} H^{j}  \tilde{H}^{k} H^{l} \cdot \epsilon_{ij} \epsilon_{kl}$ & 49505 & 367 & 1,2,3 & \mynum{3693.99791448348} \\
$81a$ & $\mathcal{O}_{6a} \cdot \tilde{H}^{i} H^{j} \cdot \epsilon_{ij}$ & 26174 & 95 & 2,3 & \mynum{152673.357091994} \\
$81b$ & $\mathcal{O}_{6b} \cdot \tilde{H}^{i} H^{j} \cdot \epsilon_{ij}$ & 17298 & 18 & 1,2,3 & \mynum{152673.357123309} \\
$81c$ & $\mathcal{O}_{4a} \cdot \tilde{H}^{i} H^{j}  \tilde{H}^{k} H^{l} \cdot \epsilon_{ij} \epsilon_{kl}$ & 15575 & 18 & 1,2,3 & \mynum{152673.357123309} \\
$81d$ & $\mathcal{O}_{4b} \cdot \tilde{H}^{i} H^{j}  \tilde{H}^{k} H^{l} \cdot \epsilon_{ij} \epsilon_{kl}$ & 12400 & 41 & 2,3,4 & \mynum{383.145107084499} \\
$82$ & $L^{i} \tilde{L}^{j} \bar{e}^{\dagger} \bar{e}^{\dagger} \bar{u}^{\dagger} \bar{d} H^{k} H^{l} \cdot \epsilon_{ik} \epsilon_{jl}$ & 1151 & 56 & 3,4,5 & \mynum{0.169788542155929} \\
$D1$ & $(DL)^{i} L^{j} {\bar{u}^{\dagger}} \bar{d}  \cdot  \epsilon_{i j}$ & --- & --- & 3,4,5 & \mynum{0.169788542155929} \\
$D2a$ & $(DL)^{i} L^{j} (DH)^{k} H^{l}  \cdot  \epsilon_{i j} \epsilon_{k l}$ & 1 & 0 & 1 & \mynum{1519498350.87562} \\
$D2b$ & $(DL)^{i} L^{j} (DH)^{k} H^{l}  \cdot  \epsilon_{i l} \epsilon_{j k}$ & 3 & 3 & 0 & \mynum{605520000000.000} \\
$D2c$ & $(DL)^{i} L^{j} (DH)^{k} H^{l}  \cdot  \epsilon_{i k} \epsilon_{j l}$ & 3 & 3 & 0 & \mynum{605520000000.000} \\
$D3$ & $L^{i} {\bar{e}^{\dagger}} H^{j} H^{k} (DH)^{l}  \cdot  \epsilon_{i k} \epsilon_{j l}$ & 4 & 0 & 1 & \mynum{39226496.2471310} \\
$D4a$ & $L^{i} L^{j} (DL)^{k} (D\bar{e}) H^{l}  \cdot  \epsilon_{i k} \epsilon_{j l}$ & 8 & 2 & 1 & \mynum{39226496.2471310} \\
$D4b$ & $L^{i} L^{j} (DL)^{k} (D\bar{e}) H^{l}  \cdot  \epsilon_{i j} \epsilon_{k l}$ & 8 & 2 & 1 & \mynum{39226496.2471310} \\
$D5a$ & $L^{i} L^{j} (DL)^{k} \tilde{L}^{l} H^{m} H^{n}  \cdot  \epsilon_{i l} \epsilon_{j m} \epsilon_{k n}$ & 21 & 0 & 1 & \mynum{3834500194.94428} \\
$D5b$ & $L^{i} L^{j} (DL)^{k} \tilde{L}^{l} H^{m} H^{n}  \cdot  \epsilon_{i k} \epsilon_{j m} \epsilon_{l n}$ & 30 & 4 & 1 & \mynum{3834500194.94428} \\
$D5c$ & $L^{i} L^{j} (DL)^{k} \tilde{L}^{l} H^{m} H^{n}  \cdot  \epsilon_{i j} \epsilon_{k m} \epsilon_{l n}$ & 30 & 4 & 1 & \mynum{3834500194.94428} \\
$D5d$ & $L^{i} L^{j} (DL)^{k} \tilde{L}^{l} H^{m} H^{n}  \cdot  \epsilon_{i m} \epsilon_{j n} \epsilon_{k l}$ & 21 & 0 & 1 & \mynum{3834500194.94428} \\
$D6a$ & $L^{i} L^{j} \bar{e} {\bar{e}^{\dagger}} (DH)^{k} H^{l}  \cdot  \epsilon_{i k} \epsilon_{j l}$ & 30 & 2 & 1 & \mynum{3834500194.94428} \\
$D6b$ & $L^{i} L^{j} \bar{e} {\bar{e}^{\dagger}} (DH)^{k} H^{l}  \cdot  \epsilon_{i j} \epsilon_{k l}$ & 16 & 0 & 2 & \mynum{9622335.71583110} \\
$D7a$ & $(DL)^{i} L^{j} Q^{k} (D\bar{d}) H^{l}  \cdot  \epsilon_{i j} \epsilon_{k l}$ & 9 & 2 & 2 & \mynum{231157.260299850} \\
$D7b$ & $(DL)^{i} L^{j} Q^{k} (D\bar{d}) H^{l}  \cdot  \epsilon_{i k} \epsilon_{j l}$ & 14 & 5 & 1 & \mynum{92116154.1084314} \\
$D7c$ & $(DL)^{i} L^{j} Q^{k} (D\bar{d}) H^{l}  \cdot  \epsilon_{i l} \epsilon_{j k}$ & 14 & 5 & 1 & \mynum{92116154.1084314} \\
$D8a$ & $L^{i} L^{j} Q^{k} \tilde{Q}^{l} (DH)^{m} H^{n}  \cdot  \epsilon_{i n} \epsilon_{j k} \epsilon_{l m}$ & 53 & 11 & 1 & \mynum{3834500194.94428} \\
$D8b$ & $L^{i} L^{j} Q^{k} \tilde{Q}^{l} (DH)^{m} H^{n}  \cdot  \epsilon_{i n} \epsilon_{j l} \epsilon_{k m}$ & 44 & 6 & 1 & \mynum{3834500194.94428} \\
$D8c$ & $L^{i} L^{j} Q^{k} \tilde{Q}^{l} (DH)^{m} H^{n}  \cdot  \epsilon_{i k} \epsilon_{j l} \epsilon_{m n}$ & 25 & 0 & 2 & \mynum{9622335.71583110} \\
$D8d$ & $L^{i} L^{j} Q^{k} \tilde{Q}^{l} (DH)^{m} H^{n}  \cdot  \epsilon_{i m} \epsilon_{j k} \epsilon_{l n}$ & 53 & 11 & 1 & \mynum{3834500194.94428} \\
$D8e$ & $L^{i} L^{j} Q^{k} \tilde{Q}^{l} (DH)^{m} H^{n}  \cdot  \epsilon_{i m} \epsilon_{j l} \epsilon_{k n}$ & 44 & 6 & 1 & \mynum{3834500194.94428} \\
$D8f$ & $L^{i} L^{j} Q^{k} \tilde{Q}^{l} (DH)^{m} H^{n}  \cdot  \epsilon_{i m} \epsilon_{j n} \epsilon_{k l}$ & 30 & 5 & 1 & \mynum{3834500194.94428} \\
$D8g$ & $L^{i} L^{j} Q^{k} \tilde{Q}^{l} (DH)^{m} H^{n}  \cdot  \epsilon_{i j} \epsilon_{k m} \epsilon_{l n}$ & 35 & 7 & 2 & \mynum{9622335.71583110} \\
$D8h$ & $L^{i} L^{j} Q^{k} \tilde{Q}^{l} (DH)^{m} H^{n}  \cdot  \epsilon_{i j} \epsilon_{k n} \epsilon_{l m}$ & 35 & 7 & 2 & \mynum{9622335.71583110} \\
$D8i$ & $L^{i} L^{j} Q^{k} \tilde{Q}^{l} (DH)^{m} H^{n}  \cdot  \epsilon_{i j} \epsilon_{k l} \epsilon_{m n}$ & 16 & 3 & 2 & \mynum{9622335.71583110} \\
$D9a$ & $L^{i} L^{j} \bar{d} {\bar{d}^{\dagger}} (DH)^{k} H^{l}  \cdot  \epsilon_{i k} \epsilon_{j l}$ & 30 & 5 & 1 & \mynum{3834500194.94428} \\
$D9b$ & $L^{i} L^{j} \bar{d} {\bar{d}^{\dagger}} (DH)^{k} H^{l}  \cdot  \epsilon_{i j} \epsilon_{k l}$ & 16 & 4 & 2 & \mynum{9622335.71583110} \\
$D10a$ & $(DL)^{i} L^{j} {\bar{u}^{\dagger}} \bar{d} H^{k} \tilde{H}^{l}  \cdot  \epsilon_{i l} \epsilon_{j k}$ & 56 & 13 & 2,3 & \mynum{1453.39191694777} \\
$D10b$ & $(DL)^{i} L^{j} {\bar{u}^{\dagger}} \bar{d} H^{k} \tilde{H}^{l}  \cdot  \epsilon_{i j} \epsilon_{k l}$ & 36 & 7 & 2,3 & \mynum{1453.39191694777} \\
$D10c$ & $(DL)^{i} L^{j} {\bar{u}^{\dagger}} \bar{d} H^{k} \tilde{H}^{l}  \cdot  \epsilon_{i k} \epsilon_{j l}$ & 56 & 13 & 2,3 & \mynum{1453.39191694777} \\
$D11$ & $(DL)^{i} L^{j} (D\bar{u}^{\dagger}) (D\bar{d})  \cdot  \epsilon_{i j}$ & --- & --- & 2,3 & \mynum{1453.39191694777} \\
$D12a$ & $L^{i} L^{j} \bar{u} {\bar{u}^{\dagger}} (DH)^{k} H^{l}  \cdot  \epsilon_{i k} \epsilon_{j l}$ & 30 & 5 & 1 & \mynum{3834500194.94428} \\
$D12b$ & $L^{i} L^{j} \bar{u} {\bar{u}^{\dagger}} (DH)^{k} H^{l}  \cdot  \epsilon_{i j} \epsilon_{k l}$ & 16 & 4 & 2 & \mynum{9622335.71583110} \\
$D13a$ & $(DL)^{i} L^{j} \tilde{Q}^{k} (D{\bar{u}^{\dagger}}) H^{l}  \cdot  \epsilon_{i j} \epsilon_{k l}$ & 4 & 2 & 2 & \mynum{9553762.74866082} \\
$D13b$ & $(DL)^{i} L^{j} \tilde{Q}^{k} (D{\bar{u}^{\dagger}}) H^{l}  \cdot  \epsilon_{i k} \epsilon_{j l}$ & 5 & 0 & 1 & \mynum{3807173871.71594} \\
$D14a$ & $L^{i} {\bar{e}^{\dagger}} Q^{j} \bar{d} (DH)^{k} H^{l}  \cdot  \epsilon_{i k} \epsilon_{j l}$ & 53 & 0 & 2 & \mynum{5967.42299748072} \\
$D14b$ & $L^{i} {\bar{e}^{\dagger}} Q^{j} \bar{d} (DH)^{k} H^{l}  \cdot  \epsilon_{i l} \epsilon_{j k}$ & 53 & 0 & 2 & \mynum{5967.42299748072} \\
$D14c$ & $L^{i} {\bar{e}^{\dagger}} Q^{j} \bar{d} (DH)^{k} H^{l}  \cdot  \epsilon_{i j} \epsilon_{k l}$ & 27 & 0 & 2 & \mynum{5967.42299748072} \\
$D15$ & $(DL)^{i} {\bar{e}^{\dagger}} (D \bar{u}^{\dagger}) \bar{d} H^{j}  \cdot  \epsilon_{i j}$ & 5 & 1 & 2,3 & \mynum{15.1766163003309} \\
$D16a$ & $L^{i} {\bar{e}^{\dagger}} \tilde{Q}^{j} {\bar{u}^{\dagger}} (DH)^{k} H^{l}  \cdot  \epsilon_{i k} \epsilon_{j l}$ & 58 & 8 & 2 & \mynum{246634.449053772} \\
$D16b$ & $L^{i} {\bar{e}^{\dagger}} \tilde{Q}^{j} {\bar{u}^{\dagger}} (DH)^{k} H^{l}  \cdot  \epsilon_{i l} \epsilon_{j k}$ & 58 & 8 & 2 & \mynum{246634.449053772} \\
$D16c$ & $L^{i} {\bar{e}^{\dagger}} \tilde{Q}^{j} {\bar{u}^{\dagger}} (DH)^{k} H^{l}  \cdot  \epsilon_{i j} \epsilon_{k l}$ & 27 & 4 & 2 & \mynum{246634.449053772} \\
$D17$ & ${\bar{e}^{\dagger}} {\bar{e}^{\dagger}} {\bar{u}^{\dagger}} \bar{d} (DH)^{i} H^{j}  \cdot  \epsilon_{i j}$ & 16 & 7 & 3,4 & \mynum{0.166691257659305} \\
$D18a$ & $(DL)^{i} L^{j} H^{k} H^{l} (DH)^{m} \tilde{H}^{n}  \cdot  \epsilon_{i k} \epsilon_{j m} \epsilon_{l n}$ & 53 & 1 & 0,1 & \mynum{3834500194.94427} \\
$D18b$ & $(DL)^{i} L^{j} H^{k} H^{l} (DH)^{m} \tilde{H}^{n}  \cdot  \epsilon_{i k} \epsilon_{j l} \epsilon_{m n}$ & 53 & 1 & 0,1 & \mynum{3834500194.94427} \\
$D18c$ & $(DL)^{i} L^{j} H^{k} H^{l} (DH)^{m} \tilde{H}^{n}  \cdot  \epsilon_{i m} \epsilon_{j l} \epsilon_{k n}$ & 53 & 1 & 0,1 & \mynum{3834500194.94427} \\
$D18d$ & $(DL)^{i} L^{j} H^{k} H^{l} (DH)^{m} \tilde{H}^{n}  \cdot  \epsilon_{i j} \epsilon_{k m} \epsilon_{l n}$ & 24 & 1 & 1,2 & \mynum{9622335.71583160} \\
$D18e$ & $(DL)^{i} L^{j} H^{k} H^{l} (DH)^{m} \tilde{H}^{n}  \cdot  \epsilon_{i n} \epsilon_{j l} \epsilon_{k m}$ & 34 & 0 & 1 & \mynum{3834500194.94428} \\
$D18f$ & $(DL)^{i} L^{j} H^{k} H^{l} (DH)^{m} \tilde{H}^{n}  \cdot  \epsilon_{i l} \epsilon_{j n} \epsilon_{k m}$ & 34 & 0 & 1 & \mynum{3834500194.94428} \\
$D19a$ & $(D^{2}L)^{i} L^{j} (D^{2} H)^{k} H^{l}  \cdot  \epsilon_{i j} \epsilon_{k l}$ & 1 & 0 & 1 & \mynum{1519498350.87562} \\
$D19b$ & $(D^{2}L)^{i} L^{j} (D^{2} H)^{k} H^{l}  \cdot  \epsilon_{i l} \epsilon_{j k}$ & 3 & 3 & 0 & \mynum{605520000000.000} \\
$D19c$ & $(D^{2}L)^{i} L^{j} (D^{2} H)^{k} H^{l}  \cdot  \epsilon_{i k} \epsilon_{j l}$ & 3 & 3 & 0 & \mynum{605520000000.000} \\
$D20$ & $L^{i} {\bar{e}^{\dagger}} H^{j} H^{k} H^{l} (DH)^{m} \tilde{H}^{n}  \cdot  \epsilon_{i l} \epsilon_{j m} \epsilon_{k n}$ & 129 & 0 & 1,2 & \mynum{248404.689388061} \\
$D21$ & $(DL)^{i} (D\bar{e}^{\dagger}) H^{j} H^{k} (DH)^{l}  \cdot  \epsilon_{i k} \epsilon_{j l}$ & 2 & 0 & 1 & \mynum{39226496.2471310} \\
$D22$ & ${\bar{e}^{\dagger}} {\bar{e}^{\dagger}} (DH)^{i} (DH)^{j} H^{k} H^{l}  \cdot  \epsilon_{i k} \epsilon_{j l}$ & 9 & 0 & 2 & \mynum{2541.15141997984} \\
  \hline
\end{longtable}

\acknowledgments

We thank Matthew J. Dolan for suggesting the term `exploding' for our
model-building procedure. We are grateful to Innes Bigaran, Yi Cai, Peter Cox,
Tomasz Dutka, Joshua Ellis, Leon Friedrich, Juan Herrero-Garc\'{i}a, Iulia
Popa-Mateiu and Michael A. Schmidt for useful discussions. Most Feynman diagrams
were generated using the Ti\textit{k}Z-Feynman package for
\LaTeX~\cite{Ellis:2016jkw}. This work was supported in part by the Australian
Research Council.

\bibliography{main}

\providecommand{\href}[2]{#2}\begingroup\raggedright\begin{thebibliography}{100}

\bibitem{MINKOWSKI1977421}
P.~Minkowski, \emph{$\mu \to e \gamma$ at a rate of one out of $10^9$ muon
  decays?},
  \href{http://dx.doi.org/https://doi.org/10.1016/0370-2693(77)90435-X}{\emph{Physics
  Letters B} {\bf 67} (1977) 421 -- 428}.

\bibitem{Yanagida:1979as}
T.~Yanagida, \emph{{Horizontal gauge symmetry and masses of neutrinos}},
  {\emph{Conf. Proc.} {\bf C7902131} (1979) 95--99}.

\bibitem{GellMann:1980vs}
M.~Gell-Mann, P.~Ramond and R.~Slansky, \emph{{Complex Spinors and Unified
  Theories}}, {\emph{Conf. Proc.} {\bf C790927} (1979) 315--321},
  [\href{http://arxiv.org/abs/1306.4669}{{\tt 1306.4669}}].

\bibitem{PhysRevLett.44.912}
R.~N. Mohapatra and G.~Senjanovi\ifmmode~\acute{c}\else \'{c}\fi{},
  \emph{Neutrino mass and spontaneous parity nonconservation},
  \href{http://dx.doi.org/10.1103/PhysRevLett.44.912}{\emph{Phys. Rev. Lett.}
  {\bf 44} (Apr, 1980) 912--915}.

\bibitem{Glashow:1979nm}
S.~L. Glashow, \emph{{The Future of Elementary Particle Physics}},
  \href{http://dx.doi.org/10.1007/978-1-4684-7197-7_15}{\emph{NATO Sci. Ser. B}
  {\bf 61} (1980) 687}.

\bibitem{Magg:1980ut}
M.~Magg and C.~Wetterich, \emph{{Neutrino Mass Problem and Gauge Hierarchy}},
  \href{http://dx.doi.org/10.1016/0370-2693(80)90825-4}{\emph{Phys. Lett.} {\bf
  94B} (1980) 61--64}.

\bibitem{PhysRevD.22.2227}
J.~Schechter and J.~W.~F. Valle, \emph{Neutrino masses in su(2)
  \ensuremath{\bigotimes} u(1) theories},
  \href{http://dx.doi.org/10.1103/PhysRevD.22.2227}{\emph{Phys. Rev. D} {\bf
  22} (Nov, 1980) 2227--2235}.

\bibitem{LAZARIDES1981287}
G.~Lazarides, Q.~Shafi and C.~Wetterich, \emph{Proton lifetime and fermion
  masses in an so(10) model},
  \href{http://dx.doi.org/https://doi.org/10.1016/0550-3213(81)90354-0}{\emph{Nuclear
  Physics B} {\bf 181} (1981) 287 -- 300}.

\bibitem{Wetterich:1981bx}
C.~Wetterich, \emph{{Neutrino Masses and the Scale of B-L Violation}},
  \href{http://dx.doi.org/10.1016/0550-3213(81)90279-0}{\emph{Nucl. Phys.} {\bf
  B187} (1981) 343--375}.

\bibitem{PhysRevD.23.165}
R.~N. Mohapatra and G.~Senjanovi\ifmmode~\acute{c}\else \'{c}\fi{},
  \emph{Neutrino masses and mixings in gauge models with spontaneous parity
  violation}, \href{http://dx.doi.org/10.1103/PhysRevD.23.165}{\emph{Phys. Rev.
  D} {\bf 23} (Jan, 1981) 165--180}.

\bibitem{Foot:1988aq}
R.~Foot, H.~Lew, X.~G. He and G.~C. Joshi, \emph{{Seesaw Neutrino Masses
  Induced by a Triplet of Leptons}},
  \href{http://dx.doi.org/10.1007/BF01415558}{\emph{Z. Phys.} {\bf C44} (1989)
  441}.

\bibitem{Zee:1980ai}
A.~Zee, \emph{{A Theory of Lepton Number Violation, Neutrino Majorana Mass, and
  Oscillation}}, \href{http://dx.doi.org/10.1016/0370-2693(80)90349-4,
  10.1016/0370-2693(80)90193-8}{\emph{Phys. Lett.} {\bf 93B} (1980) 389}.

\bibitem{Zee:1985id}
A.~Zee, \emph{{Quantum Numbers of Majorana Neutrino Masses}},
  \href{http://dx.doi.org/10.1016/0550-3213(86)90475-X}{\emph{Nucl. Phys.} {\bf
  B264} (1986) 99--110}.

\bibitem{Babu:1988ki}
K.~S. Babu, \emph{{Model of `Calculable' Majorana Neutrino Masses}},
  \href{http://dx.doi.org/10.1016/0370-2693(88)91584-5}{\emph{Phys. Lett.} {\bf
  B203} (1988) 132--136}.

\bibitem{Boucenna:2014zba}
S.~M. Boucenna, S.~Morisi and J.~W.~F. Valle, \emph{{The low-scale approach to
  neutrino masses}}, \href{http://dx.doi.org/10.1155/2014/831598}{\emph{Adv.
  High Energy Phys.} {\bf 2014} (2014) 831598},
  [\href{http://arxiv.org/abs/1404.3751}{{\tt 1404.3751}}].

\bibitem{Cai:2017jrq}
Y.~Cai, J.~Herrero-Garc\'{i}a, M.~A. Schmidt, A.~Vicente and R.~R. Volkas,
  \emph{{From the trees to the forest: a review of radiative neutrino mass
  models}}, \href{http://dx.doi.org/10.3389/fphy.2017.00063}{\emph{Front.in
  Phys.} {\bf 5} (2017) 63}, [\href{http://arxiv.org/abs/1706.08524}{{\tt
  1706.08524}}].

\bibitem{Cai:2017mow}
Y.~Cai, T.~Han, T.~Li and R.~Ruiz, \emph{{Lepton Number Violation: Seesaw
  Models and Their Collider Tests}},
  \href{http://dx.doi.org/10.3389/fphy.2018.00040}{\emph{Front.in Phys.} {\bf
  6} (2018) 40}, [\href{http://arxiv.org/abs/1711.02180}{{\tt 1711.02180}}].

\bibitem{deGouvea:2007qla}
A.~de~Gouvea and J.~Jenkins, \emph{{A Survey of Lepton Number Violation Via
  Effective Operators}},
  \href{http://dx.doi.org/10.1103/PhysRevD.77.013008}{\emph{Phys. Rev.} {\bf
  D77} (2008) 013008}, [\href{http://arxiv.org/abs/0708.1344}{{\tt
  0708.1344}}].

\bibitem{ATLAS:2012hi}
{\scshape ATLAS} collaboration, G.~Aad et~al., \emph{{Search for doubly-charged
  Higgs bosons in like-sign dilepton final states at $\sqrt{s}=7$ TeV with the
  ATLAS detector}},
  \href{http://dx.doi.org/10.1140/epjc/s10052-012-2244-2}{\emph{Eur. Phys. J.
  C} {\bf 72} (2012) 2244}, [\href{http://arxiv.org/abs/1210.5070}{{\tt
  1210.5070}}].

\bibitem{ATLAS:2014kca}
{\scshape ATLAS} collaboration, G.~Aad et~al., \emph{{Search for anomalous
  production of prompt same-sign lepton pairs and pair-produced doubly charged
  Higgs bosons with $ \sqrt{s}=8 $ TeV $pp$ collisions using the ATLAS
  detector}}, \href{http://dx.doi.org/10.1007/JHEP03(2015)041}{\emph{JHEP} {\bf
  03} (2015) 041}, [\href{http://arxiv.org/abs/1412.0237}{{\tt 1412.0237}}].

\bibitem{Aaboud:2017qph}
{\scshape ATLAS} collaboration, M.~Aaboud et~al., \emph{{Search for doubly
  charged Higgs boson production in multi-lepton final states with the ATLAS
  detector using proton--proton collisions at $\sqrt{s}=13\,\text {TeV}$}},
  \href{http://dx.doi.org/10.1140/epjc/s10052-018-5661-z}{\emph{Eur. Phys. J.
  C} {\bf 78} (2018) 199}, [\href{http://arxiv.org/abs/1710.09748}{{\tt
  1710.09748}}].

\bibitem{Chatrchyan:2012ya}
{\scshape CMS} collaboration, S.~Chatrchyan et~al., \emph{{A Search for a
  Doubly-Charged Higgs Boson in $pp$ Collisions at $\sqrt{s}=7$ TeV}},
  \href{http://dx.doi.org/10.1140/epjc/s10052-012-2189-5}{\emph{Eur. Phys. J.
  C} {\bf 72} (2012) 2189}, [\href{http://arxiv.org/abs/1207.2666}{{\tt
  1207.2666}}].

\bibitem{CMS:2016cpz}
{\scshape CMS} collaboration, \emph{{Search for a doubly-charged Higgs boson
  with $\sqrt{s}=8~\mathrm{TeV}$ $pp$ collisions at the CMS experiment}}, .

\bibitem{CMS:2017pet}
{\scshape CMS} collaboration, \emph{{A search for doubly-charged Higgs boson
  production in three and four lepton final states at
  $\sqrt{s}=13~\mathrm{TeV}$}}, .

\bibitem{Farzan:2012ev}
Y.~Farzan, S.~Pascoli and M.~A. Schmidt, \emph{{Recipes and Ingredients for
  Neutrino Mass at Loop Level}},
  \href{http://dx.doi.org/10.1007/JHEP03(2013)107}{\emph{JHEP} {\bf 03} (2013)
  107}, [\href{http://arxiv.org/abs/1208.2732}{{\tt 1208.2732}}].

\bibitem{Bonnet:2012kz}
F.~Bonnet, M.~Hirsch, T.~Ota and W.~Winter, \emph{{Systematic study of the d=5
  Weinberg operator at one-loop order}},
  \href{http://dx.doi.org/10.1007/JHEP07(2012)153}{\emph{JHEP} {\bf 07} (2012)
  153}, [\href{http://arxiv.org/abs/1204.5862}{{\tt 1204.5862}}].

\bibitem{Sierra:2014rxa}
D.~Aristizabal~Sierra, A.~Degee, L.~Dorame and M.~Hirsch, \emph{{Systematic
  classification of two-loop realizations of the Weinberg operator}},
  \href{http://dx.doi.org/10.1007/JHEP03(2015)040}{\emph{JHEP} {\bf 03} (2015)
  040}, [\href{http://arxiv.org/abs/1411.7038}{{\tt 1411.7038}}].

\bibitem{Cepedello:2018rfh}
R.~Cepedello, R.~M. Fonseca and M.~Hirsch, \emph{{Systematic classification of
  three-loop realizations of the Weinberg operator}},
  \href{http://dx.doi.org/10.1007/JHEP10(2018)197}{\emph{JHEP} {\bf 10} (2018)
  197}, [\href{http://arxiv.org/abs/1807.00629}{{\tt 1807.00629}}].

\bibitem{Cepedello:2017eqf}
R.~Cepedello, M.~Hirsch and J.~Helo, \emph{{Loop neutrino masses from $d = 7$
  operator}}, \href{http://dx.doi.org/10.1007/JHEP07(2017)079}{\emph{JHEP} {\bf
  07} (2017) 079}, [\href{http://arxiv.org/abs/1705.01489}{{\tt 1705.01489}}].

\bibitem{Babu:2001ex}
K.~S. Babu and C.~N. Leung, \emph{{Classification of effective neutrino mass
  operators}},
  \href{http://dx.doi.org/10.1016/S0550-3213(01)00504-1}{\emph{Nucl. Phys.}
  {\bf B619} (2001) 667--689}, [\href{http://arxiv.org/abs/hep-ph/0106054}{{\tt
  hep-ph/0106054}}].

\bibitem{PhysRevD.87.073007}
P.~W. Angel, N.~L. Rodd and R.~R. Volkas, \emph{Origin of neutrino masses at
  the lhc: $\ensuremath{\Delta}l=2$ effective operators and their ultraviolet
  completions}, \href{http://dx.doi.org/10.1103/PhysRevD.87.073007}{\emph{Phys.
  Rev. D} {\bf 87} (Apr, 2013) 073007}.

\bibitem{Cai:2014kra}
Y.~Cai, J.~D. Clarke, M.~A. Schmidt and R.~R. Volkas, \emph{{Testing Radiative
  Neutrino Mass Models at the LHC}},
  \href{http://dx.doi.org/10.1007/JHEP02(2015)161}{\emph{JHEP} {\bf 02} (2015)
  161}, [\href{http://arxiv.org/abs/1410.0689}{{\tt 1410.0689}}].

\bibitem{Bonnet:2009ej}
F.~Bonnet, D.~Hernandez, T.~Ota and W.~Winter, \emph{{Neutrino masses from
  higher than d=5 effective operators}},
  \href{http://dx.doi.org/10.1088/1126-6708/2009/10/076}{\emph{JHEP} {\bf 10}
  (2009) 076}, [\href{http://arxiv.org/abs/0907.3143}{{\tt 0907.3143}}].

\bibitem{Anamiati:2018cuq}
G.~Anamiati, O.~Castillo-Felisola, R.~M. Fonseca, J.~C. Helo and M.~Hirsch,
  \emph{{High-dimensional neutrino masses}},
  \href{http://dx.doi.org/10.1007/JHEP12(2018)066}{\emph{JHEP} {\bf 12} (2018)
  066}, [\href{http://arxiv.org/abs/1806.07264}{{\tt 1806.07264}}].

\bibitem{Klein:2019iws}
C.~Klein, M.~Lindner and S.~Ohmer, \emph{{Minimal Radiative Neutrino Masses}},
  \href{http://dx.doi.org/10.1007/JHEP03(2019)018}{\emph{JHEP} {\bf 03} (2019)
  018}, [\href{http://arxiv.org/abs/1901.03225}{{\tt 1901.03225}}].

\bibitem{Klein:2019jgb}
C.~Klein, M.~Lindner and S.~Vogl, \emph{{Radiative neutrino masses and
  successful $SU(5)$ unification}},
  \href{http://dx.doi.org/10.1103/PhysRevD.100.075024}{\emph{Phys. Rev. D} {\bf
  100} (2019) 075024}, [\href{http://arxiv.org/abs/1907.05328}{{\tt
  1907.05328}}].

\bibitem{neutrinomass2020}
J.~Gargalionis, ``\texttt{neutrinomass}.''
  \url{https://github.com/johngarg/neutrinomass}, 2020.

\bibitem{deBlas:2017xtg}
J.~de~Blas, J.~C. Criado, M.~Perez-Victoria and J.~Santiago, \emph{{Effective
  description of general extensions of the Standard Model: the complete
  tree-level dictionary}},
  \href{http://dx.doi.org/10.1007/JHEP03(2018)109}{\emph{JHEP} {\bf 03} (2018)
  109}, [\href{http://arxiv.org/abs/1711.10391}{{\tt 1711.10391}}].

\bibitem{Dreiner:2008tw}
H.~K. Dreiner, H.~E. Haber and S.~P. Martin, \emph{{Two-component spinor
  techniques and Feynman rules for quantum field theory and supersymmetry}},
  \href{http://dx.doi.org/10.1016/j.physrep.2010.05.002}{\emph{Phys. Rept.}
  {\bf 494} (2010) 1--196}, [\href{http://arxiv.org/abs/0812.1594}{{\tt
  0812.1594}}].

\bibitem{Fonseca:2019yya}
R.~M. Fonseca, \emph{{Enumerating the operators of an effective field theory}},
   \href{http://arxiv.org/abs/1907.12584}{{\tt 1907.12584}}.

\bibitem{Arzt:1993gz}
C.~Arzt, \emph{{Reduced effective Lagrangians}},
  \href{http://dx.doi.org/10.1016/0370-2693(94)01419-D}{\emph{Phys. Lett.} {\bf
  B342} (1995) 189--195}, [\href{http://arxiv.org/abs/hep-ph/9304230}{{\tt
  hep-ph/9304230}}].

\bibitem{Lehman:2015via}
L.~Lehman and A.~Martin, \emph{{Hilbert Series for Constructing Lagrangians:
  expanding the phenomenologist's toolbox}},
  \href{http://dx.doi.org/10.1103/PhysRevD.91.105014}{\emph{Phys. Rev.} {\bf
  D91} (2015) 105014}, [\href{http://arxiv.org/abs/1503.07537}{{\tt
  1503.07537}}].

\bibitem{Henning:2015daa}
B.~Henning, X.~Lu, T.~Melia and H.~Murayama, \emph{{Hilbert series and operator
  bases with derivatives in effective field theories}},
  \href{http://dx.doi.org/10.1007/s00220-015-2518-2}{\emph{Commun. Math. Phys.}
  {\bf 347} (2016) 363--388}, [\href{http://arxiv.org/abs/1507.07240}{{\tt
  1507.07240}}].

\bibitem{Lehman:2015coa}
L.~Lehman and A.~Martin, \emph{{Low-derivative operators of the Standard Model
  effective field theory via Hilbert series methods}},
  \href{http://dx.doi.org/10.1007/JHEP02(2016)081}{\emph{JHEP} {\bf 02} (2016)
  081}, [\href{http://arxiv.org/abs/1510.00372}{{\tt 1510.00372}}].

\bibitem{Henning:2015alf}
B.~Henning, X.~Lu, T.~Melia and H.~Murayama, \emph{{2, 84, 30, 993, 560, 15456,
  11962, 261485, ...: Higher dimension operators in the SM EFT}},
  \href{http://dx.doi.org/10.1007/JHEP09(2019)019,
  10.1007/JHEP08(2017)016}{\emph{JHEP} {\bf 08} (2017) 016},
  [\href{http://arxiv.org/abs/1512.03433}{{\tt 1512.03433}}].

\bibitem{Henning:2017fpj}
B.~Henning, X.~Lu, T.~Melia and H.~Murayama, \emph{{Operator bases,
  $S$-matrices, and their partition functions}},
  \href{http://dx.doi.org/10.1007/JHEP10(2017)199}{\emph{JHEP} {\bf 10} (2017)
  199}, [\href{http://arxiv.org/abs/1706.08520}{{\tt 1706.08520}}].

\bibitem{Lehman:2014jma}
L.~Lehman, \emph{{Extending the Standard Model Effective Field Theory with the
  Complete Set of Dimension-7 Operators}},
  \href{http://dx.doi.org/10.1103/PhysRevD.90.125023}{\emph{Phys. Rev.} {\bf
  D90} (2014) 125023}, [\href{http://arxiv.org/abs/1410.4193}{{\tt
  1410.4193}}].

\bibitem{Kobach:2016ami}
A.~Kobach, \emph{{Baryon Number, Lepton Number, and Operator Dimension in the
  Standard Model}},
  \href{http://dx.doi.org/10.1016/j.physletb.2016.05.050}{\emph{Phys. Lett.}
  {\bf B758} (2016) 455--457}, [\href{http://arxiv.org/abs/1604.05726}{{\tt
  1604.05726}}].

\bibitem{Gargalionis:2019drk}
J.~Gargalionis, I.~Popa-Mateiu and R.~R. Volkas, \emph{{Radiative neutrino mass
  model from a mass dimension-11 $\Delta L =2 $ effective operator}},
  \href{http://arxiv.org/abs/1912.12386}{{\tt 1912.12386}}.

\bibitem{deGouvea:2019xzm}
A.~De~Gouvêa, W.-C. Huang, J.~König and M.~Sen, \emph{{Accessible
  Lepton-Number-Violating Models and Negligible Neutrino Masses}},
  \href{http://dx.doi.org/10.1103/PhysRevD.100.075033}{\emph{Phys. Rev. D} {\bf
  100} (2019) 075033}, [\href{http://arxiv.org/abs/1907.02541}{{\tt
  1907.02541}}].

\bibitem{Li:2020xlh}
H.-L. Li, Z.~Ren, M.-L. Xiao, J.-H. Yu and Y.-H. Zheng, \emph{{Complete Set of
  Dimension-9 Operators in the Standard Model Effective Field Theory}},
  \href{http://arxiv.org/abs/2007.07899}{{\tt 2007.07899}}.

\bibitem{Cata:2019wbu}
O.~Catà and T.~Mannel, \emph{{Linking lepton number violation with $B$
  anomalies}},  \href{http://arxiv.org/abs/1903.01799}{{\tt 1903.01799}}.

\bibitem{Gripaios:2018zrz}
B.~Gripaios and D.~Sutherland, \emph{{DEFT: A program for operators in EFT}},
  \href{http://dx.doi.org/10.1007/JHEP01(2019)128}{\emph{JHEP} {\bf 01} (2019)
  128}, [\href{http://arxiv.org/abs/1807.07546}{{\tt 1807.07546}}].

\bibitem{Criado:2019ugp}
J.~C. Criado, \emph{{BasisGen: automatic generation of operator bases}},
  \href{http://dx.doi.org/10.1140/epjc/s10052-019-6769-5}{\emph{Eur. Phys. J.}
  {\bf C79} (2019) 256}, [\href{http://arxiv.org/abs/1901.03501}{{\tt
  1901.03501}}].

\bibitem{Fonseca:2011sy}
R.~M. Fonseca, \emph{{Calculating the renormalisation group equations of a SUSY
  model with Susyno}},
  \href{http://dx.doi.org/10.1016/j.cpc.2012.05.017}{\emph{Comput. Phys.
  Commun.} {\bf 183} (2012) 2298--2306},
  [\href{http://arxiv.org/abs/1106.5016}{{\tt 1106.5016}}].

\bibitem{Fonseca:2017lem}
R.~M. Fonseca, \emph{{The Sym2Int program: going from symmetries to
  interactions}},
  \href{http://dx.doi.org/10.1088/1742-6596/873/1/012045}{\emph{J. Phys. Conf.
  Ser.} {\bf 873} (2017) 012045}, [\href{http://arxiv.org/abs/1703.05221}{{\tt
  1703.05221}}].

\bibitem{Banerjee:2020bym}
U.~Banerjee, J.~Chakrabortty, S.~Prakash and S.~U. Rahaman, \emph{{Characters
  and Group Invariant Polynomials of (Super)fields: Road to ''Lagrangian''}},
  \href{http://arxiv.org/abs/2004.12830}{{\tt 2004.12830}}.

\bibitem{Arzt:1994gp}
C.~Arzt, M.~B. Einhorn and J.~Wudka, \emph{{Patterns of deviation from the
  standard model}},
  \href{http://dx.doi.org/10.1016/0550-3213(94)00336-D}{\emph{Nucl. Phys.} {\bf
  B433} (1995) 41--66}, [\href{http://arxiv.org/abs/hep-ph/9405214}{{\tt
  hep-ph/9405214}}].

\bibitem{Einhorn:2013kja}
M.~B. Einhorn and J.~Wudka, \emph{{The Bases of Effective Field Theories}},
  \href{http://dx.doi.org/10.1016/j.nuclphysb.2013.08.023}{\emph{Nucl. Phys.}
  {\bf B876} (2013) 556--574}, [\href{http://arxiv.org/abs/1307.0478}{{\tt
  1307.0478}}].

\bibitem{Angel:2012ug}
P.~W. Angel, N.~L. Rodd and R.~R. Volkas, \emph{{Origin of neutrino masses at
  the LHC: $\Delta L = 2$ effective operators and their ultraviolet
  completions}},
  \href{http://dx.doi.org/10.1103/PhysRevD.87.073007}{\emph{Phys. Rev.} {\bf
  D87} (2013) 073007}, [\href{http://arxiv.org/abs/1212.6111}{{\tt
  1212.6111}}].

\bibitem{Henning:2014wua}
B.~Henning, X.~Lu and H.~Murayama, \emph{{How to use the Standard Model
  effective field theory}},
  \href{http://dx.doi.org/10.1007/JHEP01(2016)023}{\emph{JHEP} {\bf 01} (2016)
  023}, [\href{http://arxiv.org/abs/1412.1837}{{\tt 1412.1837}}].

\bibitem{Herrero-Garcia:2019czj}
J.~Herrero-Garc\'\i{}a and M.~A. Schmidt, \emph{{Neutrino mass models: New
  classification and model-independent upper limits on their scale}},
  \href{http://dx.doi.org/10.1140/epjc/s10052-019-7465-1}{\emph{Eur. Phys. J.
  C} {\bf 79} (2019) 938}, [\href{http://arxiv.org/abs/1903.10552}{{\tt
  1903.10552}}].

\bibitem{Banerjee:2020jun}
U.~Banerjee, J.~Chakrabortty, S.~Prakash, S.~U. Rahaman and M.~Spannowsky,
  \emph{{Effective Operator Bases for Beyond Standard Model Scenarios: An EFT
  compendium for discoveries}},  \href{http://arxiv.org/abs/2008.11512}{{\tt
  2008.11512}}.

\bibitem{delAguila:2011gr}
F.~del Aguila, A.~Aparici, S.~Bhattacharya, A.~Santamaria and J.~Wudka,
  \emph{{A realistic model of neutrino masses with a large neutrinoless double
  beta decay rate}},
  \href{http://dx.doi.org/10.1007/JHEP05(2012)133}{\emph{JHEP} {\bf 05} (2012)
  133}, [\href{http://arxiv.org/abs/1111.6960}{{\tt 1111.6960}}].

\bibitem{delAguila:2012nu}
F.~del Aguila, A.~Aparici, S.~Bhattacharya, A.~Santamaria and J.~Wudka,
  \emph{{Effective Lagrangian approach to neutrinoless double beta decay and
  neutrino masses}},
  \href{http://dx.doi.org/10.1007/JHEP06(2012)146}{\emph{JHEP} {\bf 06} (2012)
  146}, [\href{http://arxiv.org/abs/1204.5986}{{\tt 1204.5986}}].

\bibitem{Herrero-Garcia:2016uab}
J.~Herrero-Garcia, N.~Rius and A.~Santamaria, \emph{{Higgs lepton flavour
  violation: UV completions and connection to neutrino masses}},
  \href{http://dx.doi.org/10.1007/JHEP11(2016)084}{\emph{JHEP} {\bf 11} (2016)
  084}, [\href{http://arxiv.org/abs/1605.06091}{{\tt 1605.06091}}].

\bibitem{Craig:2019wmo}
N.~Craig, M.~Jiang, Y.-Y. Li and D.~Sutherland, \emph{{Loops and Trees in
  Generic EFTs}},  \href{http://arxiv.org/abs/2001.00017}{{\tt 2001.00017}}.

\bibitem{Law:2013dya}
S.~S. Law and K.~L. McDonald, \emph{{The simplest models of radiative neutrino
  mass}}, \href{http://dx.doi.org/10.1142/S0217751X1450064X}{\emph{Int. J. Mod.
  Phys. A} {\bf 29} (2014) 1450064},
  [\href{http://arxiv.org/abs/1303.6384}{{\tt 1303.6384}}].

\bibitem{Wolfenstein:1980sy}
L.~Wolfenstein, \emph{{A Theoretical Pattern for Neutrino Oscillations}},
  \href{http://dx.doi.org/10.1016/0550-3213(80)90004-8}{\emph{Nucl. Phys. B}
  {\bf 175} (1980) 93--96}.

\bibitem{He:2003ih}
X.-G. He, \emph{{Is the Zee model neutrino mass matrix ruled out?}},
  \href{http://dx.doi.org/10.1140/epjc/s2004-01669-8}{\emph{Eur. Phys. J. C}
  {\bf 34} (2004) 371--376}, [\href{http://arxiv.org/abs/hep-ph/0307172}{{\tt
  hep-ph/0307172}}].

\bibitem{10.7717/peerj-cs.103}
A.~Meurer, C.~P. Smith, M.~Paprocki, O.~\v{C}ert\'{i}k, S.~B. Kirpichev,
  M.~Rocklin et~al., \emph{Sympy: symbolic computing in python},
  \href{http://dx.doi.org/10.7717/peerj-cs.103}{\emph{PeerJ Computer Science}
  {\bf 3} (Jan., 2017) e103}.

\bibitem{butler1991}
G.~Butler, ed., \emph{Fundamental Algorithms for Permutation Groups}.
\newblock Springer Berlin Heidelberg, 1991.
\newblock 10.1007/3-540-54955-2.

\bibitem{MANSSUR_2002}
L.~R.~U. Manssur, R.~Portugal and B.~F. Svaiter, \emph{Group-theoretic approach
  for symbolic tensor manipulation},
  \href{http://dx.doi.org/10.1142/s0129183102004571}{\emph{International
  Journal of Modern Physics C} {\bf 13} (Sep, 2002) 859–879}.

\bibitem{Hahn:2000kx}
T.~Hahn, \emph{{Generating Feynman diagrams and amplitudes with FeynArts 3}},
  \href{http://dx.doi.org/10.1016/S0010-4655(01)00290-9}{\emph{Comput. Phys.
  Commun.} {\bf 140} (2001) 418--431},
  [\href{http://arxiv.org/abs/hep-ph/0012260}{{\tt hep-ph/0012260}}].

\bibitem{igraph2006}
G.~Csardi and T.~Nepusz, \emph{The igraph software package for complex network
  research}, {\emph{InterJournal} {\bf Complex Systems} (2006) 1695}.

\bibitem{the_igraph_core_team_2020_3774399}
T.~igraph Core~Team, \emph{igraph},  Apr., 2020.
\newblock 10.5281/zenodo.3774399.

\bibitem{szabolcs_horvat_2020_3739056}
S.~Horvát, \emph{Igraph/m},  Apr., 2020.
\newblock 10.5281/zenodo.3739056.

\bibitem{SciPyProceedings_11}
A.~A. Hagberg, D.~A. Schult and P.~J. Swart, \emph{Exploring network structure,
  dynamics, and function using networkx},  in \emph{Proceedings of the 7th
  Python in Science Conference} (G.~Varoquaux, T.~Vaught and J.~Millman, eds.),
  (Pasadena, CA USA), pp.~11 -- 15, 2008.

\bibitem{Criado:2017khh}
J.~C. Criado, \emph{{MatchingTools: a Python library for symbolic effective
  field theory calculations}},
  \href{http://dx.doi.org/10.1016/j.cpc.2018.02.016}{\emph{Comput. Phys.
  Commun.} {\bf 227} (2018) 42--50},
  [\href{http://arxiv.org/abs/1710.06445}{{\tt 1710.06445}}].

\bibitem{Bakshi:2018ics}
S.~Das~Bakshi, J.~Chakrabortty and S.~K. Patra, \emph{{CoDEx: Wilson
  coefficient calculator connecting SMEFT to UV theory}},
  \href{http://dx.doi.org/10.1140/epjc/s10052-018-6444-2}{\emph{Eur. Phys. J.}
  {\bf C79} (2019) 21}, [\href{http://arxiv.org/abs/1808.04403}{{\tt
  1808.04403}}].

\bibitem{krebber2018}
M.~Krebber and H.~Barthels, \emph{{M}atch{P}y: {P}attern {M}atching in
  {P}ython}, \href{http://dx.doi.org/10.21105/joss.00670}{\emph{Journal of Open
  Source Software} {\bf 3} (June, 2018) 2}.

\bibitem{krebber2017nonlinear}
M.~Krebber, \emph{Non-linear associative-commutative many-to-one pattern
  matching with sequence variables},  2017.

\bibitem{nufitweb}
``Nufit 5.0 (2020).'' \url{www.nu-fit.org}.

\bibitem{Esteban:2018azc}
I.~Esteban, M.~Gonzalez-Garcia, A.~Hernandez-Cabezudo, M.~Maltoni and
  T.~Schwetz, \emph{{Global analysis of three-flavour neutrino oscillations:
  synergies and tensions in the determination of $\theta_{23}$, $\delta_{CP}$,
  and the mass ordering}},
  \href{http://dx.doi.org/10.1007/JHEP01(2019)106}{\emph{JHEP} {\bf 01} (2019)
  106}, [\href{http://arxiv.org/abs/1811.05487}{{\tt 1811.05487}}].

\bibitem{Babu:2010vp}
K.~S. Babu and J.~Julio, \emph{{Two-Loop Neutrino Mass Generation through
  Leptoquarks}},
  \href{http://dx.doi.org/10.1016/j.nuclphysb.2010.07.022}{\emph{Nucl. Phys.}
  {\bf B841} (2010) 130--156}, [\href{http://arxiv.org/abs/1006.1092}{{\tt
  1006.1092}}].

\bibitem{Babu:2019mfe}
K.~Babu, P.~B. Dev, S.~Jana and A.~Thapa, \emph{{Non-Standard Interactions in
  Radiative Neutrino Mass Models}},
  \href{http://dx.doi.org/10.1007/JHEP03(2020)006}{\emph{JHEP} {\bf 03} (2020)
  006}, [\href{http://arxiv.org/abs/1907.09498}{{\tt 1907.09498}}].

\bibitem{Duerr:2011zd}
M.~Duerr, M.~Lindner and A.~Merle, \emph{{On the Quantitative Impact of the
  Schechter-Valle Theorem}},
  \href{http://dx.doi.org/10.1007/JHEP06(2011)091}{\emph{JHEP} {\bf 06} (2011)
  091}, [\href{http://arxiv.org/abs/1105.0901}{{\tt 1105.0901}}].

\bibitem{Babu:2009aq}
K.~Babu, S.~Nandi and Z.~Tavartkiladze, \emph{{New Mechanism for Neutrino Mass
  Generation and Triply Charged Higgs Bosons at the LHC}},
  \href{http://dx.doi.org/10.1103/PhysRevD.80.071702}{\emph{Phys. Rev. D} {\bf
  80} (2009) 071702}, [\href{http://arxiv.org/abs/0905.2710}{{\tt 0905.2710}}].

\bibitem{Angel:2013hla}
P.~W. Angel, Y.~Cai, N.~L. Rodd, M.~A. Schmidt and R.~R. Volkas,
  \emph{{Testable two-loop radiative neutrino mass model based on an
  $LLQd^cQd^c$ effective operator}},
  \href{http://dx.doi.org/10.1007/JHEP11(2014)092,
  10.1007/JHEP10(2013)118}{\emph{JHEP} {\bf 10} (2013) 118},
  [\href{http://arxiv.org/abs/1308.0463}{{\tt 1308.0463}}].

\bibitem{Popov:2019tyc}
O.~Popov, M.~A. Schmidt and G.~White, \emph{{$R_2$ as a single leptoquark
  solution to $R_{D^{(*)}}$ and $R_{K^{(*)}}$}},
  \href{http://dx.doi.org/10.1103/PhysRevD.100.035028}{\emph{Phys. Rev. D} {\bf
  100} (2019) 035028}, [\href{http://arxiv.org/abs/1905.06339}{{\tt
  1905.06339}}].

\bibitem{Babu:2020hun}
K.~Babu, P.~B. Dev, S.~Jana and A.~Thapa, \emph{{Unified Framework for
  $B$-Anomalies, Muon $g-2$, and Neutrino Masses}},
  \href{http://arxiv.org/abs/2009.01771}{{\tt 2009.01771}}.

\bibitem{gargalionis_john_2020_4054618}
J.~Gargalionis and R.~R. Volkas, \emph{{Database of tree-level completions of
  lepton- number-violating effective operators}},  Sept., 2020.
\newblock 10.5281/zenodo.4054618.

\bibitem{Dorsner:2016wpm}
I.~Dor\v{s}ner, S.~Fajfer, A.~Greljo, J.~F. Kamenik and N.~Ko\v{s}nik,
  \emph{{Physics of leptoquarks in precision experiments and at particle
  colliders}},
  \href{http://dx.doi.org/10.1016/j.physrep.2016.06.001}{\emph{Phys. Rept.}
  {\bf 641} (2016) 1--68}, [\href{http://arxiv.org/abs/1603.04993}{{\tt
  1603.04993}}].

\bibitem{Sakaki:2013bfa}
Y.~Sakaki, M.~Tanaka, A.~Tayduganov and R.~Watanabe, \emph{{Testing leptoquark
  models in $\bar B \to D^{(*)} \tau \bar\nu$}},
  \href{http://dx.doi.org/10.1103/PhysRevD.88.094012}{\emph{Phys. Rev. D} {\bf
  88} (2013) 094012}, [\href{http://arxiv.org/abs/1309.0301}{{\tt 1309.0301}}].

\bibitem{Angelescu:2018tyl}
A.~Angelescu, D.~Be\v{c}irevi\'c, D.~Faroughy and O.~Sumensari, \emph{{Closing
  the window on single leptoquark solutions to the $B$-physics anomalies}},
  \href{http://dx.doi.org/10.1007/JHEP10(2018)183}{\emph{JHEP} {\bf 10} (2018)
  183}, [\href{http://arxiv.org/abs/1808.08179}{{\tt 1808.08179}}].

\bibitem{Becirevic:2018uab}
D.~Be\v{c}irevi\'c, B.~Panes, O.~Sumensari and R.~Zukanovich~Funchal,
  \emph{{Seeking leptoquarks in IceCube}},
  \href{http://dx.doi.org/10.1007/JHEP06(2018)032}{\emph{JHEP} {\bf 06} (2018)
  032}, [\href{http://arxiv.org/abs/1803.10112}{{\tt 1803.10112}}].

\bibitem{Becirevic:2017jtw}
D.~Be\v{c}irevi\'c and O.~Sumensari, \emph{{A leptoquark model to accommodate
  $R_K^\mathrm{exp} < R_K^\mathrm{SM}$ and $R_{K^\ast}^\mathrm{exp} <
  R_{K^\ast}^\mathrm{SM}$}},
  \href{http://dx.doi.org/10.1007/JHEP08(2017)104}{\emph{JHEP} {\bf 08} (2017)
  104}, [\href{http://arxiv.org/abs/1704.05835}{{\tt 1704.05835}}].

\bibitem{Bigaran:2020jil}
I.~Bigaran and R.~R. Volkas, \emph{{Getting chirality right: single scalar
  leptoquark solution/s to the $(g-2)_{e,\mu}$ puzzle}},
  \href{http://arxiv.org/abs/2002.12544}{{\tt 2002.12544}}.

\bibitem{Dorsner:2020aaz}
I.~Dor\v{s}ner, S.~Fajfer and S.~Saad, \emph{{$\mu \to e \gamma$ selecting
  scalar leptoquark solutions for the $(g-2)_{e,\mu}$ puzzles}},
  \href{http://arxiv.org/abs/2006.11624}{{\tt 2006.11624}}.

\bibitem{Hiller:2014yaa}
G.~Hiller and M.~Schmaltz, \emph{{$R_K$ and future $b \to s \ell \ell$ physics
  beyond the standard model opportunities}},
  \href{http://dx.doi.org/10.1103/PhysRevD.90.054014}{\emph{Phys. Rev.} {\bf
  D90} (2014) 054014}, [\href{http://arxiv.org/abs/1408.1627}{{\tt
  1408.1627}}].

\bibitem{Gripaios:2014tna}
B.~Gripaios, M.~Nardecchia and S.~A. Renner, \emph{{Composite leptoquarks and
  anomalies in $B$-meson decays}},
  \href{http://dx.doi.org/10.1007/JHEP05(2015)006}{\emph{JHEP} {\bf 05} (2015)
  006}, [\href{http://arxiv.org/abs/1412.1791}{{\tt 1412.1791}}].

\bibitem{Hiller:2017bzc}
G.~Hiller and I.~Nisandzic, \emph{{$R_K$ and $R_{K^{\ast}}$ beyond the standard
  model}}, \href{http://dx.doi.org/10.1103/PhysRevD.96.035003}{\emph{Phys. Rev.
  D} {\bf 96} (2017) 035003}, [\href{http://arxiv.org/abs/1704.05444}{{\tt
  1704.05444}}].

\bibitem{Dorsner:2017ufx}
I.~Doršner, S.~Fajfer, D.~A. Faroughy and N.~Košnik, \emph{{Saga of the two
  GUT leptoquarks in flavor universality and collider searches}},
  \href{http://arxiv.org/abs/1706.07779}{{\tt 1706.07779}}.

\bibitem{Arnold:2012sd}
J.~M. Arnold, B.~Fornal and M.~B. Wise, \emph{{Simplified models with baryon
  number violation but no proton decay}},
  \href{http://dx.doi.org/10.1103/PhysRevD.87.075004}{\emph{Phys. Rev. D} {\bf
  87} (2013) 075004}, [\href{http://arxiv.org/abs/1212.4556}{{\tt 1212.4556}}].

\bibitem{Gustafsson:2012vj}
M.~Gustafsson, J.~M. No and M.~A. Rivera, \emph{{Predictive Model for
  Radiatively Induced Neutrino Masses and Mixings with Dark Matter}},
  \href{http://dx.doi.org/10.1103/PhysRevLett.110.211802}{\emph{Phys. Rev.
  Lett.} {\bf 110} (2013) 211802}, [\href{http://arxiv.org/abs/1212.4806}{{\tt
  1212.4806}}].

\bibitem{Lees:2012xj}
{\scshape BaBar} collaboration, J.~Lees et~al., \emph{{Evidence for an excess
  of $\bar{B} \to D^{(*)} \tau^-\bar{\nu}_\tau$ decays}},
  \href{http://dx.doi.org/10.1103/PhysRevLett.109.101802}{\emph{Phys. Rev.
  Lett.} {\bf 109} (2012) 101802}, [\href{http://arxiv.org/abs/1205.5442}{{\tt
  1205.5442}}].

\bibitem{Lees:2013uzd}
{\scshape BaBar} collaboration, J.~Lees et~al., \emph{{Measurement of an Excess
  of $\bar{B} \to D^{(*)}\tau^- \bar{\nu}_\tau$ Decays and Implications for
  Charged Higgs Bosons}},
  \href{http://dx.doi.org/10.1103/PhysRevD.88.072012}{\emph{Phys. Rev. D} {\bf
  88} (2013) 072012}, [\href{http://arxiv.org/abs/1303.0571}{{\tt 1303.0571}}].

\bibitem{Huschle:2015rga}
{\scshape Belle} collaboration, M.~Huschle et~al., \emph{{Measurement of the
  branching ratio of $\bar{B} \to D^{(\ast)} \tau^- \bar{\nu}_\tau$ relative to
  $\bar{B} \to D^{(\ast)} \ell^- \bar{\nu}_\ell$ decays with hadronic tagging
  at Belle}}, \href{http://dx.doi.org/10.1103/PhysRevD.92.072014}{\emph{Phys.
  Rev.} {\bf D92} (2015) 072014}, [\href{http://arxiv.org/abs/1507.03233}{{\tt
  1507.03233}}].

\bibitem{Hirose:2016wfn}
{\scshape Belle} collaboration, S.~Hirose et~al., \emph{{Measurement of the
  $\tau$ lepton polarization and $R(D^*)$ in the decay $\bar{B} \to D^* \tau^-
  \bar{\nu}_\tau$}},
  \href{http://dx.doi.org/10.1103/PhysRevLett.118.211801}{\emph{Phys. Rev.
  Lett.} {\bf 118} (2017) 211801}, [\href{http://arxiv.org/abs/1612.00529}{{\tt
  1612.00529}}].

\bibitem{Abdesselam:2016cgx}
{\scshape Belle} collaboration, A.~Abdesselam et~al., \emph{{Measurement of the
  branching ratio of $\bar{B}^0 \rightarrow D^{*+} \tau^- \bar{\nu}_{\tau}$
  relative to $\bar{B}^0 \rightarrow D^{*+} \ell^- \bar{\nu}_{\ell}$ decays
  with a semileptonic tagging method}},  in \emph{{51st Rencontres de Moriond
  on EW Interactions and Unified Theories}}, 3, 2016.
\newblock \href{http://arxiv.org/abs/1603.06711}{{\tt 1603.06711}}.

\bibitem{Aaij:2017tyk}
{\scshape LHCb} collaboration, R.~Aaij et~al., \emph{{Measurement of the ratio
  of branching fractions
  $\mathcal{B}(B_c^+\,\to\,J/\psi\tau^+\nu_\tau)$/$\mathcal{B}(B_c^+\,\to\,J/\psi\mu^+\nu_\mu)$}},
  \href{http://dx.doi.org/10.1103/PhysRevLett.120.121801}{\emph{Phys. Rev.
  Lett.} {\bf 120} (2018) 121801}, [\href{http://arxiv.org/abs/1711.05623}{{\tt
  1711.05623}}].

\bibitem{Aaij:2017uff}
{\scshape LHCb} collaboration, R.~Aaij et~al., \emph{{Measurement of the ratio
  of the $B^0 \to D^{*-} \tau^+ \nu_{\tau}$ and $B^0 \to D^{*-} \mu^+
  \nu_{\mu}$ branching fractions using three-prong $\tau$-lepton decays}},
  \href{http://dx.doi.org/10.1103/PhysRevLett.120.171802}{\emph{Phys. Rev.
  Lett.} {\bf 120} (2018) 171802}, [\href{http://arxiv.org/abs/1708.08856}{{\tt
  1708.08856}}].

\bibitem{Amhis:2019ckw}
{\scshape HFLAV} collaboration, Y.~S. Amhis et~al., \emph{{Averages of
  $b$-hadron, $c$-hadron, and $\tau$-lepton properties as of 2018}},
  \href{http://arxiv.org/abs/1909.12524}{{\tt 1909.12524}}.

\bibitem{Aaij:2019wad}
{\scshape LHCb} collaboration, R.~Aaij et~al., \emph{{Search for
  lepton-universality violation in $B^+\to K^+\ell^+\ell^-$ decays}},
  \href{http://dx.doi.org/10.1103/PhysRevLett.122.191801}{\emph{Phys. Rev.
  Lett.} {\bf 122} (2019) 191801}, [\href{http://arxiv.org/abs/1903.09252}{{\tt
  1903.09252}}].

\bibitem{Aaij:2017vbb}
{\scshape LHCb} collaboration, R.~Aaij et~al., \emph{{Test of lepton
  universality with $B^{0} \rightarrow K^{*0}\ell^{+}\ell^{-}$ decays}},
  \href{http://dx.doi.org/10.1007/JHEP08(2017)055}{\emph{JHEP} {\bf 08} (2017)
  055}, [\href{http://arxiv.org/abs/1705.05802}{{\tt 1705.05802}}].

\bibitem{Aaij:2015oid}
{\scshape LHCb} collaboration, R.~Aaij et~al., \emph{{Angular analysis of the
  $B^{0} \to K^{*0} \mu^{+} \mu^{-}$ decay using 3 fb$^{-1}$ of integrated
  luminosity}}, \href{http://dx.doi.org/10.1007/JHEP02(2016)104}{\emph{JHEP}
  {\bf 02} (2016) 104}, [\href{http://arxiv.org/abs/1512.04442}{{\tt
  1512.04442}}].

\bibitem{ATLAS-CONF-2017-023}
{\scshape ATLAS} collaboration, \emph{{Angular analysis of $B^0_d \to
  K^{*}\mu^+\mu^-$ decays in $pp$ collisions at $\sqrt{s}= 8$ TeV with the
  ATLAS detector}},  Tech. Rep. ATLAS-CONF-2017-023, CERN, Geneva, Apr, 2017.

\bibitem{CMS-PAS-BPH-15-008}
{\scshape CMS} collaboration, \emph{{Measurement of the $P_1$ and $P_5'$
  angular parameters of the decay $\mathrm{B}^0 \to \mathrm{K}^{*0} \mu^+
  \mu^-$ in proton-proton collisions at $\sqrt{s}=8~\mathrm{TeV}$}},  Tech.
  Rep. CMS-PAS-BPH-15-008, CERN, Geneva, 2017.

\bibitem{Khachatryan:2015isa}
{\scshape CMS} collaboration, V.~Khachatryan et~al., \emph{{Angular analysis of
  the decay $B^0 \to K^{*0} \mu^+ \mu^-$ from pp collisions at $\sqrt s = 8$
  TeV}}, \href{http://dx.doi.org/10.1016/j.physletb.2015.12.020}{\emph{Phys.
  Lett. B} {\bf 753} (2016) 424--448},
  [\href{http://arxiv.org/abs/1507.08126}{{\tt 1507.08126}}].

\bibitem{Aaij:2014pli}
{\scshape LHCb} collaboration, R.~Aaij et~al., \emph{{Differential branching
  fractions and isospin asymmetries of $B \to K^{(*)} \mu^+ \mu^-$ decays}},
  \href{http://dx.doi.org/10.1007/JHEP06(2014)133}{\emph{JHEP} {\bf 06} (2014)
  133}, [\href{http://arxiv.org/abs/1403.8044}{{\tt 1403.8044}}].

\bibitem{Aaij:2015esa}
{\scshape LHCb} collaboration, R.~Aaij et~al., \emph{{Angular analysis and
  differential branching fraction of the decay $B^0_s\to\phi\mu^+\mu^-$}},
  \href{http://dx.doi.org/10.1007/JHEP09(2015)179}{\emph{JHEP} {\bf 09} (2015)
  179}, [\href{http://arxiv.org/abs/1506.08777}{{\tt 1506.08777}}].

\bibitem{Aebischer:2017ugx}
J.~Aebischer et~al., \emph{{WCxf: an exchange format for Wilson coefficients
  beyond the Standard Model}},
  \href{http://dx.doi.org/10.1016/j.cpc.2018.05.022}{\emph{Comput. Phys.
  Commun.} {\bf 232} (2018) 71--83},
  [\href{http://arxiv.org/abs/1712.05298}{{\tt 1712.05298}}].

\bibitem{Straub:2018kue}
D.~M. Straub, \emph{{flavio: a Python package for flavour and precision
  phenomenology in the Standard Model and beyond}},
  \href{http://arxiv.org/abs/1810.08132}{{\tt 1810.08132}}.

\bibitem{Cai:2017wry}
Y.~Cai, J.~Gargalionis, M.~A. Schmidt and R.~R. Volkas, \emph{{Reconsidering
  the One Leptoquark solution: flavor anomalies and neutrino mass}},
  \href{http://dx.doi.org/10.1007/JHEP10(2017)047}{\emph{JHEP} {\bf 10} (2017)
  047}, [\href{http://arxiv.org/abs/1704.05849}{{\tt 1704.05849}}].

\bibitem{Aebischer:2019mlg}
J.~Aebischer, W.~Altmannshofer, D.~Guadagnoli, M.~Reboud, P.~Stangl and D.~M.
  Straub, \emph{{$B$-decay discrepancies after Moriond 2019}},
  \href{http://dx.doi.org/10.1140/epjc/s10052-020-7817-x}{\emph{Eur. Phys. J.
  C} {\bf 80} (2020) 252}, [\href{http://arxiv.org/abs/1903.10434}{{\tt
  1903.10434}}].

\bibitem{Ellis:2016jkw}
J.~Ellis, \emph{{TikZ-Feynman: Feynman diagrams with TikZ}},
  \href{http://dx.doi.org/10.1016/j.cpc.2016.08.019}{\emph{Comput. Phys.
  Commun.} {\bf 210} (2017) 103--123},
  [\href{http://arxiv.org/abs/1601.05437}{{\tt 1601.05437}}].

\end{thebibliography}\endgroup

\end{document}